\documentclass[twocolumn,prd,preprintnumbers,superscriptaddress,landscape,nofootinbib]{revtex4}
\usepackage[utf8]{inputenc}
\usepackage{amsmath}
\usepackage{amsfonts}
\usepackage{amsthm}
\usepackage{float}
\usepackage{braket}
\usepackage{graphicx}
\usepackage{url}
\usepackage{csquotes}
\usepackage[colorlinks=true, pdfstartview=FitV, linkcolor=red, citecolor=blue, urlcolor=blue]{hyperref}
\usepackage{slashed}
\usepackage[normalem]{ulem}
\usepackage{bbold}
\usepackage{comment}
\usepackage{caption}

\usepackage{subfig}
\graphicspath{{./plots/}}

\newcommand{\be}{\begin{equation}}      
\newcommand{\ee}{\end{equation}}      
\newcommand{\bea}{\begin{eqnarray}}      
\newcommand{\eea}{\end{eqnarray}}

\newcommand{\tr}{\mathrm{tr}}
\newcommand{\e}{\mathrm{e}}

\newcommand{\Tr}{\mathrm{Tr}}

\begin{document}


\title{Robustness of quantum correlation in quantum energy teleportation}

\author{Kazuki Ikeda}
\email[]{kazuki7131@gmail.com}
\affiliation{Co-design Center for Quantum Advantage (C2QA), Stony Brook University, USA}
\affiliation{Center For Nuclear Theory, Department of Physics and Astronomy, Stony Brook University, USA}

\author{Adam Lowe}
\email[]{a.lowe3@aston.ac.uk}
\affiliation{College of Engineering and Physical Sciences, Department of Applied Mathematics and Data Science, Aston University, Birmingham B4 7ET, United Kingdom}




\bibliographystyle{unsrt}

\begin{abstract}
We present the evolution of quantum correlation in the quantum energy teleportation (QET) protocol using quantum discord, instead of the traditionally used entanglement entropy. In the QET protocol, where local observations and conditional operations are repeated, quantum correlations become nontrivial because of the statistical creation of mixed states. In this paper, we use quantum discord as a measure of quantum correlation in mixed states and investigate its relationship to teleported energy and phase transitions. During the process of Alice and Bob performing QET, one would expect that the entanglement between Alice and Bob is completely broken by Alice's measurement of the quantum state, and thus the quantum correlation disappears. However, contrary to this expectation, it is shown using quantum discord that the quantum correlation does not disappear during the entire process of QET. To demonstrate the robustness of the quantum correlation in QET at various phase structures, we perform the numerical analysis using several benchmark models including the Nambu-Jona-Lasino (NJL) model with both the chiral chemical potential and the chemical potential, which are useful to study the phase structures mimicking the chiral 
imbalance between left- and right- quarks coupled to the chirality density operator. In all cases we studied, the quantum discord behaved as an order parameter of the phase transition.
\end{abstract}
\maketitle

\section{Introduction}
Elucidating the nature of quantum correlations in quantum many-body systems is not only of paramount importance for understanding the physics of quantum many-body systems, but is also useful for the development of quantum communication technology and quantum devices~\cite{RevModPhys.81.865,L_Henderson_2001,PhysRevLett.98.140402,PhysRevLett.80.2245,PhysRevLett.110.240402,PhysRevLett.112.210401,PhysRevLett.115.020403,PhysRevX.10.041012}. Quantum energy teleportation (QET) is one of the best protocols to investigate this problem~\cite{Hotta_2010,HOTTA20085671}. This is because QET is the simplest protocol between two remote players performing local operation and classical communication (LOCC), including the measurement of entangled ground state, and can be realised in a wide range of models, including various artificial quantum network models, quantum field theories and condensed matter systems with impurities~\cite{2023arXiv230111884I,Ikeda:2023uni,Ikeda:2023xmf,2023arXiv230608242I,Ikeda:2023ljh,PhysRevA.82.042329,PhysRevA.84.032336}.

It should be emphasised that QET is also a useful protocol for investigating the time evolution of quantum correlations. Although the entanglement in the ground state is important in QET, the time evolution of quantum correlations is quite nontrivial because LOCC, especially the conditional operations according to local measurement, allows the statistical creation of a mixed state from the initial pure state. Here it is important that LOCC dose not increase quantum entanglement, but entanglement entropy is no longer a proper measure of quantum correlation in a mixed state. It is also non-trivial whether there are non-zero quantum correlations in the quantum mixed states created by QET, since the entanglement is lost by the measurement of the ground state, which happens when the energy is injected to the system by a supplier.

To address this question, we employ the quantum discord~\cite{PhysRevLett.88.017901,PhysRevA.54.3824,PhysRevA.80.024103}, which has been proposed as a measure of general quantum states including both pure states and mixed states. The quantum discord in any pure state is equivalent to the entanglement entropy. Given most realistic states are mixtures, this implies that quantum discord is an interesting tool in quantum information science.

Qualitatively, quantum discord takes the difference in quantum mutual information, before and after measurement, where the resulting effect of the measurement is minimised. If this is non-zero, this implies that the measurement has changed the state, which will only occur if the system possesses quantum correlation. The main drawback of quantum discord is the optimisation procedure is generally a discrete task, and scales with the dimensionality of the system. Therefore, generally speaking the optimisation is an NP-hard problem \cite{Huang_2014}. There have been various attempts to circumvent this issue by utilising geometric discord \cite{PhysRevLett.105.190502} as an approximation to quantum discord, and by extension Fisher information \cite{lowe}. However, these techniques have drawbacks, for example geometric discord can increase under local reversible operations \cite{PhysRevA.86.034101}.

Additionally, determining how measurement can affect quantum correlations is an increasingly active area of research. It has been found that measurement can induce phase transitions \cite{PhysRevX.9.031009,PhysRevB.98.205136}, and even more interestingly, it has been found to induce quantum correlation into the system \cite{PhysRevResearch.5.L042031,meas_ent}. Therefore there is a clear benefit to optimally choosing the measurement in order to steer the system into a desired state \cite{PhysRevResearch.2.042014,PhysRevResearch.2.033347}. It has recently been found that different types of measurement can introduce an asymmetry into the system, which is not present through projective measurements \cite{arXiv:2401.09304}.

Alongside this, there has been significant work in utilising measures of quantum correlation as a tool for witnessing and detecting phase transitions \cite{De_Chiara_2018,PhysRevB.78.224413,PhysRevA.66.032110}. It has already been shown that both entanglement entropy and quantum discord exhibit non-trivial behaviour around the critical points. However, recently it was suggested that QET could be used as a technique to witness and detect phase transitions due to its inherent link with entanglement entropy \cite{PhysRevD.107.L071502,2023arXiv230500996I}. However, the link between quantum discord and QET has been less studied, despite the QET protocol necessarily resulting in mixed states. Understanding and investigating whether quantum discord is able to accurately reflect phase transitions throughout the QET protocol is the motivation for this work, whilst also considering the fundamental question of how quantum correlation evolves through the process. In \cite{trevison2015quantum}, quantum discord of a state coupled to a thermal bath is studied, however in our paper, we address the states in QET without a heat bath. We are interested in studying the origin of the quantum resources of QET in an idealized setup at zero temperature.  

To demonstrate the robustness of the quantum discord in QET, we address several benchmark models including a 2d QED (Quantum Electrodynamics) model~\cite{schwinger1962gauge,Schwinger:1962tp} and 2d Nambu-Jona-Lasino (NJL) model~\cite{PhysRev.122.345,PhysRev.124.246}, which is sometimes called the Thirring model~\cite{Thirring:1958in} or Gross-Nevue model~\cite{PhysRevD.10.3235}. They are recognised as useful effective theories of 4d QCD (Quantum Chromodynamics) and actively studied from a perspective of quantum simulation/computation~\cite{wallraff2004strong,majer2007coupling,Jordan:2011ne,Jordan:2011ci,Zohar:2012ay,Zohar:2012xf,Banerjee:2012xg,Banerjee:2012pg,Wiese:2013uua,Wiese:2014rla,Jordan:2014tma,Garcia-Alvarez:2014uda,Marcos:2014lda,Bazavov:2015kka,Zohar:2015hwa, Mezzacapo:2015bra, Dalmonte:2016alw, Zohar:2016iic, Martinez:2016yna, Bermudez:2017yrq, gambetta2017building, krinner2018spontaneous, Macridin:2018gdw, Zache:2018jbt, Zhang:2018ufj, Klco:2018kyo, Klco:2018zqz, Lu:2018pjk, Klco:2019xro, Lamm:2018siq, Gustafson:2019mpk, Klco:2019evd, Alexandru:2019ozf, Alexandru:2019nsa, Mueller:2019qqj, Lamm:2019uyc, Magnifico:2019kyj, Chakraborty:2019, Kharzeev:2020kgc, PhysRevLett.131.021902,Shaw:2020udc, sahinoslu2020hamiltonian, Paulson:2020zjd, Mathis:2020fuo, Ji:2020kjk, Raychowdhury:2019iki, Davoudi:2020yln, Dasgupta:2020itb, Magnifico:2018wek, PhysRevD.107.L071502, Rico:2013qya, Buyens:2013yza, Buyens:2016ecr, Banuls:2013jaa, Banuls:2015sta, Banuls:2019bmf, Kokail:2018eiw,Ikeda:2023vfk,Ikeda:2023uqy,Grieninger:2023ufa,Mishra:2019xbh,Czajka:2022plx,Farrell:2023fgd,Farrell:2024fit}. In all cases we tested, we detected phase transition points by the quantum discord and confirmed the its robustness throughout the procedure of QET. 

This article is organised as follows. In Sec.~\ref{sec:QET} we briefly review the QET protocol in a general spin chain model. In Sec.~\ref{sec:discord}, we give a short review of quantum discord and describe its relation with QET. In Sec.~\ref{sec:min}, we demonstrate the robustness of the quantum discord in the minimal QET model. We also demonstrate QET with the state-of-the-art quantum computer of IBM (\texttt{ibm\_osaka}), which a cloud-based quantum computer with 127 qubits that is available to the general public. We report that the experimental data showed significant improvement from the previous results conducted with some IBM quantum computers~\cite{Ikeda:2023uni}, which are now retired. Furthermore, in Sec.~\ref{sec:QED} we work on two (1+1)-dimensional models of effective theory of QCD. In Sec.~\ref{sec:Schwinger}, we study the simplest fermion model that undergoes a phase transition in the entangled ground state and, in Sec.~\ref{sec:NJL}, we consider the NJL model with a finite chemical potential $\mu$ and finite chiral chemical potential $\mu_5$, which are useful to mimic the chiral imbalance between
right- and left-chirality quarks coupled with the chirality charge density operator. We report that in both cases, quantum discord is robust in QET and shoows an abrupt change at the corresponding phase transition points.  

\section{\label{sec:QET}Quantum energy teleportation}
We initially clarify the QET protocol \cite{Hotta_2010,HOTTA20085671} on a general locally interacting $N$-site system. By considering a local Hamiltonian with $H=\sum_{n=0}^{N-1}H_n$,
where $H_n$ represents the local Hamiltonian on a given site which is then interacting with other qubits in the system. It is important that this Hamiltonian, and each individual local Hamiltonian are constrained by
\begin{align}
\begin{aligned}
\label{eq:condition}
\bra{g}H\ket{g}=\bra{g}H_n\ket{g}=0,~\forall n\in \{1,\cdots,N\},
\end{aligned}    
\end{align}
where $\ket{g}$ is the ground state of the full Hamiltonian $H$ and $N$ is the number of sites. For the QET protocol to work, the initial ground state must be entangled to allow for non-local effects in the system to occur. Given the system is in the ground state, any local operations which occur on the system will necessarily increase the local energy.

Suppose that Alice is a supplier of energy and Bob is a receiver of this energy. Alice measures her one-qubit unitary operator $\sigma_{A}$, using the projective measurement operator $P_{A}(\mu)=\frac{1}{2}(1+\mu \sigma_{A})$ and obtains either $\mu=-1$ or $\mu=+1$. In a quantum circuit, Alice's measurement can be interpreted as $\mu=+1$ corresponding to $\ket{0}$ and $\mu=-1$ corresponding to $\ket{1}$ in the basis of $Z=\ket{0}\bra{0}-\ket{1}\bra{1}=\sum_\mu\mu\ket{\frac{-\mu+1}{2}}\bra{\frac{-\mu+1}{2}}$). At this point, Alice can measure either $\{\sigma_x, \sigma_y, \sigma_z\}$, as long as it is on Alice's subsystem. Given this measurement, the injected energy $E_A$ is localised around subsystem $A$. However, Alice cannot extract this energy at $A$. In order to circumvent this problem local operations and classical communications (LOCC) can be utilised, by communicating to the receiver of energy, Bob. By utilising local operations, Bob can extract some energy from his local system. 

The classical communication aspect of the protocol requires, Alice sending her measurement result $\mu$ to Bob. At this state, Bob performs a conditional operation $U_{B}(\mu)$ to his state and measures his local Hamiltonian $H_{B}$, where $U_{B}(\mu)$ is given by
\begin{equation}
\label{eq:operation}
    U_{B}(\mu)=\cos\theta I-i\mu\sin\theta\sigma_{B},
\end{equation}
where $\theta$ obeys 
\begin{align}
\label{eq:theta}
    \cos(2\theta)=\frac{\xi}{\sqrt{\xi^2+\eta^2}},~
    \sin(2\theta)=\frac{\eta}{\sqrt{\xi^2+\eta^2}},
\end{align}
where
\begin{align}
\label{eq:params}
\xi=\bra{g}\sigma_{B}H\sigma_{B}\ket{g},~\eta=\bra{g}\sigma_{A}\dot{\sigma}_{B}\ket{g},
\end{align}
with $\dot{\sigma}_{B}=i[H,\sigma_{B}]$. It is important that the local Hamiltonian satisfies $[H,\sigma_{B}]=[H_{B},\sigma_{B}]$. 
The average quantum state $\rho_\text{QET}$, which crucially is a mixed state, is obtained after Bob operates $U_{B}(\mu)$ to $\frac{1}{\sqrt{p(\mu)}}P_{A}(\mu)\ket{g}$, where $p(\mu)$ is the normalisation factor. 

The resulting density matrix, $\rho_\text{QET}$ after these operations is given by 
\begin{equation}
\label{eq:rho_QET}
    \rho_\text{QET}=\sum_{\mu\in\{-1,1\}}U_{B}(\mu)P_{A}(\mu)\ket{g}\bra{g}P_{A}(\mu)U^\dagger_{B}(\mu). 
\end{equation}
Using this result, the expected energy at Bob's local system can be evaluated as 
\begin{equation}
\label{eq:QET}
    \langle E_{B}\rangle=\Tr[\rho_\text{QET}H_{B}]=\frac{1}{2}\left[\xi-\sqrt{\xi^2+\eta^2}\right], 
\end{equation}
which is negative if $\eta\neq 0$. 
To emphasise, the strictly positive energy $-\langle E_{B}\rangle$ is then activated at Bob's device to due energy conservation in the system. This makes the crucial assumption that there is no energy dissipation in the system.

\section{\label{sec:discord}Quantum discord}
It is considered that quantum discord captures quantum correlations which exist beyond entanglement. This is particularly important when considering mixed states, which can possess quantum correlation, despite being inherently local. Importantly, quantum discord can be used to capture the quantum correlation in states which are not pure, subsequently it doesn't possess the limitations of using entanglement entropy as a measure for quantum correlation. This makes it an adequate tool for determining how quantum correlation evolves through the QET process, due to the mixed state which occurs after measurement. Formally, quantum discord is given by
\begin{equation}
   D_A \equiv D_A (\rho_{AB}) = I_{AB} (\rho_{AB}) - J_{A|B} (\rho_{AB}),
\end{equation}
where $I_{AB} (\rho_{AB})$ is the quantum mutual information defined by
\begin{equation}
    I_{AB} (\rho_{AB}) = S(\rho_A) + S(\rho_B) - S(\rho_{AB}),
\end{equation}
where $\rho_{AB}$ is the shared state between Alice and Bob, $S(\rho)=-\tr(\rho \ln \rho)$ is the von-Neumann entropy, and $\rho_i = \tr_j \rho_{ij}$, for different $i,j \in \{A,B\}$. $J_{A|B} (\rho_{AB})$ is the minimized quantum conditional entropy, where
\begin{equation}
    J_{A|B} (\rho_{AB}) = S(\rho_A) - S_{A|B} (\rho_{AB}),
\end{equation}
with
\begin{equation}
    S_{A|B} (\rho_{AB}) = \min\limits_{\Pi_{k}^{B}} \sum_k p_k S(\rho_{A|k}),
    \label{eq:10}
\end{equation}
where the minimisation is performed over all measurements, and $p_k = \tr (\mathbb{1}^{A} \otimes \Pi_k \rho_{AB} \mathbb{1}^{A} \otimes \Pi_{k}^{\dagger})$, and $\rho_{A|k}=\tr_B (\mathbb{1}^{A} \otimes \Pi_k \rho_{AB} \mathbb{1}^{A} \otimes \Pi_{k}^{\dagger})/p_k$. Note the measurement is performed in Bob's subsystem, whilst Alice's subsystem is given by $\mathbb{1}^{A}$. The measurement is a general projector given by $\Pi_k = (1/2)(\mathbb{1} + k {\bf{n}}\cdot{\boldsymbol{\sigma}})$, where $k \in \{+1,-1\}$ denotes measuring spin up or spin down, $\textbf{n}=(n_x,n_y,n_z)$ is the Bloch vector, and $\boldsymbol{\sigma}=(X,Y,Z)$ is the vector of Pauli matrices. This measurement is performed on each qubit in whichever subsystem is being optimised over. An analogous definition can be given where the measurement is performed instead in Alice's subsystem. This is denoted by $D_B (\rho_{AB})$, which is given by
\begin{equation}
   D_B \equiv D_B (\rho_{AB}) = I_{AB} (\rho_{AB}) - J_{B|A} (\rho_{AB}),
\end{equation}
where
\begin{equation}
    J_{B|A} (\rho_{AB}) = S(\rho_B) - S_{B|A} (\rho_{AB}),
\end{equation}
with
\begin{equation}
    S_{B|A} (\rho_{AB}) = \min\limits_{\Pi_{k}^{A}} \sum_k p_k S(\rho_{B|k}),
    \label{eq:10}
\end{equation}
where $p_k = \tr (\Pi_k \otimes \mathbb{1}^{B} \rho_{AB} \Pi_{k}^{\dagger} \otimes \mathbb{1}^{B})$, and $\rho_{B|k}=\tr_A (\Pi_k \otimes \mathbb{1}^{B} \rho_{AB}\Pi_{k}^{\dagger} \otimes \mathbb{1}^{B})/p_k$. In general, $D_A (\rho_{AB}) \neq D_B (\rho_{AB})$, revealing inherent asymmetry in the measure which is also not apparent in entanglement entropy. The numerical computation of the quantum discord makes use of QuTiP \cite{JOHANSSON20121760}. 

Here we give an important remark on a relation between the discord and the entanglement consumption
\begin{equation}
    \Delta S_{AB} = - \tr_B \rho_B \ln \rho_B - \sum_{\mu} P_A (\mu) \Big( -\tr_B \rho_B (\mu) \ln \rho_B (\mu) \Big), 
\end{equation}
which ``measures” the difference in the entropy before and after the QET, therefore it indicates ``consumption" of the entropy in QET. The key difference between the definitions is found the second term in each. In the quantum discord, it is a conditioned measurement only on one subsystem which is then optimised over. Whereas, in $\Delta S_{AB}$, there is no optimisation, and there have been measurements/operations in both subsystems. At this point, it is important to highlight the strength of quantum discord over $\Delta S_{AB}$, in the sense it can be used with mixed states. It is also important to note that entanglement entropy is not a measure of mixed state entanglement.

\section{\label{sec:min}Minimal model}
\begin{figure*}
  \centering
  \subfloat[a][$D_B(k)$ for $h=1$ throughout the QET protocol.] {\includegraphics[width=0.47\linewidth]{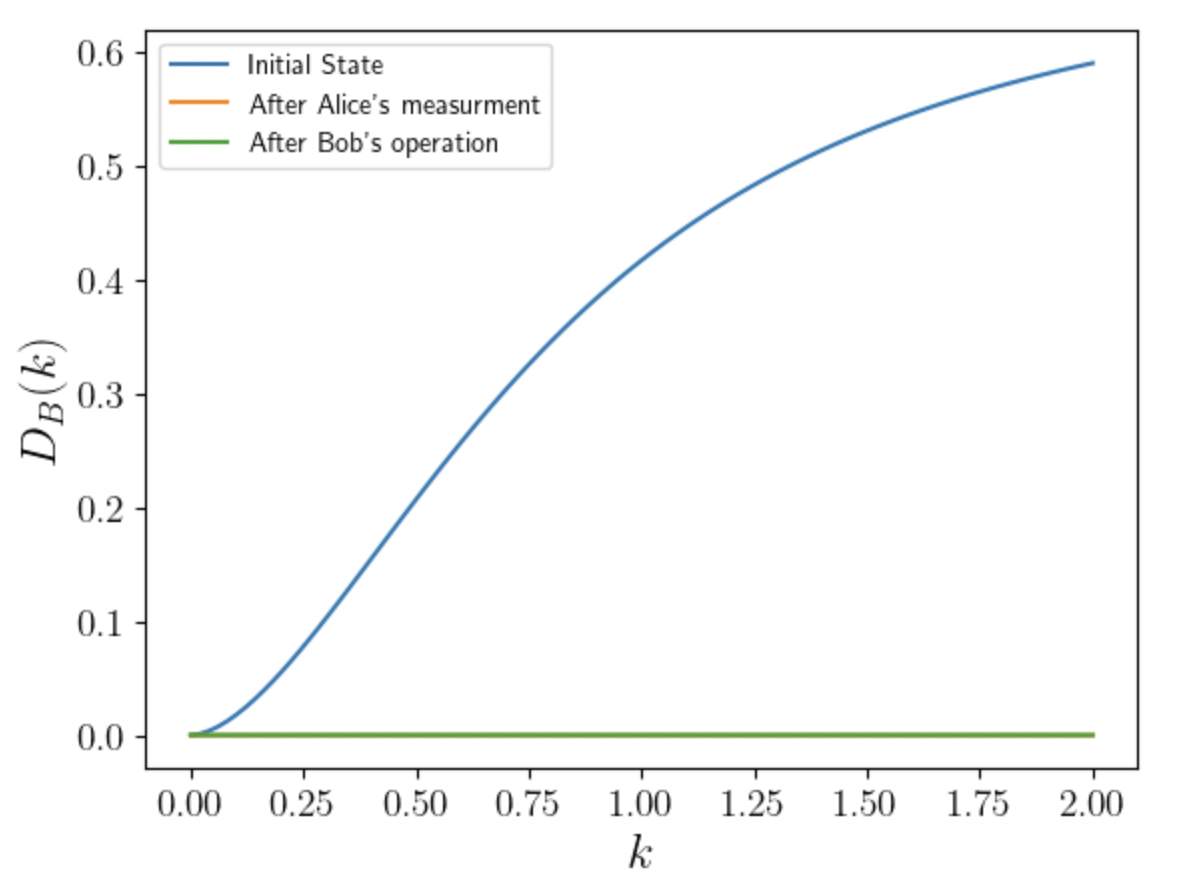}} \label{fig:a} \subfloat[a][$D_B(h)$ for $k=1$ throughout the QET protocol.] {\includegraphics[width=0.47\linewidth]{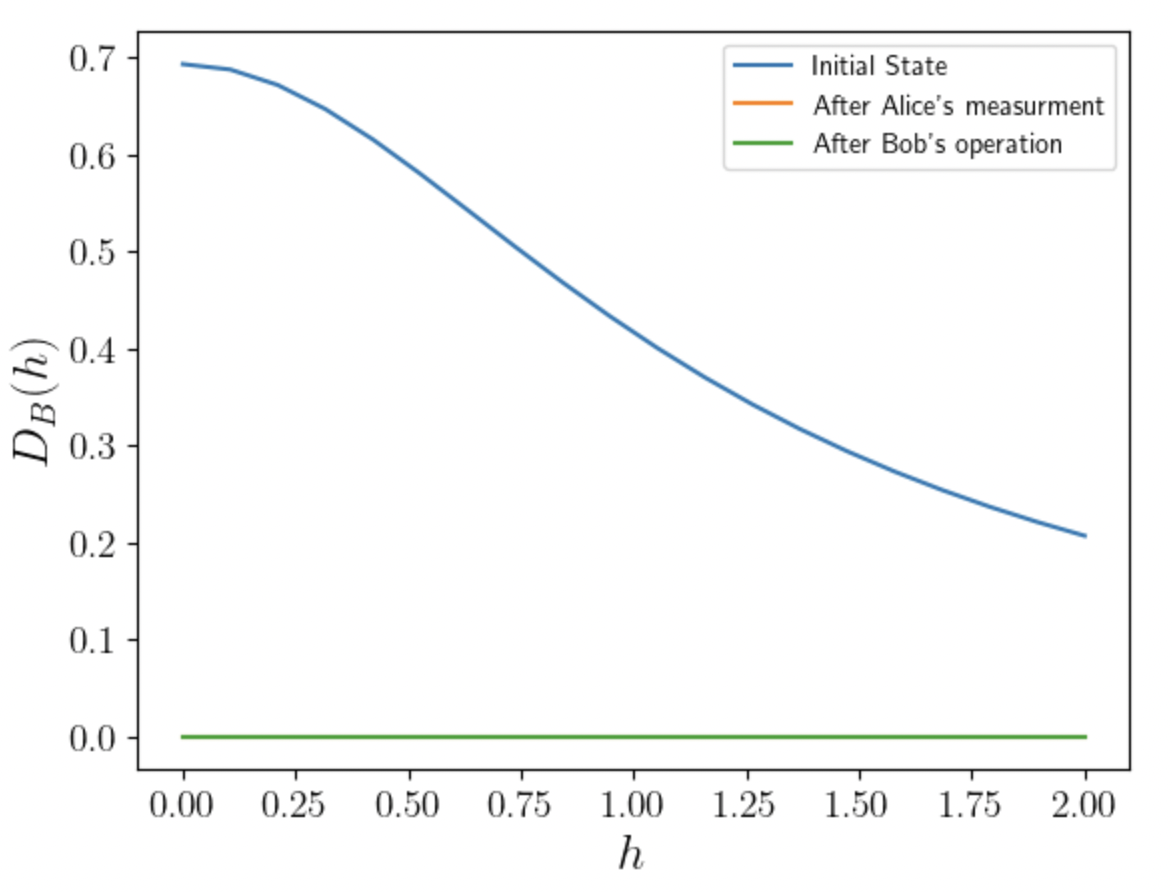}} \label{fig:g} \\\subfloat[a][Confirming the entanglement entropy and quantum discord correspond for the initial pure state.] {\includegraphics[width=0.47\linewidth]{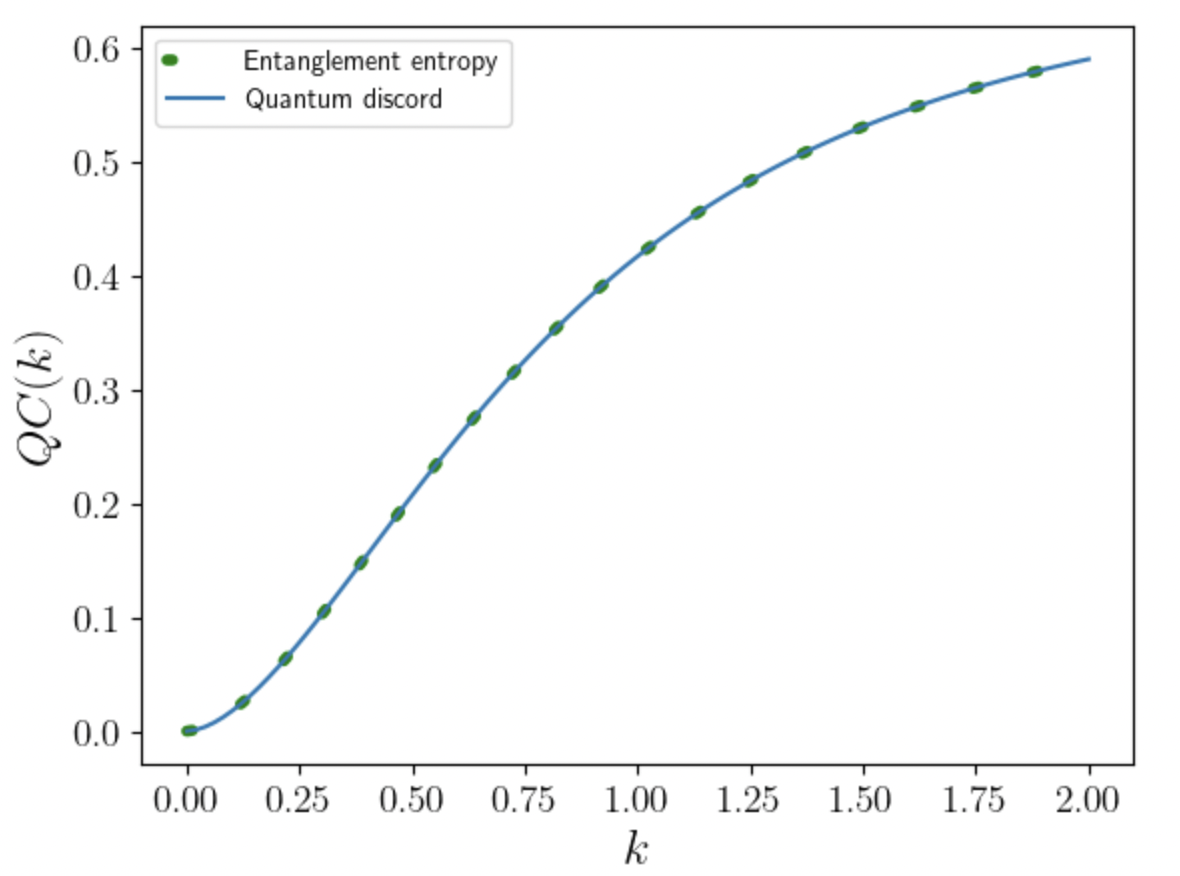}} \label{fig:g} 
 \subfloat[a][$D_A(h)$ for $k=1$ throughout the QET protocol.]{\includegraphics[width=0.47\linewidth]{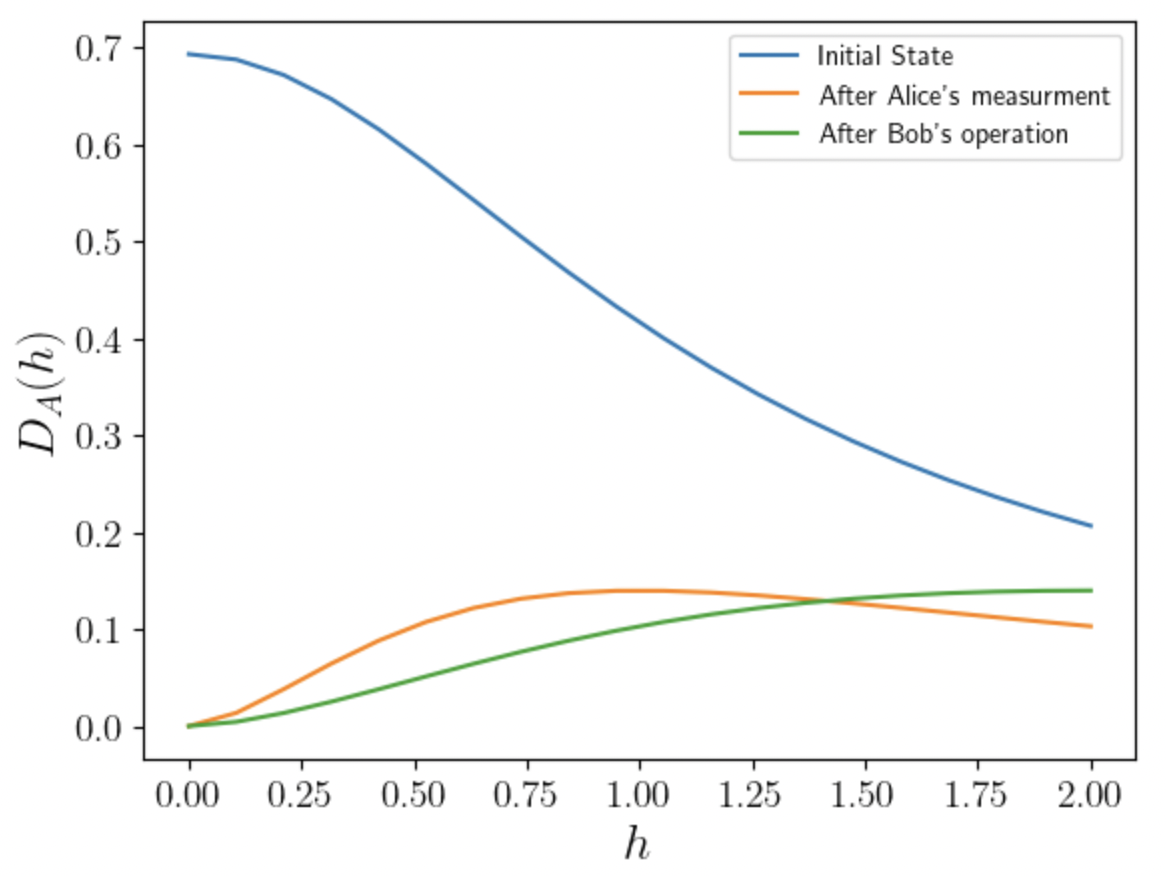}} \\
  \caption{This figure shows the quantum discord for differing parameters where the quantum discord has been computed for both $D_A$ and $D_B$ in order to investigate the asymmetry. For $D_B$, it is clear that the quantum correlation is destroyed by the QET protocol. However, for $D_A$, it is clear it persists. In particular, Bob's operation creates some quantum discord. } \label{fig:1}
\end{figure*}
The minimal model consists of interacting two qubits, whose ground state is analytically solvable. The corresponding Hamiltonian is given by
\begin{align}
\begin{aligned}
    H &= H_A + H_B + V, \\
    H_n &= h Z_n + \frac{h^2}{\sqrt{h^2 + k^2}}, \\
    V &= 2 k X_A X_B + \frac{2k^2}{\sqrt{h^2 + k^2}},
\end{aligned}
\end{align}
where $k,h$ are positive constants, and $n=A,B$ denoting Alice's and Bob's sites. The ground state is 
\begin{equation}
\ket{g} = \frac{1}{\sqrt{2}} \sqrt{1 - \frac{h}{\sqrt{h^2 + k^2}}} \ket{00} - \frac{1}{\sqrt{2}} \sqrt{1 + \frac{h}{\sqrt{h^2 + k^2}}} \ket{11}.
\end{equation}

Using this model, the quantum discord is computed at different stages of the QET protocol, in order to determine the relationship between energy teleportation and quantum correlation. It is important to stress that Alice's projective measurement renders the initial state no longer pure, therefore entanglement entropy can not be used to capture the quantum correlation. 
\begin{figure*}
    \centering
    \includegraphics[width=0.34\linewidth]{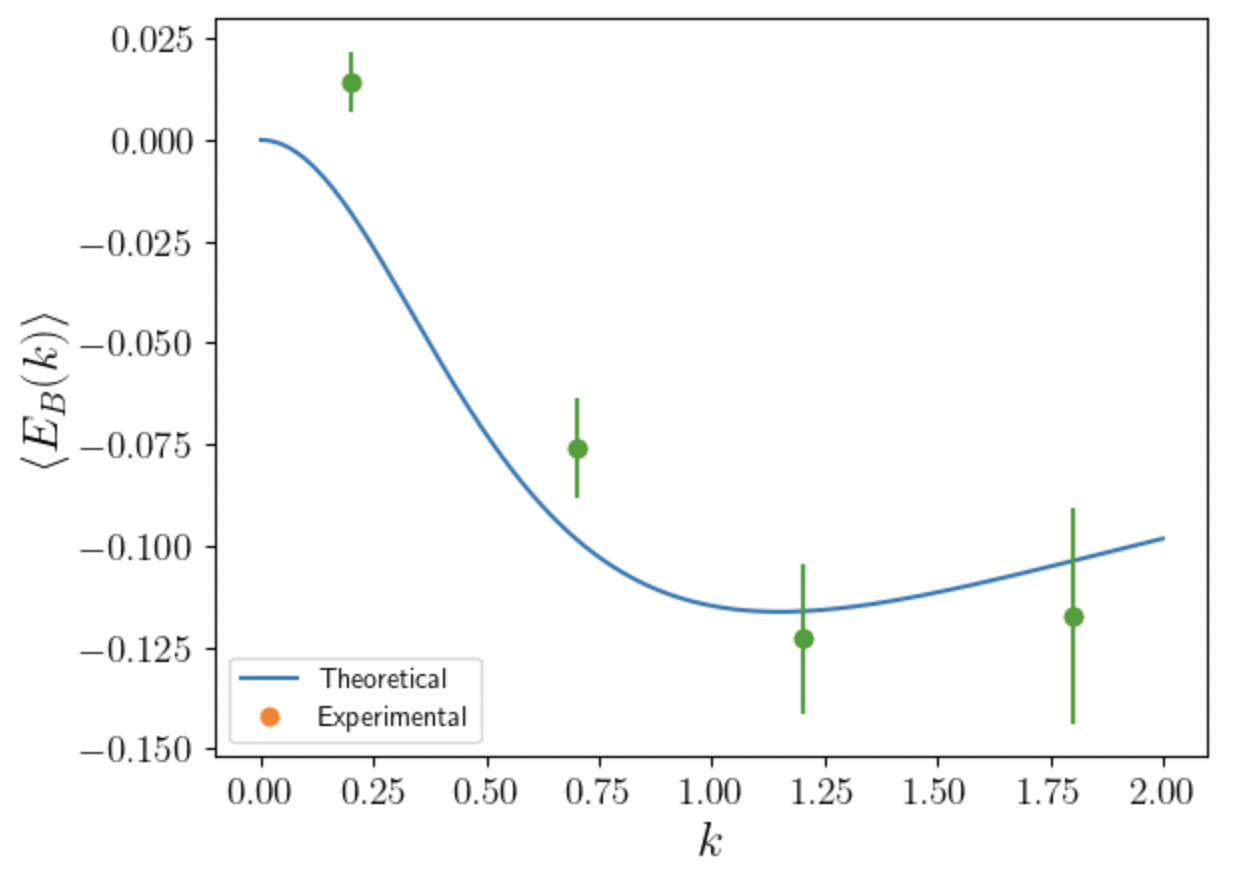}
    \centering
    \includegraphics[width=0.34\linewidth]{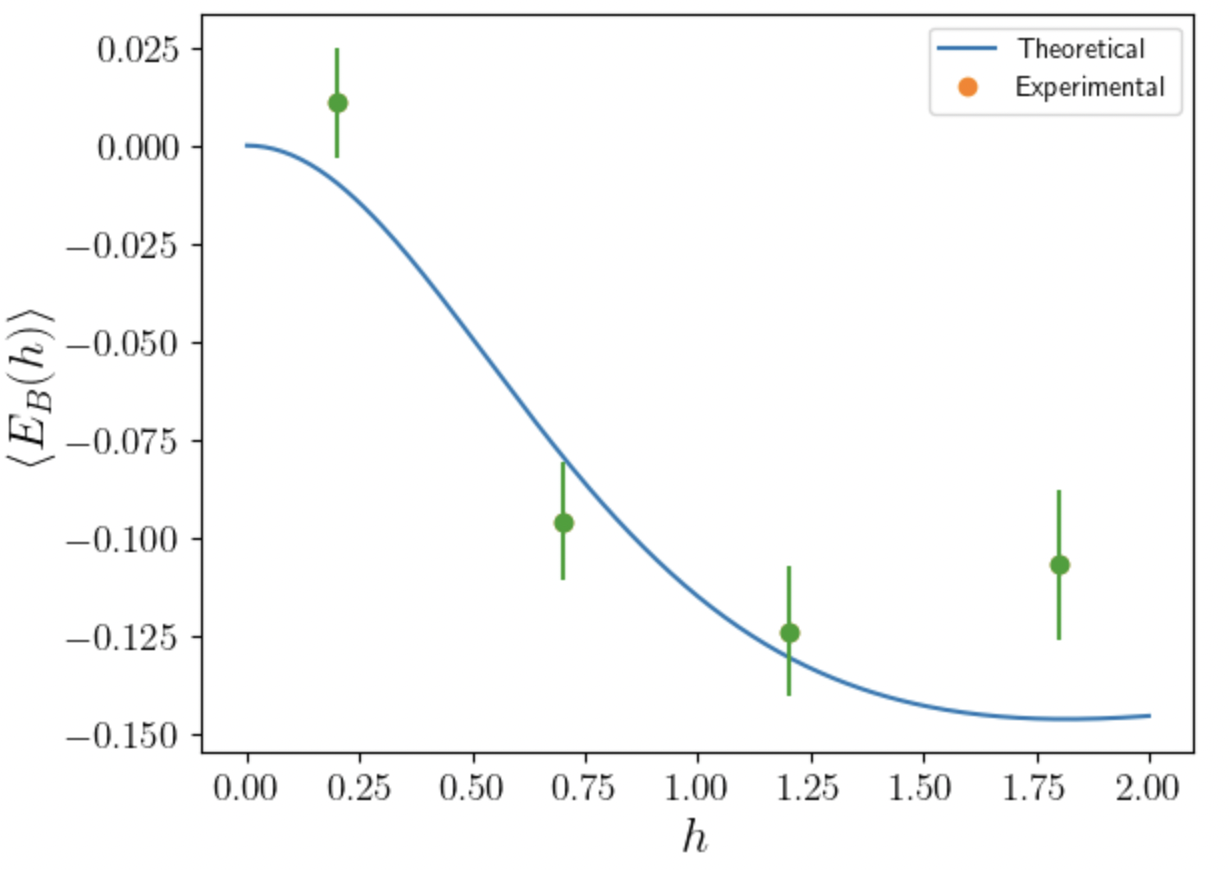}
    \centering
    \includegraphics[width=0.3\linewidth]{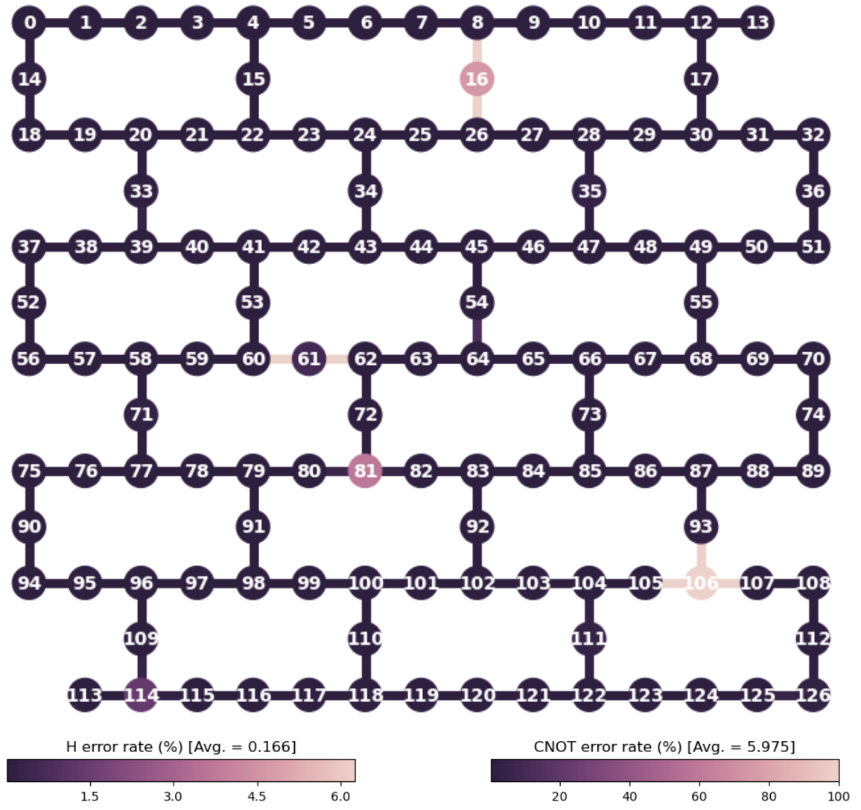}
    \caption{Teleported energy for minimal model using {\texttt{ibm\_osaka}}. left: $h=1$ middle: $k=1$. This is architecture of the quantum hardware that was used to perform the experimental implementation. The error rates are shown in the figure, and the specific gate errors of qubits 39 and 40, which we used for one of the runs of our experiment, are given in the text. The detected teleported energy results become closer to the theoretical values than the previous results performed with some retired IBM quantum computers~\cite{Ikeda:2023uni}.}
    \label{fig:2}
\end{figure*}

In order to account for all measurements in computing the discord, the sum over $\mu$ is taken, and then the measurements are minimised based on this. Given the measurement is performed on a qubit, the measurement can be parameterised on the Bloch sphere such that the optimisation is performed over $\theta,\phi$ which denotes the respective angles in the Bloch sphere. This allows the optimisation to be continuous, rather than discrete, drastically simplifying the optimisation.

Fig.~\ref{fig:1} confirms that the quantum discord matches the entanglement entropy for the initial state. This is to be expected as the initial state is pure, so for this case the quantum discord should yield the same results as entanglement entropy. It is also seen in Fig.~\ref{fig:1} that the quantum discord captures the quantum correlations as the QET develops. It is clear that for and $D_A(h)$, there is still quantum correlation which persists when tracing out Bob's subsystem in the minimisation. As expected, Alice's measurement destroys some quantum correlation, as the measurement changes the state from pure-entangled into a mixed state. However, it is interesting to see that there still remains quantum correlation after measurement. Additionally, it is clear that there are regions where Bob's operation creates some quantum discord. This is not surprising as it is known that unitary operation can increase quantum discord \cite{PhysRevLett.107.170502,doi:10.1142/S123016121440006X,PhysRevA.83.010301}. We highlight that this does not imply that entanglement is created through LOCC. This suggests the QET protocol could be used as a mechanism for increasing quantum discord if an initial pure state has its quantum entanglement reduced by measurement.

In order to investigate how the difference in measurement and operation between Alice and Bob affects the quantum correlation, it is instructive to compute $D_A(h)$. When computing the discord for this case, the measurement and optimisation is performed on Alice's subsystem, and then this is traced out leaving only Bob's subsystem. It is clear that the QET protocol introduces asymmetry in the quantum correlation, as all quantum correlation is destroyed after Alice's measurement. Furthermore, Bob's operation is not able to create any quantum entanglement. Subsequently, the QET protocol is one-way quantum correlated, as whether quantum correlation persists entirely depends which system is traced out.

In order to study the relationship between the amount of energy teleported and quantum discord, the theoretical teleported energy is plotted over the same regions in Fig~\ref{fig:2}. It is seen that for the initial state, there is qualitatively similar behaviour. This matches our current understanding of the relationship between entanglement entropy and the amount of energy teleported.

The experimental results were performed using two implementations. In one implementation, qubits 39 and 40 were chosen (see map in appendix) on {\texttt{ibm\_osaka}}, which had gate errors of $\sim 0.0002$ and $\sim 0.0004$ respectively. It can be seen this matches the theoretical result quite well.

The result of comparing the theoretical teleported energy with the experimental results using the best current available quantum computers, allows a qualitative check that there is a link between, QET and quantum correlation.  Whilst, the experimental results do not fully lie within the theoretical result, this is not too much of a concern due the unreliability of current quantum computers due to noise. It is actually impressive that some of the experimental points match up to the theory as well as they do. This gives sufficient motivation that using the theoretical teleported energy corresponds to what is expected in experiment. Whilst there is no clear phase transition in the minimal model, the fact that QET is shown to be a real experimental phenomenon opens up the possibility that it could be used to detect and witness phase transitions in more complicated models. This is one of the motivations for the rest of the paper, in particular determining whether the phase transitions can be still be detected throughout the QET protocol.

\begin{figure*}
    \centering
    \includegraphics[width=0.31\linewidth]{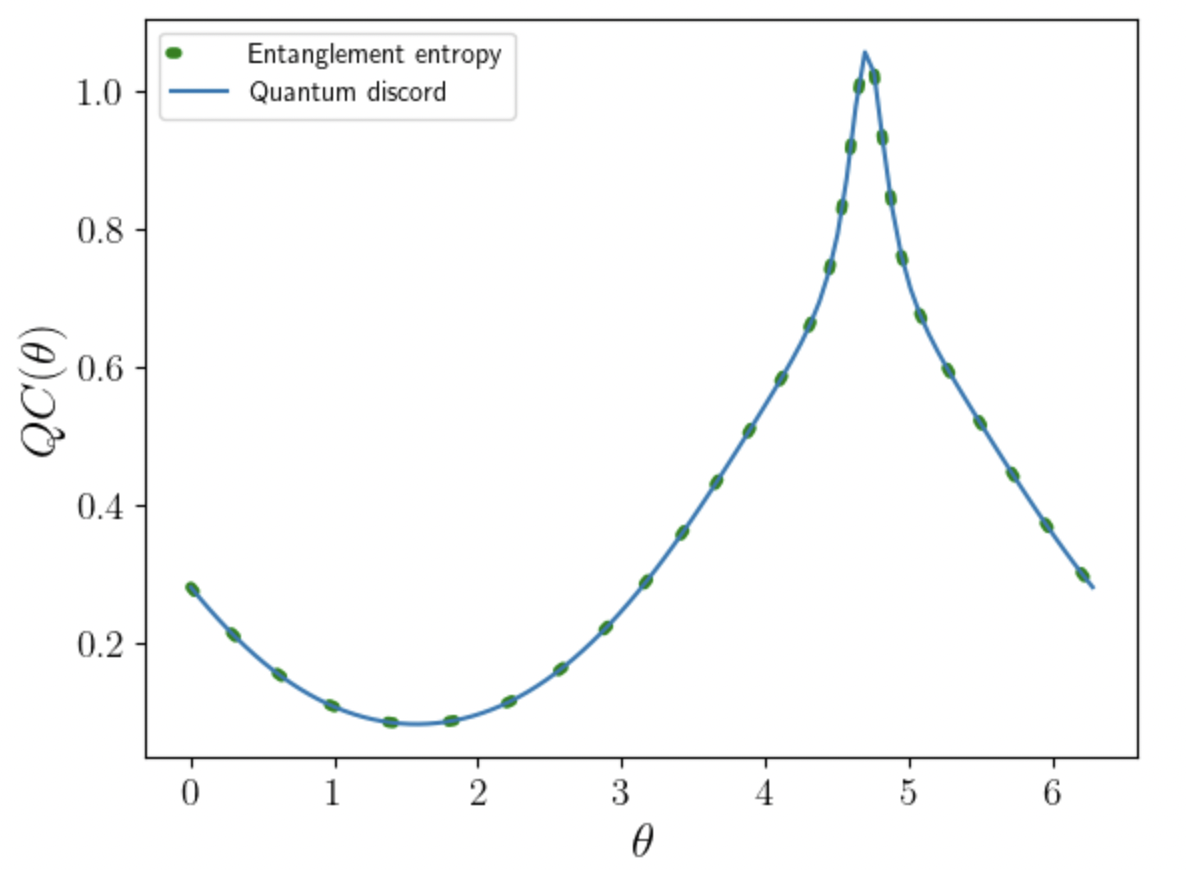}
    \centering
    \includegraphics[width=0.31\linewidth]{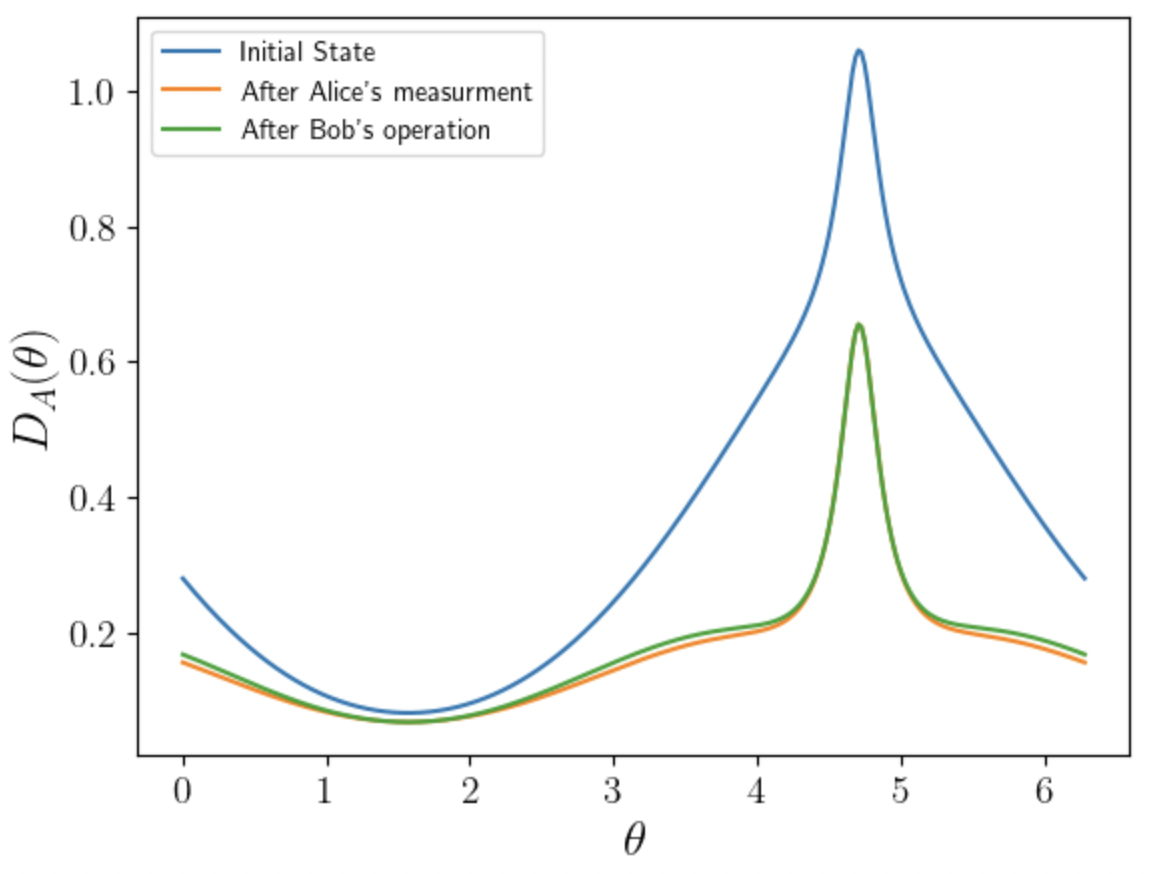}
    \centering
    \includegraphics[width=0.31\linewidth]{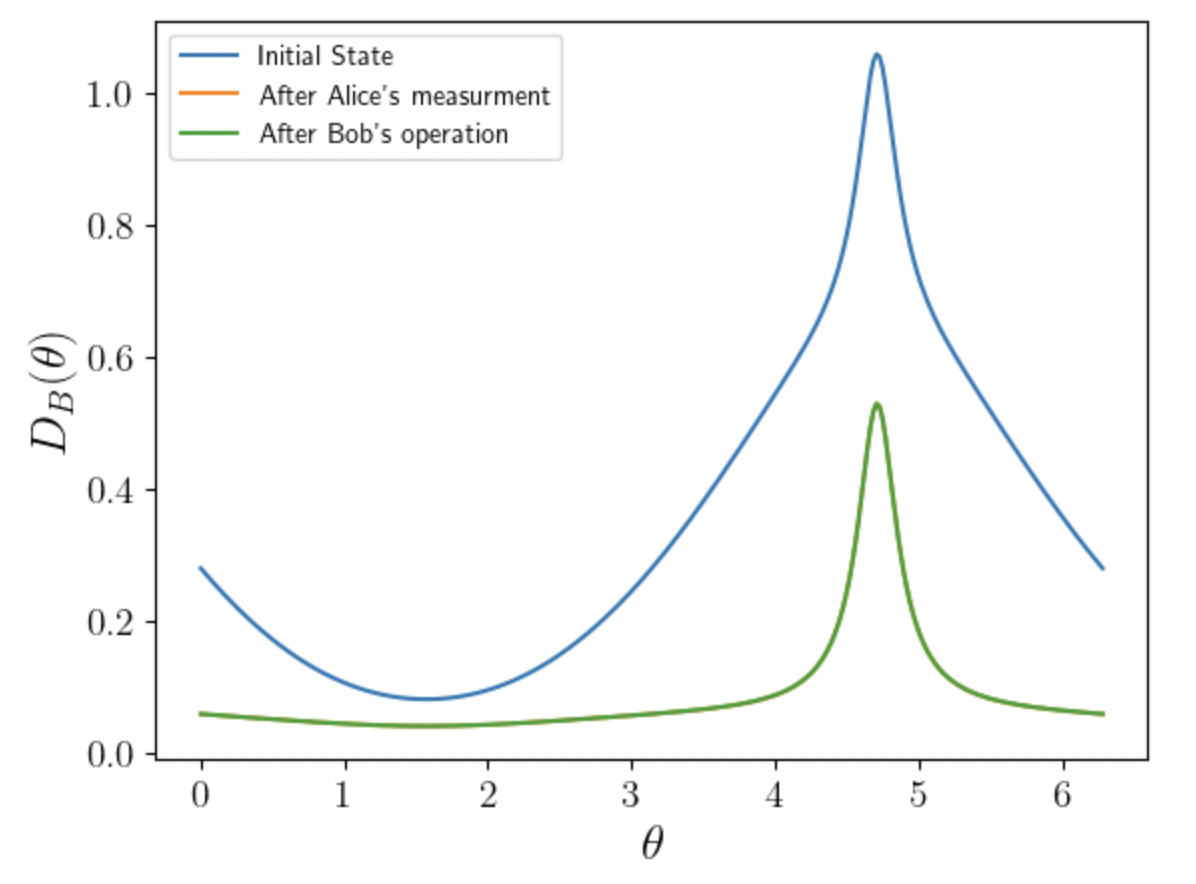}
    \caption{$QC(\theta)$ denotes the quantum correlation as a function of $\theta$. The left figure shows the comparison between entanglement entropy and discord before any operations have taken place. As expected they match, due to the initial state being pure. The middle figure is the discord ($D_A(\theta)$) for the Schwinger model through the QET protocol. The parameters set are described in the text. Notice there are significant peaks at the phase transition. Also it can be seen that Bob's operation increases the quantum discord. The right figure is the discord ($D_B(\theta)$) when measurement is performed on the other subsystem. The main difference between $D_A(\theta)$ and $D_B(\theta)$ is the Bob's measurement does not increase the quantum discord in $D_B(\theta)$.}
    \label{fig:Schwinger1}
\end{figure*}

\begin{figure*}
    \centering
    \includegraphics[width=0.31\linewidth]{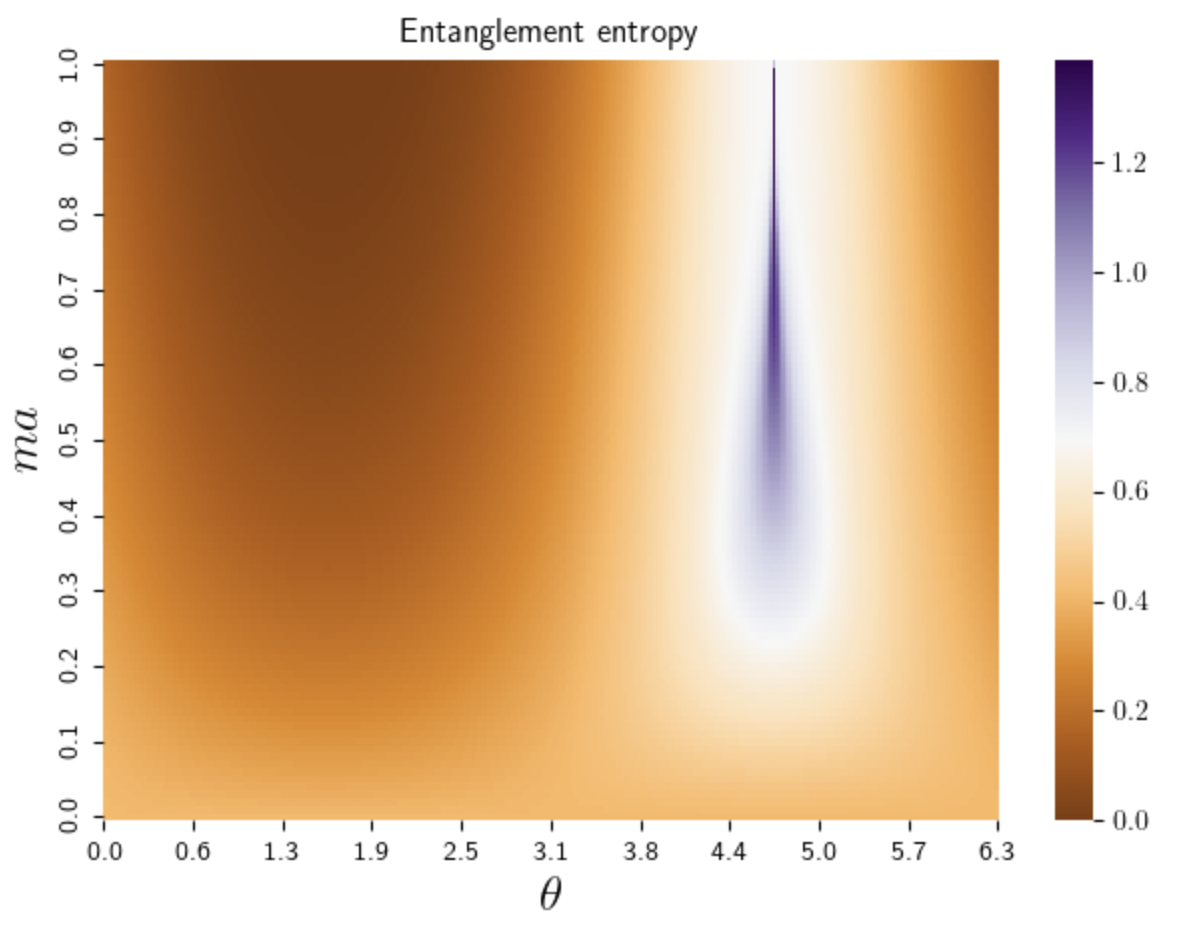}
    \centering
    \includegraphics[width=0.31\linewidth]{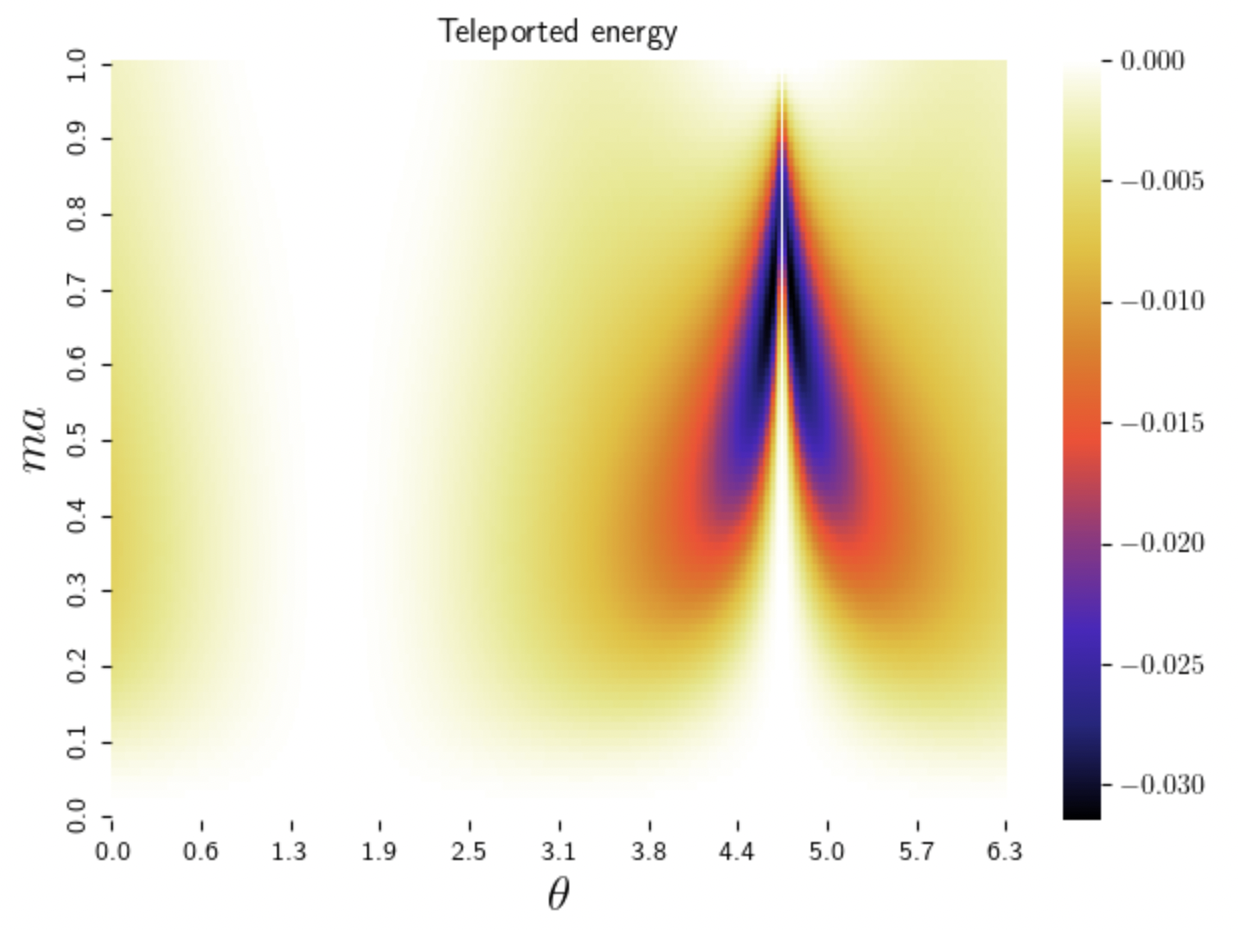}
    \centering
    \includegraphics[width=0.31\linewidth]{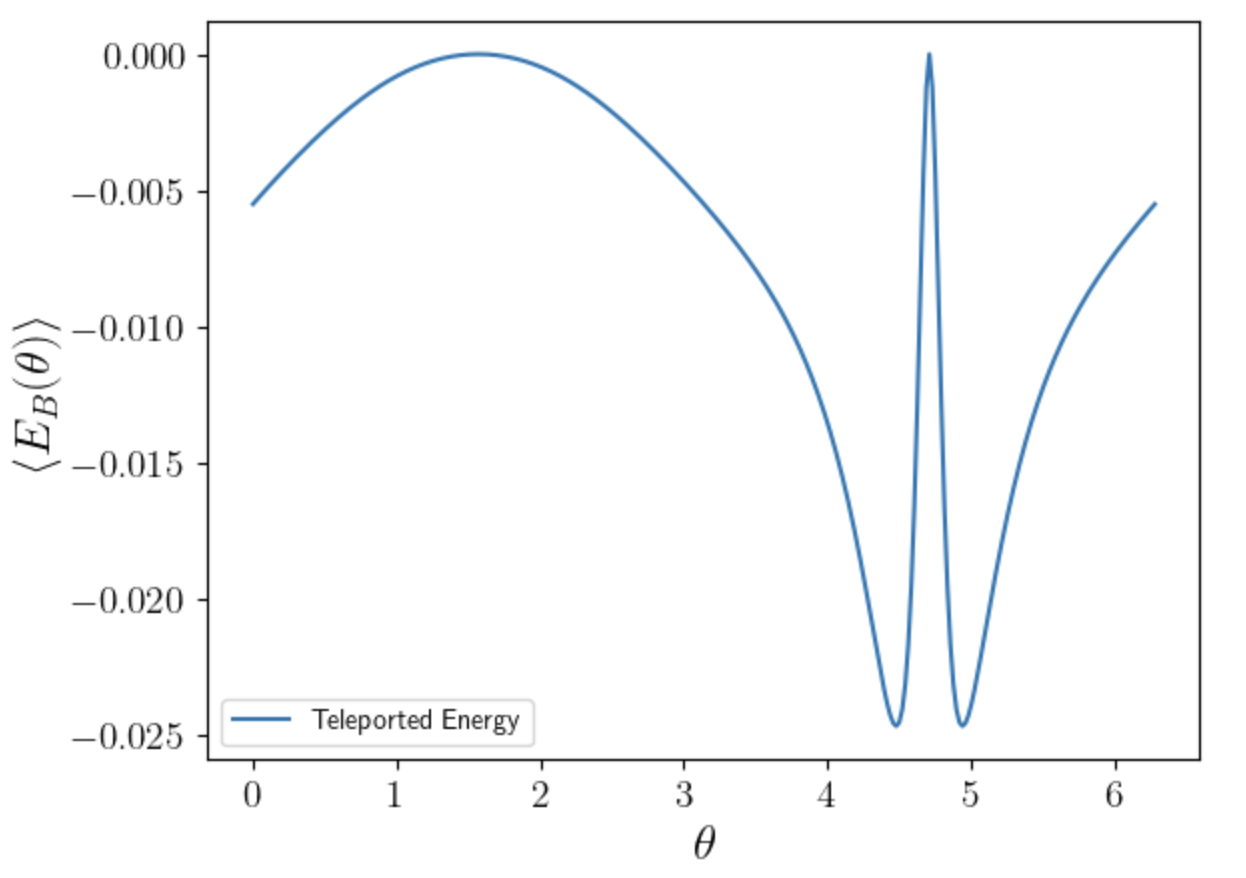}
    \caption{This figure looks at capturing the effects of the entanglement entropy and teleported energy. The entropy was computed for the ground state of the Hamiltonian. It is seen that there is a qualitative similarity between the two. It is clear that the regions where there is a significant change in behaviour is replicated in all three figures. In order to make qualitative comparisons with Fig.~\ref{fig:Schwinger1}, the plot on the right was computed for $ma=0.5$.}
    \label{fig:Schwinger2}
\end{figure*}
\section{\label{sec:QED}QET in Quantum Field Theory}
\subsection{\label{sec:Schwinger}Schwinger model}
As a concrete low dimensional model, we work on the Schwinger model~\cite{schwinger1962gauge,Schwinger:1962tp}. The Hamiltonian of the massive Schwinger model with the topological $\theta$ term in (1+1)-dimensional Minkowski space in temporal gauge $A_0=0$ is
\begin{align} \label{eq:Ham}
H &=\int dz \Big[    
    \frac{E^2}{2}
    -\bar{\psi}(i\gamma^1\partial_1 - g\gamma^1A_1 - m\,\e^{i\gamma_5\theta} 
    + \mu \gamma^0)\psi \Big] \,.
\end{align}
Here, $A_\mu$ is the $U(1)$ gauge potential, $E=\dot{A}_1$ is the electric field, $\psi$ is a two-component fermion field, $m$ is the fermion mass, $g$ is the coupling constant and $\gamma^\mu$ are the two-dimensional $\gamma$-matrices $\{\gamma^\mu,\gamma^\nu\}=2\eta^{\mu\nu}$ defined as follows:
\begin{equation}
    \gamma_0=Z,~\gamma_1=-iY,~\gamma_5=\gamma_0\gamma_1=-X.
\end{equation}

In the spin representation, the Hamiltonian becomes 
\begin{align}
\begin{aligned}
\label{eq:Hamiltonian}
H&=\sum_{n=1}^{N-1}
    \left(\frac{1}{4a}-\frac{m}{4}(-1)^n\sin\theta\right)
    \left(X_n X_{n+1}+Y_n Y_{n+1}\right)
\\&+\sum_{n=1}^{N}\frac{m(-1)^n\cos\theta}{2} Z_n+\frac{ag^2}{2}\sum_{n=1}^{N}L^2_n,
\end{aligned}
\end{align}
where $L_n=\sum_{i=1}^n\frac{Z_i+(-1)^i}{2}$. To perform QET, the Hamiltonian needs to be locally interacting, however the electric field contains the all-to-all interactions. For this, we decouple the gauge dynamics from the Hamiltonian by taking the weak coupling limit. Throughout this paper, whenever we refer to the Schwinger model, we mean the model at the weak coupling limit. The interacting fermion model will be studied in Sec.~\ref{sec:NJL}. The corresponding phase diagram is presented in Figs.~\ref{fig:Schwinger1} and ~\ref{fig:Schwinger2}, where the entanglement entropy between the left and right subsystems was used as an order parameter. For our numerical study we use $ma=0.5$ and consider a small system $N=4$ because computing discord is numerically expensive.

To describe the phase structure, let us consider the heavy mass limit, where $H_m(\theta)=\frac{m\cos\theta}{2}\sum_{n=1}^N(-1)^nZ_n$ is dominant. At $\theta=0$, the Neel state $\ket{01\cdots01}$ is the ground state of $H_m(\theta)$ since $Z\ket{0}=\ket{0},~Z\ket{1}=-\ket{1}$. The mass term at a a general $\theta$ is related with $H_m(\theta)=-H_m(\pi+\theta)$, therefore the ground state of $H_m(\theta)$ undergoes the transition from $\ket{01\cdots01}$ to $\ket{10\cdots01}$ as $\theta$ changes from $0$ to $\pi$. In a small mass regime, the numerical phase transition point would be shifted because of the kinetic term, which creates the entanglement in the ground state. In the rest of this subsection, we discuss this phase transition point using QET and the quantum discord. 

Computing the discord in this scenario has added layers of complexity due to there being two main issues. The first issue is how is the system divided, since it must be bipartite, there are three different valid choices for how to split the system up for $N=4$. The second issue is related to the choice of splitting the system, as the measurements available differ. For example, in the minimal model, there are only two measurement possibilities, either measuring $\mu=\pm 1$. However, suppose the system is split up such that Alice's subsystem contains two sites, and Bob's subsystem contains two sites, and the measurement and optimisation takes place in Bob's subsystem. This represents $D_A(\theta)$. For this case, Bob has four possible measurement possibilities, Bob can measure $\mu=\pm 1$ in both site 3 and site 4, but crucially, if Bob measures $\mu=1$ in site 3, then site 4 can still take values $\pm 1$. The same logic applies if the sites are switched. Therefore to correctly compute the discord, all possible measurements must be considered. 

For the $N=4$ Schwinger model, it is chosen to split the system such that Alice's subsystem has sites 1 and 2, and Bob's subsystem has sites 3 and 4. This means the possible measurements that can occur when computing the discord are described above. Once again, it is seen that the discord reproduced the result from entanglement entropy. Furthermore, the natural deduction from this is that discord is a good tool for witnessing and detecting phase transitions.

As expected, the quantum correlation decreases after Alice's measurement, however it is interesting to note that there is still a significant peak at the phase transition, even though this is now a mixed state. This in particular highlights the benefit of using discord as it demonstrates the phase transition can still be detected even if the system is not in a pure state. Additionally this peak persists even after Bob's operation. There is a slight increase in quantum discord for $D_A (\theta)$ after Bob's operation, but not as significant as in the minimal model scenario. When comparing against $D_B (\theta)$, the results are reasonably similar to $D_A (\theta)$. The slight difference that Bob's operation does not increase the quantum correlation. This is consistence with what was witnessed in the minimal model. However, unlike the minimal model, some non-trivial quantum correlation is shown to survive. This could be due to the fact that there is a phase transition in the Schwinger model.

In order to confirm the link between the teleported energy and discord, the separate plots in Figs.~\ref{fig:Schwinger1} and ~\ref{fig:Schwinger2}, can be compared. It is clear that around $\theta \sim 4.7$, there is a clear change in behaviour in both the discord and the teleported energy. This is as a result of the phase transition occurring around this region. This suggests that any significant change in behaviour in both the discord and teleported energy is indicative of a phase transition.

\subsection{\label{sec:NJL}Nambu-Jona-Lasinio model}

Here we consider the Nambu-Jona-Lasinio (NJL) model \cite{PhysRev.122.345,PhysRev.124.246}, which has been widely studied in high energy physics and condensed matter physics as an effective model of QCD \cite{PhysRevD.77.114028,COSTA2007431,Korepin:1979qq,Klevansky:1992qe,PhysRevD.36.819}. Recently the (1+1)-dimensional NJL model has been actively studied as a practically useful model for quantum simulation and quantum computation~\cite{Mishra:2019xbh,Czajka:2022plx,PhysRevD.107.L071502,PhysRevD.107.114505},

In this study, we consider the (1+1)-dimensional NJL model with chemical potential $\mu$ and chiral chemical potential $\mu_5$, whose Lagrangian density $\mathcal{L}$ and Hamiltonian densitiy $\mathcal{H}=i\psi^\dagger\partial_0\psi-\mathcal{L}$ are given as follows:
\begin{align}
\begin{aligned}
    \mathcal{L}&=\bar{\psi}(i\slashed{\partial}-m)\psi+g(\bar{\psi}\psi)^2+\mu\bar{\psi}\gamma_0\psi+\mu_5\bar{\psi}\gamma_0\gamma_5\psi,\\
\mathcal{H}&=\bar{\psi}(i\gamma_1\partial_1+m)\psi-g(\bar{\psi}\psi)^2-\mu\bar{\psi}\gamma_0\psi-\mu_5\bar{\psi}\gamma_0\gamma_5\psi. 
\end{aligned}
\end{align}
The finite chemical potential $\mu$ and chiral chemical potential $\mu_5$ are useful to study the phase structure of quarks in the presence of chiral 
imbalance. Moreover, finite $g$ introduces interactions between the particles. 

The Hamiltonian $H=\int \mathcal{H}dx$ can be approximated by the lattice Hamiltonian whose qubit representation under the open boundary condition becomes 
\begin{align}
\begin{aligned}
    H=&\frac{1}{4a}\sum_{n=1}^{N-1}\left(X_nX_{n+1}+Y_nY_{n+1}\right)+\frac{m}{2}\sum_{n=1}^N(-1)^nZ_n \\
    &+\frac{g}{2a}\sum_{n=1}^{N-1}(Z_n+1)(Z_{n+1}+1)-\frac{\mu}{2}\sum_{n=1}^NZ_n\\
    &-\frac{\mu_5}{2}\sum_{n=1}^{N-1}(X_nY_{n+1}-Y_nX_{n+1}), 
\end{aligned}
\end{align}
where $a$ is the lattice spacing constant. The QET in the NJL model without the chemical potential $\mu$ and the chiral chemical potential $\mu_5$ was studied in the previous work by one of the authors~\cite{PhysRevD.107.L071502}. 

The entanglement entropy and teleported energy are computed for this model where $N=4$. The entanglement entropy traces over the last two sites, whereas for the teleported energy, Alice performs the measurement on site 1, and Bob performs the unitary operation on site 4. The parameters are chosen such that $a=g=1$. Additionally, it would be interesting to study how quantum discord changes for different partitions of the subsystems, and subsequently how this is related to QET.

\begin{figure*}
  \centering
  \subfloat[a][This is the entanglement entropy of the ground state of the NJL model.] {\includegraphics[width=0.49\linewidth]{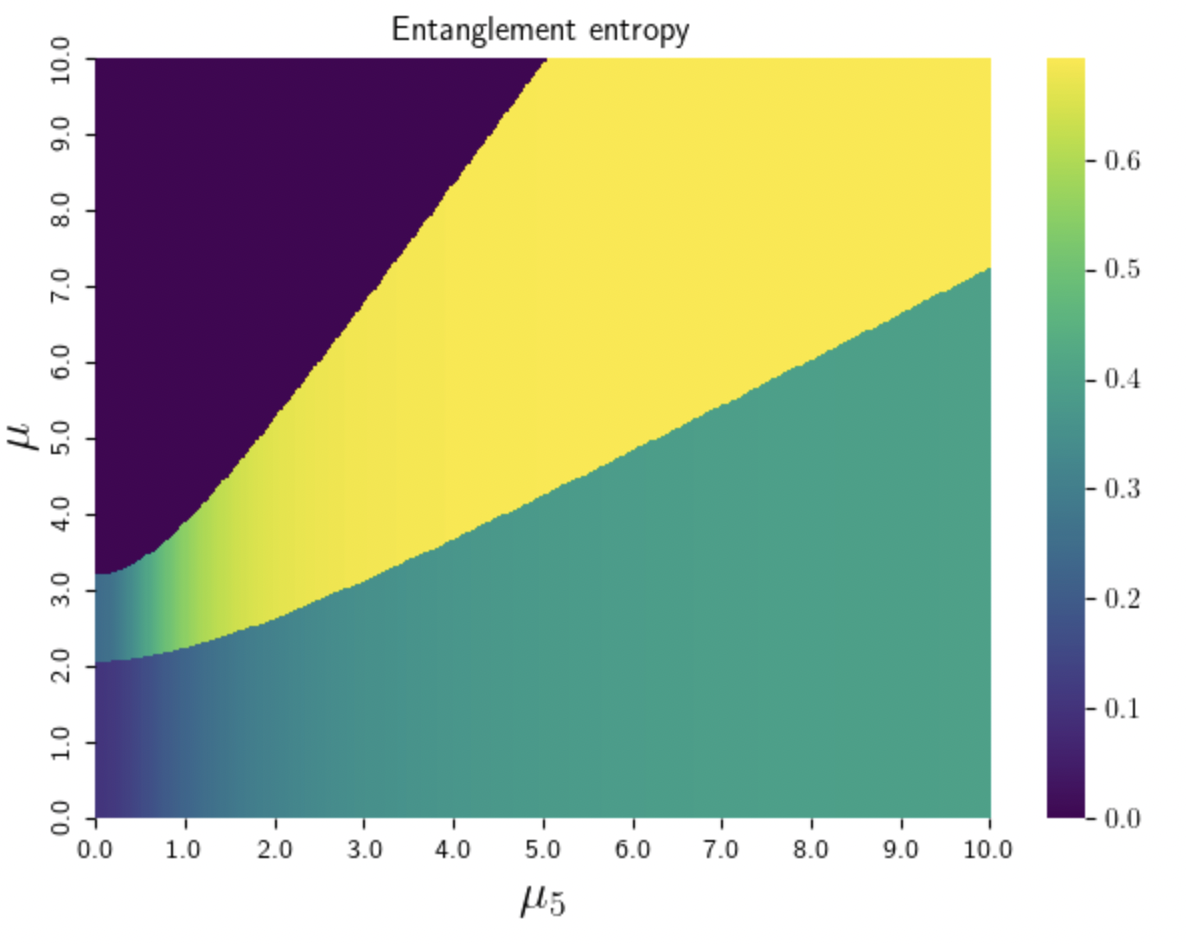}} \label{fig:a} \subfloat[a][This is the theoretical teleported energy for the NJL model, for the given parameters in the text, with $m=1$.] {\includegraphics[width=0.49\linewidth]{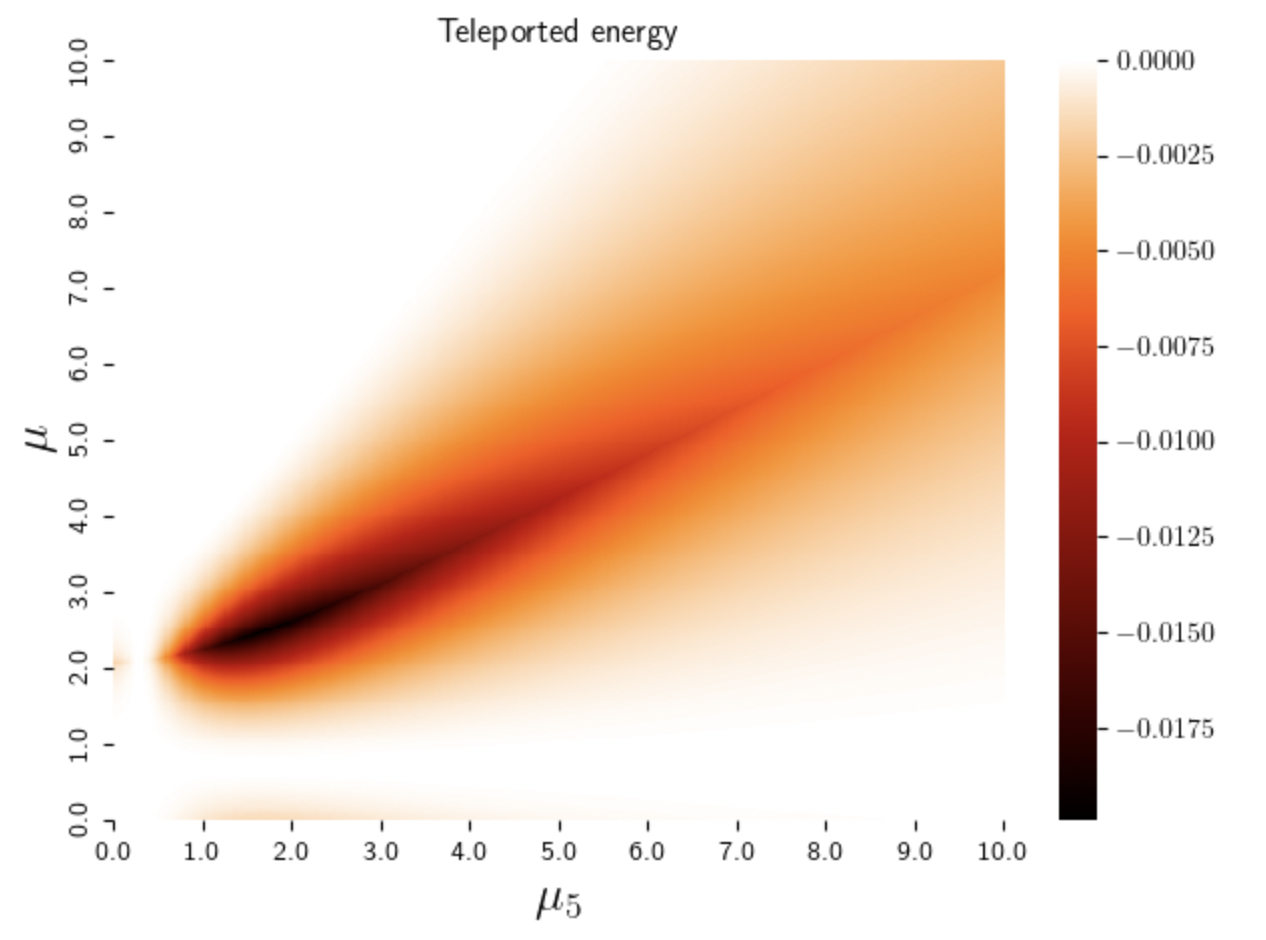}} \label{fig:g} \\
  \subfloat[a][$D_A(\mu_5)$ for the NJL model using $\mu=5$.]{\includegraphics[width=0.49\linewidth]{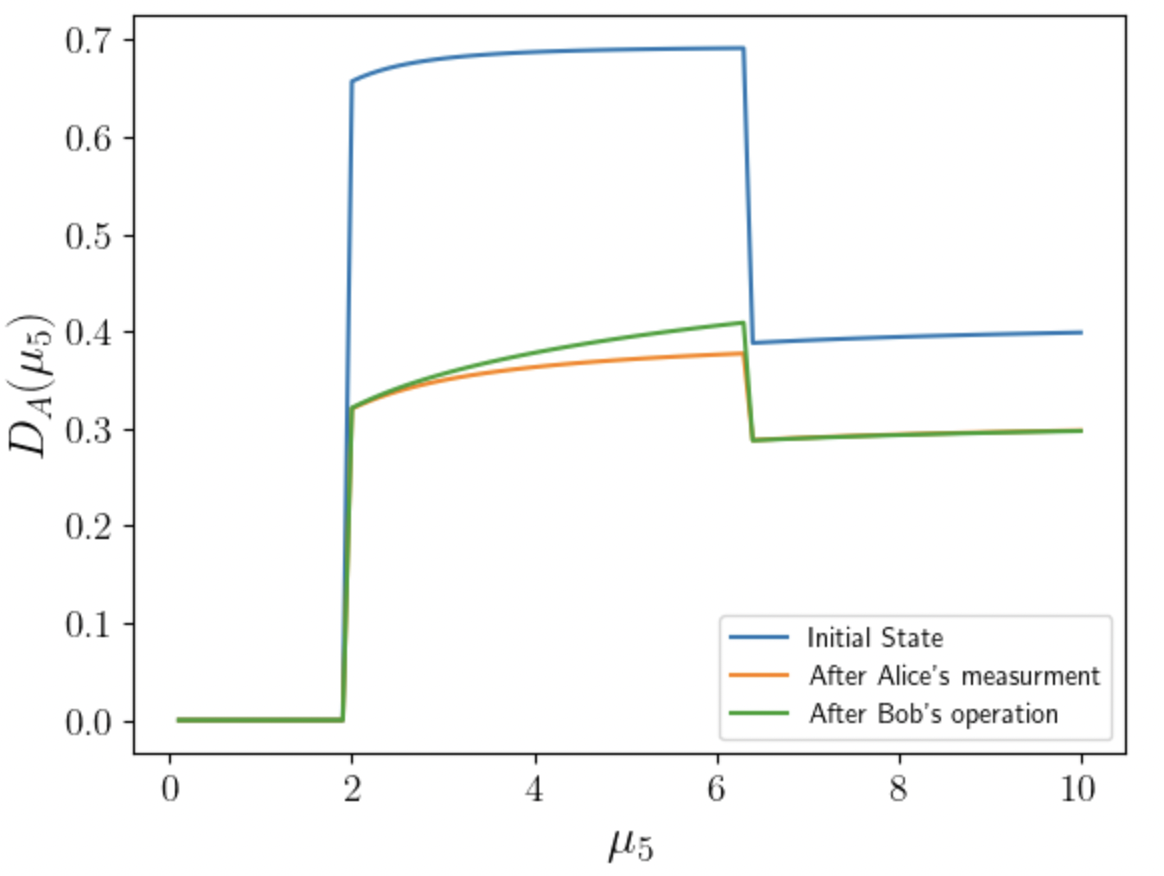}} \label{fig:a} \subfloat[a][$D_B(\mu_5)$ for the NJL model using $\mu=5$.]{\includegraphics[width=0.49\linewidth]{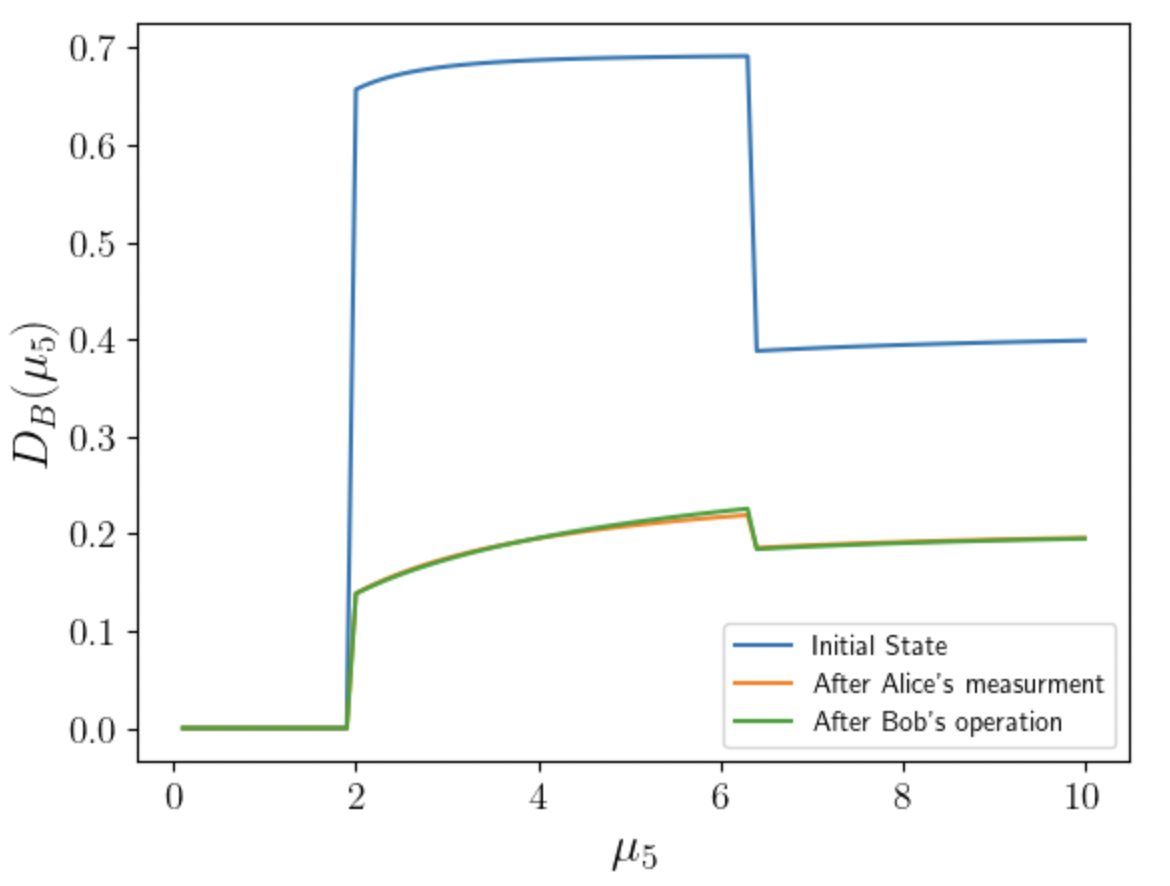}} \\
  \caption{This figure shows the quantum discord for the NJL model, where the quantum discord has been computed for both $D_A(\mu_5)$ and $D_B(\mu_5)$ in order to investigate the asymmetry. There is a clear asymmetry in the quantum discord, however quantum correlation persists, irrespective of which subsystem is measured on when computing the quantum discord. The behaviour of the phase transition is qualitatively seen in the quantum discord throughout the QET protocol.} \label{fig:NJL}
\end{figure*}

Computing the quantum discord for this model is performed in the same way as the Schwinger model, as the NJL model is chosen to have four sites, and the subsystems are chosen such that subsystem A consists of sites 1 and 2. Whereas, subsystem B consists of sites 3 and 4. The only difference is due to the discontinuity in the phase transition, therefore the condition that $x S(\hat{\rho}/x) \rightarrow 0$ was imposed when $x$ is zero. This condition emerges from Eq. (\ref{eq:10}) and the relevant definitions, where it is crucial to note that $\rho_{A|k}$ contains a factor of $1/p_k$. Therefore, we need to take into account the possibility for $p_k=0$, otherwise the numerical simulations fail due to a division of zero. However, by noting Eq. (\ref{eq:10}) contains $p_k S(\rho_{A|k})$ allows us to impose the condition that this term goes to zero whenever $p_k=0$. It is this discontinuity which demonstrates the difference in the phase transition between the Schwinger model and the NJL model, as in the former, the transition is continuous so imposing this condition is not necessary.

Moreover, it is interesting that the asymmetry persists in the quantum discord, depending on which subsystem is measured. Given this occurs for all three models considered, this implies that the QET protocol introduces an asymmetric quantum discord. This is seen by considering the ground state where the quantum discord is the same, irrespective over which subsystem is measured. Whether this asymmetry could be used for practical benefit in future protocol design is an open question.

\section{Conclusion and Discussion}
The main focus of this work has been to demonstrate the existence of non-trivial quantum correlation in QET. This is especially prevalent in the QET protocol when mixed states emerge due to measurements and the conditional local operations. To show this, we have demonstrated the complex interplay between QET and quantum discord, both theoretically and numerically to justify our claims. It is shown that the QET protocol introduces asymmetry into the quantum discord, which shows an abrupt change around the phase transition point. This has been demonstrated for various models, including the massive fermion with a topological term (the Schwinger model at the weak coupling limit) and the NJL model with both chiral chemical potential $\mu_5$ and chemical potential $\mu$.

Our work provided the first analysis of the quantum correlation in the process of QET. Despite the computational hardness of the quantum discord in a many-body quantum system, we gave a significant first step to understanding the relation between the quantum discord in the post-measurement mixed state and the teleported energy. We speculate that our approach to the problems will shed a new light into a lot of interesting properties of quantum many body systems, especially from perspectives of quantum information theory and technology. 

\section*{Acknowledgement}
The work was supported by the U.S. Department of Energy, Office of Science, National Quantum Information Science Research Centers, Co-design Center for Quantum Advantage (C2QA) under Contract No.DE-SC0012704 and from UKRI under contract No.ST/Y004965/1. KI thanks Masahiro Hotta for useful communication at Stony Brook, especially for the discussion about his work on quantum discord~\cite{trevison2015quantum}. AL also is grateful to Prof. Dmitry Kharzeev for his hospitality at Stony Brook University.

\section*{Data availability}The quantum computing code of QET used for this work is available in GitHub~\cite{Ikeda_Quantum_Energy_Teleportation_2023}, and the other data and codes will be available upon reasonable request to the authors.

\bibliographystyle{apsrev4-1}
\bibliography{ref}

\begin{thebibliography}{117}%
\makeatletter
\providecommand \@ifxundefined [1]{%
 \@ifx{#1\undefined}
}%
\providecommand \@ifnum [1]{%
 \ifnum #1\expandafter \@firstoftwo
 \else \expandafter \@secondoftwo
 \fi
}%
\providecommand \@ifx [1]{%
 \ifx #1\expandafter \@firstoftwo
 \else \expandafter \@secondoftwo
 \fi
}%
\providecommand \natexlab [1]{#1}%
\providecommand \enquote  [1]{``#1''}%
\providecommand \bibnamefont  [1]{#1}%
\providecommand \bibfnamefont [1]{#1}%
\providecommand \citenamefont [1]{#1}%
\providecommand \href@noop [0]{\@secondoftwo}%
\providecommand \href [0]{\begingroup \@sanitize@url \@href}%
\providecommand \@href[1]{\@@startlink{#1}\@@href}%
\providecommand \@@href[1]{\endgroup#1\@@endlink}%
\providecommand \@sanitize@url [0]{\catcode `\\12\catcode `\$12\catcode
  `\&12\catcode `\#12\catcode `\^12\catcode `\_12\catcode `\%12\relax}%
\providecommand \@@startlink[1]{}%
\providecommand \@@endlink[0]{}%
\providecommand \url  [0]{\begingroup\@sanitize@url \@url }%
\providecommand \@url [1]{\endgroup\@href {#1}{\urlprefix }}%
\providecommand \urlprefix  [0]{URL }%
\providecommand \Eprint [0]{\href }%
\providecommand \doibase [0]{http://dx.doi.org/}%
\providecommand \selectlanguage [0]{\@gobble}%
\providecommand \bibinfo  [0]{\@secondoftwo}%
\providecommand \bibfield  [0]{\@secondoftwo}%
\providecommand \translation [1]{[#1]}%
\providecommand \BibitemOpen [0]{}%
\providecommand \bibitemStop [0]{}%
\providecommand \bibitemNoStop [0]{.\EOS\space}%
\providecommand \EOS [0]{\spacefactor3000\relax}%
\providecommand \BibitemShut  [1]{\csname bibitem#1\endcsname}%
\let\auto@bib@innerbib\@empty
\bibitem [{\citenamefont {Horodecki}\ \emph {et~al.}(2009)\citenamefont
  {Horodecki}, \citenamefont {Horodecki}, \citenamefont {Horodecki},\ and\
  \citenamefont {Horodecki}}]{RevModPhys.81.865}%
  \BibitemOpen
  \bibfield  {author} {\bibinfo {author} {\bibfnamefont {R.}~\bibnamefont
  {Horodecki}}, \bibinfo {author} {\bibfnamefont {P.}~\bibnamefont
  {Horodecki}}, \bibinfo {author} {\bibfnamefont {M.}~\bibnamefont
  {Horodecki}}, \ and\ \bibinfo {author} {\bibfnamefont {K.}~\bibnamefont
  {Horodecki}},\ }\href {\doibase 10.1103/RevModPhys.81.865} {\bibfield
  {journal} {\bibinfo  {journal} {Rev. Mod. Phys.}\ }\textbf {\bibinfo {volume}
  {81}},\ \bibinfo {pages} {865} (\bibinfo {year} {2009})}\BibitemShut
  {NoStop}%
\bibitem [{\citenamefont {Henderson}\ and\ \citenamefont
  {Vedral}(2001)}]{L_Henderson_2001}%
  \BibitemOpen
  \bibfield  {author} {\bibinfo {author} {\bibfnamefont {L.}~\bibnamefont
  {Henderson}}\ and\ \bibinfo {author} {\bibfnamefont {V.}~\bibnamefont
  {Vedral}},\ }\href {\doibase 10.1088/0305-4470/34/35/315} {\bibfield
  {journal} {\bibinfo  {journal} {Journal of Physics A: Mathematical and
  General}\ }\textbf {\bibinfo {volume} {34}},\ \bibinfo {pages} {6899}
  (\bibinfo {year} {2001})}\BibitemShut {NoStop}%
\bibitem [{\citenamefont {Wiseman}\ \emph {et~al.}(2007)\citenamefont
  {Wiseman}, \citenamefont {Jones},\ and\ \citenamefont
  {Doherty}}]{PhysRevLett.98.140402}%
  \BibitemOpen
  \bibfield  {author} {\bibinfo {author} {\bibfnamefont {H.~M.}\ \bibnamefont
  {Wiseman}}, \bibinfo {author} {\bibfnamefont {S.~J.}\ \bibnamefont {Jones}},
  \ and\ \bibinfo {author} {\bibfnamefont {A.~C.}\ \bibnamefont {Doherty}},\
  }\href {\doibase 10.1103/PhysRevLett.98.140402} {\bibfield  {journal}
  {\bibinfo  {journal} {Phys. Rev. Lett.}\ }\textbf {\bibinfo {volume} {98}},\
  \bibinfo {pages} {140402} (\bibinfo {year} {2007})}\BibitemShut {NoStop}%
\bibitem [{\citenamefont {Wootters}(1998)}]{PhysRevLett.80.2245}%
  \BibitemOpen
  \bibfield  {author} {\bibinfo {author} {\bibfnamefont {W.~K.}\ \bibnamefont
  {Wootters}},\ }\href {\doibase 10.1103/PhysRevLett.80.2245} {\bibfield
  {journal} {\bibinfo  {journal} {Phys. Rev. Lett.}\ }\textbf {\bibinfo
  {volume} {80}},\ \bibinfo {pages} {2245} (\bibinfo {year}
  {1998})}\BibitemShut {NoStop}%
\bibitem [{\citenamefont {Girolami}\ \emph {et~al.}(2013)\citenamefont
  {Girolami}, \citenamefont {Tufarelli},\ and\ \citenamefont
  {Adesso}}]{PhysRevLett.110.240402}%
  \BibitemOpen
  \bibfield  {author} {\bibinfo {author} {\bibfnamefont {D.}~\bibnamefont
  {Girolami}}, \bibinfo {author} {\bibfnamefont {T.}~\bibnamefont {Tufarelli}},
  \ and\ \bibinfo {author} {\bibfnamefont {G.}~\bibnamefont {Adesso}},\ }\href
  {\doibase 10.1103/PhysRevLett.110.240402} {\bibfield  {journal} {\bibinfo
  {journal} {Phys. Rev. Lett.}\ }\textbf {\bibinfo {volume} {110}},\ \bibinfo
  {pages} {240402} (\bibinfo {year} {2013})}\BibitemShut {NoStop}%
\bibitem [{\citenamefont {Girolami}\ \emph {et~al.}(2014)\citenamefont
  {Girolami}, \citenamefont {Souza}, \citenamefont {Giovannetti}, \citenamefont
  {Tufarelli}, \citenamefont {Filgueiras}, \citenamefont {Sarthour},
  \citenamefont {Soares-Pinto}, \citenamefont {Oliveira},\ and\ \citenamefont
  {Adesso}}]{PhysRevLett.112.210401}%
  \BibitemOpen
  \bibfield  {author} {\bibinfo {author} {\bibfnamefont {D.}~\bibnamefont
  {Girolami}}, \bibinfo {author} {\bibfnamefont {A.~M.}\ \bibnamefont {Souza}},
  \bibinfo {author} {\bibfnamefont {V.}~\bibnamefont {Giovannetti}}, \bibinfo
  {author} {\bibfnamefont {T.}~\bibnamefont {Tufarelli}}, \bibinfo {author}
  {\bibfnamefont {J.~G.}\ \bibnamefont {Filgueiras}}, \bibinfo {author}
  {\bibfnamefont {R.~S.}\ \bibnamefont {Sarthour}}, \bibinfo {author}
  {\bibfnamefont {D.~O.}\ \bibnamefont {Soares-Pinto}}, \bibinfo {author}
  {\bibfnamefont {I.~S.}\ \bibnamefont {Oliveira}}, \ and\ \bibinfo {author}
  {\bibfnamefont {G.}~\bibnamefont {Adesso}},\ }\href {\doibase
  10.1103/PhysRevLett.112.210401} {\bibfield  {journal} {\bibinfo  {journal}
  {Phys. Rev. Lett.}\ }\textbf {\bibinfo {volume} {112}},\ \bibinfo {pages}
  {210401} (\bibinfo {year} {2014})}\BibitemShut {NoStop}%
\bibitem [{\citenamefont {Streltsov}\ \emph {et~al.}(2015)\citenamefont
  {Streltsov}, \citenamefont {Singh}, \citenamefont {Dhar}, \citenamefont
  {Bera},\ and\ \citenamefont {Adesso}}]{PhysRevLett.115.020403}%
  \BibitemOpen
  \bibfield  {author} {\bibinfo {author} {\bibfnamefont {A.}~\bibnamefont
  {Streltsov}}, \bibinfo {author} {\bibfnamefont {U.}~\bibnamefont {Singh}},
  \bibinfo {author} {\bibfnamefont {H.~S.}\ \bibnamefont {Dhar}}, \bibinfo
  {author} {\bibfnamefont {M.~N.}\ \bibnamefont {Bera}}, \ and\ \bibinfo
  {author} {\bibfnamefont {G.}~\bibnamefont {Adesso}},\ }\href {\doibase
  10.1103/PhysRevLett.115.020403} {\bibfield  {journal} {\bibinfo  {journal}
  {Phys. Rev. Lett.}\ }\textbf {\bibinfo {volume} {115}},\ \bibinfo {pages}
  {020403} (\bibinfo {year} {2015})}\BibitemShut {NoStop}%
\bibitem [{\citenamefont {Morris}\ \emph {et~al.}(2020)\citenamefont {Morris},
  \citenamefont {Yadin}, \citenamefont {Fadel}, \citenamefont {Zibold},
  \citenamefont {Treutlein},\ and\ \citenamefont
  {Adesso}}]{PhysRevX.10.041012}%
  \BibitemOpen
  \bibfield  {author} {\bibinfo {author} {\bibfnamefont {B.}~\bibnamefont
  {Morris}}, \bibinfo {author} {\bibfnamefont {B.}~\bibnamefont {Yadin}},
  \bibinfo {author} {\bibfnamefont {M.}~\bibnamefont {Fadel}}, \bibinfo
  {author} {\bibfnamefont {T.}~\bibnamefont {Zibold}}, \bibinfo {author}
  {\bibfnamefont {P.}~\bibnamefont {Treutlein}}, \ and\ \bibinfo {author}
  {\bibfnamefont {G.}~\bibnamefont {Adesso}},\ }\href {\doibase
  10.1103/PhysRevX.10.041012} {\bibfield  {journal} {\bibinfo  {journal} {Phys.
  Rev. X}\ }\textbf {\bibinfo {volume} {10}},\ \bibinfo {pages} {041012}
  (\bibinfo {year} {2020})}\BibitemShut {NoStop}%
\bibitem [{\citenamefont {Hotta}(2010)}]{Hotta_2010}%
  \BibitemOpen
  \bibfield  {author} {\bibinfo {author} {\bibfnamefont {M.}~\bibnamefont
  {Hotta}},\ }\href {\doibase 10.1088/1751-8113/43/10/105305} {\bibfield
  {journal} {\bibinfo  {journal} {Journal of Physics A: Mathematical and
  Theoretical}\ }\textbf {\bibinfo {volume} {43}},\ \bibinfo {pages} {105305}
  (\bibinfo {year} {2010})}\BibitemShut {NoStop}%
\bibitem [{\citenamefont {Hotta}(2008)}]{HOTTA20085671}%
  \BibitemOpen
  \bibfield  {author} {\bibinfo {author} {\bibfnamefont {M.}~\bibnamefont
  {Hotta}},\ }\href {\doibase https://doi.org/10.1016/j.physleta.2008.07.007}
  {\bibfield  {journal} {\bibinfo  {journal} {Physics Letters A}\ }\textbf
  {\bibinfo {volume} {372}},\ \bibinfo {pages} {5671} (\bibinfo {year}
  {2008})}\BibitemShut {NoStop}%
\bibitem [{\citenamefont {{Ikeda}}(2023)}]{2023arXiv230111884I}%
  \BibitemOpen
  \bibfield  {author} {\bibinfo {author} {\bibfnamefont {K.}~\bibnamefont
  {{Ikeda}}},\ }\href {\doibase 10.48550/arXiv.2301.11884} {\bibfield
  {journal} {\bibinfo  {journal} {to appear in IET Quantum Communication}\ }
  (\bibinfo {year} {2023}),\ 10.48550/arXiv.2301.11884},\ \Eprint
  {http://arxiv.org/abs/2301.11884} {arXiv:2301.11884 [quant-ph]} \BibitemShut
  {NoStop}%
\bibitem [{\citenamefont {Ikeda}(2023{\natexlab{a}})}]{Ikeda:2023uni}%
  \BibitemOpen
  \bibfield  {author} {\bibinfo {author} {\bibfnamefont {K.}~\bibnamefont
  {Ikeda}},\ }\href {\doibase 10.1103/PhysRevApplied.20.024051} {\bibfield
  {journal} {\bibinfo  {journal} {Phys. Rev. Applied}\ }\textbf {\bibinfo
  {volume} {20}},\ \bibinfo {pages} {024051} (\bibinfo {year}
  {2023}{\natexlab{a}})},\ \Eprint {http://arxiv.org/abs/2301.02666}
  {arXiv:2301.02666 [quant-ph]} \BibitemShut {NoStop}%
\bibitem [{\citenamefont {Ikeda}(2023{\natexlab{b}})}]{Ikeda:2023xmf}%
  \BibitemOpen
  \bibfield  {author} {\bibinfo {author} {\bibfnamefont {K.}~\bibnamefont
  {Ikeda}},\ }\href {\doibase 10.1116/5.0164999} {\bibfield  {journal}
  {\bibinfo  {journal} {AVS Quantum Sci.}\ }\textbf {\bibinfo {volume} {5}},\
  \bibinfo {pages} {035002} (\bibinfo {year} {2023}{\natexlab{b}})},\ \Eprint
  {http://arxiv.org/abs/2302.09630} {arXiv:2302.09630 [quant-ph]} \BibitemShut
  {NoStop}%
\bibitem [{\citenamefont {Ikeda}\ and\ \citenamefont
  {Lowe}(2023)}]{2023arXiv230608242I}%
  \BibitemOpen
  \bibfield  {author} {\bibinfo {author} {\bibfnamefont {K.}~\bibnamefont
  {Ikeda}}\ and\ \bibinfo {author} {\bibfnamefont {A.}~\bibnamefont {Lowe}},\
  }\href@noop {} {\  (\bibinfo {year} {2023})},\ \Eprint
  {http://arxiv.org/abs/2306.08242} {arXiv:2306.08242 [quant-ph]} \BibitemShut
  {NoStop}%
\bibitem [{\citenamefont {Ikeda}\ \emph
  {et~al.}(2023{\natexlab{a}})\citenamefont {Ikeda}, \citenamefont {Singh},\
  and\ \citenamefont {Slager}}]{Ikeda:2023ljh}%
  \BibitemOpen
  \bibfield  {author} {\bibinfo {author} {\bibfnamefont {K.}~\bibnamefont
  {Ikeda}}, \bibinfo {author} {\bibfnamefont {R.}~\bibnamefont {Singh}}, \ and\
  \bibinfo {author} {\bibfnamefont {R.-J.}\ \bibnamefont {Slager}},\
  }\href@noop {} {\  (\bibinfo {year} {2023}{\natexlab{a}})},\ \Eprint
  {http://arxiv.org/abs/2310.15936} {arXiv:2310.15936 [quant-ph]} \BibitemShut
  {NoStop}%
\bibitem [{\citenamefont {Nambu}\ and\ \citenamefont
  {Hotta}(2010)}]{PhysRevA.82.042329}%
  \BibitemOpen
  \bibfield  {author} {\bibinfo {author} {\bibfnamefont {Y.}~\bibnamefont
  {Nambu}}\ and\ \bibinfo {author} {\bibfnamefont {M.}~\bibnamefont {Hotta}},\
  }\href {\doibase 10.1103/PhysRevA.82.042329} {\bibfield  {journal} {\bibinfo
  {journal} {Phys. Rev. A}\ }\textbf {\bibinfo {volume} {82}},\ \bibinfo
  {pages} {042329} (\bibinfo {year} {2010})}\BibitemShut {NoStop}%
\bibitem [{\citenamefont {Yusa}\ \emph {et~al.}(2011)\citenamefont {Yusa},
  \citenamefont {Izumida},\ and\ \citenamefont {Hotta}}]{PhysRevA.84.032336}%
  \BibitemOpen
  \bibfield  {author} {\bibinfo {author} {\bibfnamefont {G.}~\bibnamefont
  {Yusa}}, \bibinfo {author} {\bibfnamefont {W.}~\bibnamefont {Izumida}}, \
  and\ \bibinfo {author} {\bibfnamefont {M.}~\bibnamefont {Hotta}},\ }\href
  {\doibase 10.1103/PhysRevA.84.032336} {\bibfield  {journal} {\bibinfo
  {journal} {Phys. Rev. A}\ }\textbf {\bibinfo {volume} {84}},\ \bibinfo
  {pages} {032336} (\bibinfo {year} {2011})}\BibitemShut {NoStop}%
\bibitem [{\citenamefont {Ollivier}\ and\ \citenamefont
  {Zurek}(2001)}]{PhysRevLett.88.017901}%
  \BibitemOpen
  \bibfield  {author} {\bibinfo {author} {\bibfnamefont {H.}~\bibnamefont
  {Ollivier}}\ and\ \bibinfo {author} {\bibfnamefont {W.~H.}\ \bibnamefont
  {Zurek}},\ }\href {\doibase 10.1103/PhysRevLett.88.017901} {\bibfield
  {journal} {\bibinfo  {journal} {Phys. Rev. Lett.}\ }\textbf {\bibinfo
  {volume} {88}},\ \bibinfo {pages} {017901} (\bibinfo {year}
  {2001})}\BibitemShut {NoStop}%
\bibitem [{\citenamefont {Bennett}\ \emph {et~al.}(1996)\citenamefont
  {Bennett}, \citenamefont {DiVincenzo}, \citenamefont {Smolin},\ and\
  \citenamefont {Wootters}}]{PhysRevA.54.3824}%
  \BibitemOpen
  \bibfield  {author} {\bibinfo {author} {\bibfnamefont {C.~H.}\ \bibnamefont
  {Bennett}}, \bibinfo {author} {\bibfnamefont {D.~P.}\ \bibnamefont
  {DiVincenzo}}, \bibinfo {author} {\bibfnamefont {J.~A.}\ \bibnamefont
  {Smolin}}, \ and\ \bibinfo {author} {\bibfnamefont {W.~K.}\ \bibnamefont
  {Wootters}},\ }\href {\doibase 10.1103/PhysRevA.54.3824} {\bibfield
  {journal} {\bibinfo  {journal} {Phys. Rev. A}\ }\textbf {\bibinfo {volume}
  {54}},\ \bibinfo {pages} {3824} (\bibinfo {year} {1996})}\BibitemShut
  {NoStop}%
\bibitem [{\citenamefont {Werlang}\ \emph {et~al.}(2009)\citenamefont
  {Werlang}, \citenamefont {Souza}, \citenamefont {Fanchini},\ and\
  \citenamefont {Villas~Boas}}]{PhysRevA.80.024103}%
  \BibitemOpen
  \bibfield  {author} {\bibinfo {author} {\bibfnamefont {T.}~\bibnamefont
  {Werlang}}, \bibinfo {author} {\bibfnamefont {S.}~\bibnamefont {Souza}},
  \bibinfo {author} {\bibfnamefont {F.~F.}\ \bibnamefont {Fanchini}}, \ and\
  \bibinfo {author} {\bibfnamefont {C.~J.}\ \bibnamefont {Villas~Boas}},\
  }\href {\doibase 10.1103/PhysRevA.80.024103} {\bibfield  {journal} {\bibinfo
  {journal} {Phys. Rev. A}\ }\textbf {\bibinfo {volume} {80}},\ \bibinfo
  {pages} {024103} (\bibinfo {year} {2009})}\BibitemShut {NoStop}%
\bibitem [{\citenamefont {Huang}(2014)}]{Huang_2014}%
  \BibitemOpen
  \bibfield  {author} {\bibinfo {author} {\bibfnamefont {Y.}~\bibnamefont
  {Huang}},\ }\href {\doibase 10.1088/1367-2630/16/3/033027} {\bibfield
  {journal} {\bibinfo  {journal} {New Journal of Physics}\ }\textbf {\bibinfo
  {volume} {16}},\ \bibinfo {pages} {033027} (\bibinfo {year}
  {2014})}\BibitemShut {NoStop}%
\bibitem [{\citenamefont {Daki\ifmmode~\acute{c}\else \'{c}\fi{}}\ \emph
  {et~al.}(2010)\citenamefont {Daki\ifmmode~\acute{c}\else \'{c}\fi{}},
  \citenamefont {Vedral},\ and\ \citenamefont
  {Brukner}}]{PhysRevLett.105.190502}%
  \BibitemOpen
  \bibfield  {author} {\bibinfo {author} {\bibfnamefont {B.}~\bibnamefont
  {Daki\ifmmode~\acute{c}\else \'{c}\fi{}}}, \bibinfo {author} {\bibfnamefont
  {V.}~\bibnamefont {Vedral}}, \ and\ \bibinfo {author} {\bibfnamefont
  {i.~c.~v.}\ \bibnamefont {Brukner}},\ }\href {\doibase
  10.1103/PhysRevLett.105.190502} {\bibfield  {journal} {\bibinfo  {journal}
  {Phys. Rev. Lett.}\ }\textbf {\bibinfo {volume} {105}},\ \bibinfo {pages}
  {190502} (\bibinfo {year} {2010})}\BibitemShut {NoStop}%
\bibitem [{\citenamefont {Lowe}\ and\ \citenamefont {Yurkevich}(2022)}]{lowe}%
  \BibitemOpen
  \bibfield  {author} {\bibinfo {author} {\bibfnamefont {A.}~\bibnamefont
  {Lowe}}\ and\ \bibinfo {author} {\bibfnamefont {I.~V.}\ \bibnamefont
  {Yurkevich}},\ }\href {\doibase 10.1063/10.00102047} {\bibfield  {journal}
  {\bibinfo  {journal} {Low Temperature Physics}\ }\textbf {\bibinfo {volume}
  {48}},\ \bibinfo {pages} {396} (\bibinfo {year} {2022})}\BibitemShut
  {NoStop}%
\bibitem [{\citenamefont {Piani}(2012)}]{PhysRevA.86.034101}%
  \BibitemOpen
  \bibfield  {author} {\bibinfo {author} {\bibfnamefont {M.}~\bibnamefont
  {Piani}},\ }\href {\doibase 10.1103/PhysRevA.86.034101} {\bibfield  {journal}
  {\bibinfo  {journal} {Phys. Rev. A}\ }\textbf {\bibinfo {volume} {86}},\
  \bibinfo {pages} {034101} (\bibinfo {year} {2012})}\BibitemShut {NoStop}%
\bibitem [{\citenamefont {Skinner}\ \emph {et~al.}(2019)\citenamefont
  {Skinner}, \citenamefont {Ruhman},\ and\ \citenamefont
  {Nahum}}]{PhysRevX.9.031009}%
  \BibitemOpen
  \bibfield  {author} {\bibinfo {author} {\bibfnamefont {B.}~\bibnamefont
  {Skinner}}, \bibinfo {author} {\bibfnamefont {J.}~\bibnamefont {Ruhman}}, \
  and\ \bibinfo {author} {\bibfnamefont {A.}~\bibnamefont {Nahum}},\ }\href
  {\doibase 10.1103/PhysRevX.9.031009} {\bibfield  {journal} {\bibinfo
  {journal} {Phys. Rev. X}\ }\textbf {\bibinfo {volume} {9}},\ \bibinfo {pages}
  {031009} (\bibinfo {year} {2019})}\BibitemShut {NoStop}%
\bibitem [{\citenamefont {Li}\ \emph {et~al.}(2018)\citenamefont {Li},
  \citenamefont {Chen},\ and\ \citenamefont {Fisher}}]{PhysRevB.98.205136}%
  \BibitemOpen
  \bibfield  {author} {\bibinfo {author} {\bibfnamefont {Y.}~\bibnamefont
  {Li}}, \bibinfo {author} {\bibfnamefont {X.}~\bibnamefont {Chen}}, \ and\
  \bibinfo {author} {\bibfnamefont {M.~P.~A.}\ \bibnamefont {Fisher}},\ }\href
  {\doibase 10.1103/PhysRevB.98.205136} {\bibfield  {journal} {\bibinfo
  {journal} {Phys. Rev. B}\ }\textbf {\bibinfo {volume} {98}},\ \bibinfo
  {pages} {205136} (\bibinfo {year} {2018})}\BibitemShut {NoStop}%
\bibitem [{\citenamefont {Carisch}\ \emph {et~al.}(2023)\citenamefont
  {Carisch}, \citenamefont {Romito},\ and\ \citenamefont
  {Zilberberg}}]{PhysRevResearch.5.L042031}%
  \BibitemOpen
  \bibfield  {author} {\bibinfo {author} {\bibfnamefont {C.}~\bibnamefont
  {Carisch}}, \bibinfo {author} {\bibfnamefont {A.}~\bibnamefont {Romito}}, \
  and\ \bibinfo {author} {\bibfnamefont {O.}~\bibnamefont {Zilberberg}},\
  }\href {\doibase 10.1103/PhysRevResearch.5.L042031} {\bibfield  {journal}
  {\bibinfo  {journal} {Phys. Rev. Res.}\ }\textbf {\bibinfo {volume} {5}},\
  \bibinfo {pages} {L042031} (\bibinfo {year} {2023})}\BibitemShut {NoStop}%
\bibitem [{\citenamefont {Hoke}\ \emph {et~al.}(2023)\citenamefont {Hoke} \emph
  {et~al.}}]{meas_ent}%
  \BibitemOpen
  \bibfield  {author} {\bibinfo {author} {\bibfnamefont {J.~C.}\ \bibnamefont
  {Hoke}} \emph {et~al.},\ }\href {\doibase 10.1038/s41586-023-06505-7}
  {\bibfield  {journal} {\bibinfo  {journal} {Nature}\ }\textbf {\bibinfo
  {volume} {622}},\ \bibinfo {pages} {481} (\bibinfo {year}
  {2023})}\BibitemShut {NoStop}%
\bibitem [{\citenamefont {Kumar}\ \emph {et~al.}(2020)\citenamefont {Kumar},
  \citenamefont {Snizhko},\ and\ \citenamefont
  {Gefen}}]{PhysRevResearch.2.042014}%
  \BibitemOpen
  \bibfield  {author} {\bibinfo {author} {\bibfnamefont {P.}~\bibnamefont
  {Kumar}}, \bibinfo {author} {\bibfnamefont {K.}~\bibnamefont {Snizhko}}, \
  and\ \bibinfo {author} {\bibfnamefont {Y.}~\bibnamefont {Gefen}},\ }\href
  {\doibase 10.1103/PhysRevResearch.2.042014} {\bibfield  {journal} {\bibinfo
  {journal} {Phys. Rev. Res.}\ }\textbf {\bibinfo {volume} {2}},\ \bibinfo
  {pages} {042014} (\bibinfo {year} {2020})}\BibitemShut {NoStop}%
\bibitem [{\citenamefont {Roy}\ \emph {et~al.}(2020)\citenamefont {Roy},
  \citenamefont {Chalker}, \citenamefont {Gornyi},\ and\ \citenamefont
  {Gefen}}]{PhysRevResearch.2.033347}%
  \BibitemOpen
  \bibfield  {author} {\bibinfo {author} {\bibfnamefont {S.}~\bibnamefont
  {Roy}}, \bibinfo {author} {\bibfnamefont {J.~T.}\ \bibnamefont {Chalker}},
  \bibinfo {author} {\bibfnamefont {I.~V.}\ \bibnamefont {Gornyi}}, \ and\
  \bibinfo {author} {\bibfnamefont {Y.}~\bibnamefont {Gefen}},\ }\href
  {\doibase 10.1103/PhysRevResearch.2.033347} {\bibfield  {journal} {\bibinfo
  {journal} {Phys. Rev. Res.}\ }\textbf {\bibinfo {volume} {2}},\ \bibinfo
  {pages} {033347} (\bibinfo {year} {2020})}\BibitemShut {NoStop}%
\bibitem [{\citenamefont {Lowe}\ and\ \citenamefont
  {Medina-Guerra}(2024)}]{arXiv:2401.09304}%
  \BibitemOpen
  \bibfield  {author} {\bibinfo {author} {\bibfnamefont {A.}~\bibnamefont
  {Lowe}}\ and\ \bibinfo {author} {\bibfnamefont {E.}~\bibnamefont
  {Medina-Guerra}},\ }\href@noop {} {\  (\bibinfo {year} {2024})},\ \Eprint
  {http://arxiv.org/abs/2401.09304} {arXiv:2401.09304 [quant-ph]} \BibitemShut
  {NoStop}%
\bibitem [{\citenamefont {Chiara}\ and\ \citenamefont
  {Sanpera}(2018)}]{De_Chiara_2018}%
  \BibitemOpen
  \bibfield  {author} {\bibinfo {author} {\bibfnamefont {G.~D.}\ \bibnamefont
  {Chiara}}\ and\ \bibinfo {author} {\bibfnamefont {A.}~\bibnamefont
  {Sanpera}},\ }\href {\doibase 10.1088/1361-6633/aabf61} {\bibfield  {journal}
  {\bibinfo  {journal} {Reports on Progress in Physics}\ }\textbf {\bibinfo
  {volume} {81}},\ \bibinfo {pages} {074002} (\bibinfo {year}
  {2018})}\BibitemShut {NoStop}%
\bibitem [{\citenamefont {Dillenschneider}(2008)}]{PhysRevB.78.224413}%
  \BibitemOpen
  \bibfield  {author} {\bibinfo {author} {\bibfnamefont {R.}~\bibnamefont
  {Dillenschneider}},\ }\href {\doibase 10.1103/PhysRevB.78.224413} {\bibfield
  {journal} {\bibinfo  {journal} {Phys. Rev. B}\ }\textbf {\bibinfo {volume}
  {78}},\ \bibinfo {pages} {224413} (\bibinfo {year} {2008})}\BibitemShut
  {NoStop}%
\bibitem [{\citenamefont {Osborne}\ and\ \citenamefont
  {Nielsen}(2002)}]{PhysRevA.66.032110}%
  \BibitemOpen
  \bibfield  {author} {\bibinfo {author} {\bibfnamefont {T.~J.}\ \bibnamefont
  {Osborne}}\ and\ \bibinfo {author} {\bibfnamefont {M.~A.}\ \bibnamefont
  {Nielsen}},\ }\href {\doibase 10.1103/PhysRevA.66.032110} {\bibfield
  {journal} {\bibinfo  {journal} {Phys. Rev. A}\ }\textbf {\bibinfo {volume}
  {66}},\ \bibinfo {pages} {032110} (\bibinfo {year} {2002})}\BibitemShut
  {NoStop}%
\bibitem [{\citenamefont {Ikeda}(2023{\natexlab{c}})}]{PhysRevD.107.L071502}%
  \BibitemOpen
  \bibfield  {author} {\bibinfo {author} {\bibfnamefont {K.}~\bibnamefont
  {Ikeda}},\ }\href {\doibase 10.1103/PhysRevD.107.L071502} {\bibfield
  {journal} {\bibinfo  {journal} {Phys. Rev. D}\ }\textbf {\bibinfo {volume}
  {107}},\ \bibinfo {pages} {L071502} (\bibinfo {year}
  {2023}{\natexlab{c}})}\BibitemShut {NoStop}%
\bibitem [{\citenamefont {Ikeda}\ \emph
  {et~al.}(2023{\natexlab{b}})\citenamefont {Ikeda}, \citenamefont {Kharzeev},
  \citenamefont {Meyer},\ and\ \citenamefont {Shi}}]{2023arXiv230500996I}%
  \BibitemOpen
  \bibfield  {author} {\bibinfo {author} {\bibfnamefont {K.}~\bibnamefont
  {Ikeda}}, \bibinfo {author} {\bibfnamefont {D.~E.}\ \bibnamefont {Kharzeev}},
  \bibinfo {author} {\bibfnamefont {R.}~\bibnamefont {Meyer}}, \ and\ \bibinfo
  {author} {\bibfnamefont {S.}~\bibnamefont {Shi}},\ }\href {\doibase
  10.1103/PhysRevD.108.L091501} {\bibfield  {journal} {\bibinfo  {journal}
  {Phys. Rev. D}\ }\textbf {\bibinfo {volume} {108}},\ \bibinfo {pages}
  {L091501} (\bibinfo {year} {2023}{\natexlab{b}})},\ \Eprint
  {http://arxiv.org/abs/2305.00996} {arXiv:2305.00996 [hep-ph]} \BibitemShut
  {NoStop}%
\bibitem [{\citenamefont {Trevison}\ and\ \citenamefont
  {Hotta}(2015)}]{trevison2015quantum}%
  \BibitemOpen
  \bibfield  {author} {\bibinfo {author} {\bibfnamefont {J.}~\bibnamefont
  {Trevison}}\ and\ \bibinfo {author} {\bibfnamefont {M.}~\bibnamefont
  {Hotta}},\ }\href {\doibase 10.1088/1751-8113/48/17/175302} {\bibfield
  {journal} {\bibinfo  {journal} {Journal of Physics A: Mathematical and
  Theoretical}\ }\textbf {\bibinfo {volume} {48}},\ \bibinfo {pages} {175302}
  (\bibinfo {year} {2015})}\BibitemShut {NoStop}%
\bibitem [{\citenamefont {Schwinger}(1962{\natexlab{a}})}]{schwinger1962gauge}%
  \BibitemOpen
  \bibfield  {author} {\bibinfo {author} {\bibfnamefont {J.~S.}\ \bibnamefont
  {Schwinger}},\ }\href {\doibase 10.1103/PhysRev.125.397} {\bibfield
  {journal} {\bibinfo  {journal} {Phys. Rev.}\ }\textbf {\bibinfo {volume}
  {125}},\ \bibinfo {pages} {397} (\bibinfo {year} {1962}{\natexlab{a}})},\
  \bibinfo {note} {[,151(1962)]}\BibitemShut {NoStop}%
\bibitem [{\citenamefont {Schwinger}(1962{\natexlab{b}})}]{Schwinger:1962tp}%
  \BibitemOpen
  \bibfield  {author} {\bibinfo {author} {\bibfnamefont {J.~S.}\ \bibnamefont
  {Schwinger}},\ }\href {\doibase 10.1103/PhysRev.128.2425} {\bibfield
  {journal} {\bibinfo  {journal} {Phys. Rev.}\ }\textbf {\bibinfo {volume}
  {128}},\ \bibinfo {pages} {2425} (\bibinfo {year}
  {1962}{\natexlab{b}})}\BibitemShut {NoStop}%
\bibitem [{\citenamefont {Nambu}\ and\ \citenamefont
  {Jona-Lasinio}(1961{\natexlab{a}})}]{PhysRev.122.345}%
  \BibitemOpen
  \bibfield  {author} {\bibinfo {author} {\bibfnamefont {Y.}~\bibnamefont
  {Nambu}}\ and\ \bibinfo {author} {\bibfnamefont {G.}~\bibnamefont
  {Jona-Lasinio}},\ }\href {\doibase 10.1103/PhysRev.122.345} {\bibfield
  {journal} {\bibinfo  {journal} {Phys. Rev.}\ }\textbf {\bibinfo {volume}
  {122}},\ \bibinfo {pages} {345} (\bibinfo {year}
  {1961}{\natexlab{a}})}\BibitemShut {NoStop}%
\bibitem [{\citenamefont {Nambu}\ and\ \citenamefont
  {Jona-Lasinio}(1961{\natexlab{b}})}]{PhysRev.124.246}%
  \BibitemOpen
  \bibfield  {author} {\bibinfo {author} {\bibfnamefont {Y.}~\bibnamefont
  {Nambu}}\ and\ \bibinfo {author} {\bibfnamefont {G.}~\bibnamefont
  {Jona-Lasinio}},\ }\href {\doibase 10.1103/PhysRev.124.246} {\bibfield
  {journal} {\bibinfo  {journal} {Phys. Rev.}\ }\textbf {\bibinfo {volume}
  {124}},\ \bibinfo {pages} {246} (\bibinfo {year}
  {1961}{\natexlab{b}})}\BibitemShut {NoStop}%
\bibitem [{\citenamefont {Thirring}(1958)}]{Thirring:1958in}%
  \BibitemOpen
  \bibfield  {author} {\bibinfo {author} {\bibfnamefont {W.~E.}\ \bibnamefont
  {Thirring}},\ }\href {\doibase 10.1016/0003-4916(58)90015-0} {\bibfield
  {journal} {\bibinfo  {journal} {Annals Phys.}\ }\textbf {\bibinfo {volume}
  {3}},\ \bibinfo {pages} {91} (\bibinfo {year} {1958})}\BibitemShut {NoStop}%
\bibitem [{\citenamefont {Gross}\ and\ \citenamefont
  {Neveu}(1974)}]{PhysRevD.10.3235}%
  \BibitemOpen
  \bibfield  {author} {\bibinfo {author} {\bibfnamefont {D.~J.}\ \bibnamefont
  {Gross}}\ and\ \bibinfo {author} {\bibfnamefont {A.}~\bibnamefont {Neveu}},\
  }\href {\doibase 10.1103/PhysRevD.10.3235} {\bibfield  {journal} {\bibinfo
  {journal} {Phys. Rev. D}\ }\textbf {\bibinfo {volume} {10}},\ \bibinfo
  {pages} {3235} (\bibinfo {year} {1974})}\BibitemShut {NoStop}%
\bibitem [{\citenamefont {Wallraff}\ \emph {et~al.}(2004)\citenamefont
  {Wallraff}, \citenamefont {Schuster}, \citenamefont {Blais}, \citenamefont
  {Frunzio}, \citenamefont {Huang}, \citenamefont {Majer}, \citenamefont
  {Kumar}, \citenamefont {Girvin},\ and\ \citenamefont
  {Schoelkopf}}]{wallraff2004strong}%
  \BibitemOpen
  \bibfield  {author} {\bibinfo {author} {\bibfnamefont {A.}~\bibnamefont
  {Wallraff}}, \bibinfo {author} {\bibfnamefont {D.~I.}\ \bibnamefont
  {Schuster}}, \bibinfo {author} {\bibfnamefont {A.}~\bibnamefont {Blais}},
  \bibinfo {author} {\bibfnamefont {L.}~\bibnamefont {Frunzio}}, \bibinfo
  {author} {\bibfnamefont {R.~S.}\ \bibnamefont {Huang}}, \bibinfo {author}
  {\bibfnamefont {J.}~\bibnamefont {Majer}}, \bibinfo {author} {\bibfnamefont
  {S.}~\bibnamefont {Kumar}}, \bibinfo {author} {\bibfnamefont {S.~M.}\
  \bibnamefont {Girvin}}, \ and\ \bibinfo {author} {\bibfnamefont {R.~J.}\
  \bibnamefont {Schoelkopf}},\ }\href {\doibase 10.1038/nature02851} {\bibfield
   {journal} {\bibinfo  {journal} {Nature}\ }\textbf {\bibinfo {volume}
  {431}},\ \bibinfo {pages} {162} (\bibinfo {year} {2004})}\BibitemShut
  {NoStop}%
\bibitem [{\citenamefont {Majer}\ \emph {et~al.}(2007)\citenamefont {Majer},
  \citenamefont {Chow}, \citenamefont {Gambetta}, \citenamefont {Koch},
  \citenamefont {Johnson}, \citenamefont {Schreier}, \citenamefont {Frunzio},
  \citenamefont {Schuster}, \citenamefont {Houck}, \citenamefont {Wallraff},
  \citenamefont {Blais}, \citenamefont {Devoret}, \citenamefont {Girvin},\ and\
  \citenamefont {Schoelkopf}}]{majer2007coupling}%
  \BibitemOpen
  \bibfield  {author} {\bibinfo {author} {\bibfnamefont {J.}~\bibnamefont
  {Majer}}, \bibinfo {author} {\bibfnamefont {J.~M.}\ \bibnamefont {Chow}},
  \bibinfo {author} {\bibfnamefont {J.~M.}\ \bibnamefont {Gambetta}}, \bibinfo
  {author} {\bibfnamefont {J.}~\bibnamefont {Koch}}, \bibinfo {author}
  {\bibfnamefont {B.~R.}\ \bibnamefont {Johnson}}, \bibinfo {author}
  {\bibfnamefont {J.~A.}\ \bibnamefont {Schreier}}, \bibinfo {author}
  {\bibfnamefont {L.}~\bibnamefont {Frunzio}}, \bibinfo {author} {\bibfnamefont
  {D.~I.}\ \bibnamefont {Schuster}}, \bibinfo {author} {\bibfnamefont {A.~A.}\
  \bibnamefont {Houck}}, \bibinfo {author} {\bibfnamefont {A.}~\bibnamefont
  {Wallraff}}, \bibinfo {author} {\bibfnamefont {A.}~\bibnamefont {Blais}},
  \bibinfo {author} {\bibfnamefont {M.~H.}\ \bibnamefont {Devoret}}, \bibinfo
  {author} {\bibfnamefont {S.~M.}\ \bibnamefont {Girvin}}, \ and\ \bibinfo
  {author} {\bibfnamefont {R.~J.}\ \bibnamefont {Schoelkopf}},\ }\href
  {\doibase 10.1038/nature06184} {\bibfield  {journal} {\bibinfo  {journal}
  {Nature}\ }\textbf {\bibinfo {volume} {449}},\ \bibinfo {pages} {443}
  (\bibinfo {year} {2007})}\BibitemShut {NoStop}%
\bibitem [{\citenamefont {Jordan}\ \emph {et~al.}(2012)\citenamefont {Jordan},
  \citenamefont {Lee},\ and\ \citenamefont {Preskill}}]{Jordan:2011ne}%
  \BibitemOpen
  \bibfield  {author} {\bibinfo {author} {\bibfnamefont {S.~P.}\ \bibnamefont
  {Jordan}}, \bibinfo {author} {\bibfnamefont {K.~S.~M.}\ \bibnamefont {Lee}},
  \ and\ \bibinfo {author} {\bibfnamefont {J.}~\bibnamefont {Preskill}},\
  }\href {\doibase 10.1126/science.1217069} {\bibfield  {journal} {\bibinfo
  {journal} {Science}\ }\textbf {\bibinfo {volume} {336}},\ \bibinfo {pages}
  {1130} (\bibinfo {year} {2012})},\ \Eprint {http://arxiv.org/abs/1111.3633}
  {arXiv:1111.3633 [quant-ph]} \BibitemShut {NoStop}%
\bibitem [{\citenamefont {Jordan}\ \emph {et~al.}(2011)\citenamefont {Jordan},
  \citenamefont {Lee},\ and\ \citenamefont {Preskill}}]{Jordan:2011ci}%
  \BibitemOpen
  \bibfield  {author} {\bibinfo {author} {\bibfnamefont {S.~P.}\ \bibnamefont
  {Jordan}}, \bibinfo {author} {\bibfnamefont {K.~S.~M.}\ \bibnamefont {Lee}},
  \ and\ \bibinfo {author} {\bibfnamefont {J.}~\bibnamefont {Preskill}},\
  }\href@noop {} {\  (\bibinfo {year} {2011})},\ \bibinfo {note} {[Quant. Inf.
  Comput.14,1014(2014)]},\ \Eprint {http://arxiv.org/abs/1112.4833}
  {arXiv:1112.4833 [hep-th]} \BibitemShut {NoStop}%
\bibitem [{\citenamefont {Zohar}\ \emph {et~al.}(2012)\citenamefont {Zohar},
  \citenamefont {Cirac},\ and\ \citenamefont {Reznik}}]{Zohar:2012ay}%
  \BibitemOpen
  \bibfield  {author} {\bibinfo {author} {\bibfnamefont {E.}~\bibnamefont
  {Zohar}}, \bibinfo {author} {\bibfnamefont {J.~I.}\ \bibnamefont {Cirac}}, \
  and\ \bibinfo {author} {\bibfnamefont {B.}~\bibnamefont {Reznik}},\ }\href
  {\doibase 10.1103/PhysRevLett.109.125302} {\bibfield  {journal} {\bibinfo
  {journal} {Phys. Rev. Lett.}\ }\textbf {\bibinfo {volume} {109}},\ \bibinfo
  {pages} {125302} (\bibinfo {year} {2012})},\ \Eprint
  {http://arxiv.org/abs/1204.6574} {arXiv:1204.6574 [quant-ph]} \BibitemShut
  {NoStop}%
\bibitem [{\citenamefont {Zohar}\ \emph {et~al.}(2013)\citenamefont {Zohar},
  \citenamefont {Cirac},\ and\ \citenamefont {Reznik}}]{Zohar:2012xf}%
  \BibitemOpen
  \bibfield  {author} {\bibinfo {author} {\bibfnamefont {E.}~\bibnamefont
  {Zohar}}, \bibinfo {author} {\bibfnamefont {J.~I.}\ \bibnamefont {Cirac}}, \
  and\ \bibinfo {author} {\bibfnamefont {B.}~\bibnamefont {Reznik}},\ }\href
  {\doibase 10.1103/PhysRevLett.110.125304} {\bibfield  {journal} {\bibinfo
  {journal} {Phys. Rev. Lett.}\ }\textbf {\bibinfo {volume} {110}},\ \bibinfo
  {pages} {125304} (\bibinfo {year} {2013})},\ \Eprint
  {http://arxiv.org/abs/1211.2241} {arXiv:1211.2241 [quant-ph]} \BibitemShut
  {NoStop}%
\bibitem [{\citenamefont {Banerjee}\ \emph {et~al.}(2013)\citenamefont
  {Banerjee}, \citenamefont {Bögli}, \citenamefont {Dalmonte}, \citenamefont
  {Rico}, \citenamefont {Stebler}, \citenamefont {Wiese},\ and\ \citenamefont
  {Zoller}}]{Banerjee:2012xg}%
  \BibitemOpen
  \bibfield  {author} {\bibinfo {author} {\bibfnamefont {D.}~\bibnamefont
  {Banerjee}}, \bibinfo {author} {\bibfnamefont {M.}~\bibnamefont {Bögli}},
  \bibinfo {author} {\bibfnamefont {M.}~\bibnamefont {Dalmonte}}, \bibinfo
  {author} {\bibfnamefont {E.}~\bibnamefont {Rico}}, \bibinfo {author}
  {\bibfnamefont {P.}~\bibnamefont {Stebler}}, \bibinfo {author} {\bibfnamefont
  {U.~J.}\ \bibnamefont {Wiese}}, \ and\ \bibinfo {author} {\bibfnamefont
  {P.}~\bibnamefont {Zoller}},\ }\href {\doibase
  10.1103/PhysRevLett.110.125303} {\bibfield  {journal} {\bibinfo  {journal}
  {Phys. Rev. Lett.}\ }\textbf {\bibinfo {volume} {110}},\ \bibinfo {pages}
  {125303} (\bibinfo {year} {2013})},\ \Eprint {http://arxiv.org/abs/1211.2242}
  {arXiv:1211.2242 [cond-mat.quant-gas]} \BibitemShut {NoStop}%
\bibitem [{\citenamefont {Banerjee}\ \emph {et~al.}(2012)\citenamefont
  {Banerjee}, \citenamefont {Dalmonte}, \citenamefont {Muller}, \citenamefont
  {Rico}, \citenamefont {Stebler}, \citenamefont {Wiese},\ and\ \citenamefont
  {Zoller}}]{Banerjee:2012pg}%
  \BibitemOpen
  \bibfield  {author} {\bibinfo {author} {\bibfnamefont {D.}~\bibnamefont
  {Banerjee}}, \bibinfo {author} {\bibfnamefont {M.}~\bibnamefont {Dalmonte}},
  \bibinfo {author} {\bibfnamefont {M.}~\bibnamefont {Muller}}, \bibinfo
  {author} {\bibfnamefont {E.}~\bibnamefont {Rico}}, \bibinfo {author}
  {\bibfnamefont {P.}~\bibnamefont {Stebler}}, \bibinfo {author} {\bibfnamefont
  {U.~J.}\ \bibnamefont {Wiese}}, \ and\ \bibinfo {author} {\bibfnamefont
  {P.}~\bibnamefont {Zoller}},\ }\href {\doibase
  10.1103/PhysRevLett.109.175302} {\bibfield  {journal} {\bibinfo  {journal}
  {Phys. Rev. Lett.}\ }\textbf {\bibinfo {volume} {109}},\ \bibinfo {pages}
  {175302} (\bibinfo {year} {2012})},\ \Eprint {http://arxiv.org/abs/1205.6366}
  {arXiv:1205.6366 [cond-mat.quant-gas]} \BibitemShut {NoStop}%
\bibitem [{\citenamefont {Wiese}(2013)}]{Wiese:2013uua}%
  \BibitemOpen
  \bibfield  {author} {\bibinfo {author} {\bibfnamefont {U.-J.}\ \bibnamefont
  {Wiese}},\ }\href {\doibase 10.1002/andp.201300104} {\bibfield  {journal}
  {\bibinfo  {journal} {Annalen Phys.}\ }\textbf {\bibinfo {volume} {525}},\
  \bibinfo {pages} {777} (\bibinfo {year} {2013})},\ \Eprint
  {http://arxiv.org/abs/1305.1602} {arXiv:1305.1602 [quant-ph]} \BibitemShut
  {NoStop}%
\bibitem [{\citenamefont {Wiese}(2014)}]{Wiese:2014rla}%
  \BibitemOpen
  \bibfield  {author} {\bibinfo {author} {\bibfnamefont {U.-J.}\ \bibnamefont
  {Wiese}},\ }\bibfield  {booktitle} {\emph {\bibinfo {booktitle}
  {{Proceedings, 24th International Conference on Ultra-Relativistic
  Nucleus-Nucleus Collisions (Quark Matter 2014): Darmstadt, Germany, May
  19-24, 2014}}},\ }\href {\doibase 10.1016/j.nuclphysa.2014.09.102} {\bibfield
   {journal} {\bibinfo  {journal} {Nucl. Phys.}\ }\textbf {\bibinfo {volume}
  {A931}},\ \bibinfo {pages} {246} (\bibinfo {year} {2014})},\ \Eprint
  {http://arxiv.org/abs/1409.7414} {arXiv:1409.7414 [hep-th]} \BibitemShut
  {NoStop}%
\bibitem [{\citenamefont {Jordan}\ \emph {et~al.}(2014)\citenamefont {Jordan},
  \citenamefont {Lee},\ and\ \citenamefont {Preskill}}]{Jordan:2014tma}%
  \BibitemOpen
  \bibfield  {author} {\bibinfo {author} {\bibfnamefont {S.~P.}\ \bibnamefont
  {Jordan}}, \bibinfo {author} {\bibfnamefont {K.~S.~M.}\ \bibnamefont {Lee}},
  \ and\ \bibinfo {author} {\bibfnamefont {J.}~\bibnamefont {Preskill}},\
  }\href@noop {} {\  (\bibinfo {year} {2014})},\ \Eprint
  {http://arxiv.org/abs/1404.7115} {arXiv:1404.7115 [hep-th]} \BibitemShut
  {NoStop}%
\bibitem [{\citenamefont {García-Álvarez}\ \emph {et~al.}(2015)\citenamefont
  {García-Álvarez}, \citenamefont {Casanova}, \citenamefont {Mezzacapo},
  \citenamefont {Egusquiza}, \citenamefont {Lamata}, \citenamefont {Romero},\
  and\ \citenamefont {Solano}}]{Garcia-Alvarez:2014uda}%
  \BibitemOpen
  \bibfield  {author} {\bibinfo {author} {\bibfnamefont {L.}~\bibnamefont
  {García-Álvarez}}, \bibinfo {author} {\bibfnamefont {J.}~\bibnamefont
  {Casanova}}, \bibinfo {author} {\bibfnamefont {A.}~\bibnamefont {Mezzacapo}},
  \bibinfo {author} {\bibfnamefont {I.~L.}\ \bibnamefont {Egusquiza}}, \bibinfo
  {author} {\bibfnamefont {L.}~\bibnamefont {Lamata}}, \bibinfo {author}
  {\bibfnamefont {G.}~\bibnamefont {Romero}}, \ and\ \bibinfo {author}
  {\bibfnamefont {E.}~\bibnamefont {Solano}},\ }\href {\doibase
  10.1103/PhysRevLett.114.070502} {\bibfield  {journal} {\bibinfo  {journal}
  {Phys. Rev. Lett.}\ }\textbf {\bibinfo {volume} {114}},\ \bibinfo {pages}
  {070502} (\bibinfo {year} {2015})},\ \Eprint {http://arxiv.org/abs/1404.2868}
  {arXiv:1404.2868 [quant-ph]} \BibitemShut {NoStop}%
\bibitem [{\citenamefont {Marcos}\ \emph {et~al.}(2014)\citenamefont {Marcos},
  \citenamefont {Widmer}, \citenamefont {Rico}, \citenamefont {Hafezi},
  \citenamefont {Rabl}, \citenamefont {Wiese},\ and\ \citenamefont
  {Zoller}}]{Marcos:2014lda}%
  \BibitemOpen
  \bibfield  {author} {\bibinfo {author} {\bibfnamefont {D.}~\bibnamefont
  {Marcos}}, \bibinfo {author} {\bibfnamefont {P.}~\bibnamefont {Widmer}},
  \bibinfo {author} {\bibfnamefont {E.}~\bibnamefont {Rico}}, \bibinfo {author}
  {\bibfnamefont {M.}~\bibnamefont {Hafezi}}, \bibinfo {author} {\bibfnamefont
  {P.}~\bibnamefont {Rabl}}, \bibinfo {author} {\bibfnamefont {U.~J.}\
  \bibnamefont {Wiese}}, \ and\ \bibinfo {author} {\bibfnamefont
  {P.}~\bibnamefont {Zoller}},\ }\href {\doibase 10.1016/j.aop.2014.09.011}
  {\bibfield  {journal} {\bibinfo  {journal} {Annals Phys.}\ }\textbf {\bibinfo
  {volume} {351}},\ \bibinfo {pages} {634} (\bibinfo {year} {2014})},\ \Eprint
  {http://arxiv.org/abs/1407.6066} {arXiv:1407.6066 [quant-ph]} \BibitemShut
  {NoStop}%
\bibitem [{\citenamefont {Bazavov}\ \emph {et~al.}(2015)\citenamefont
  {Bazavov}, \citenamefont {Meurice}, \citenamefont {Tsai}, \citenamefont
  {Unmuth-Yockey},\ and\ \citenamefont {Zhang}}]{Bazavov:2015kka}%
  \BibitemOpen
  \bibfield  {author} {\bibinfo {author} {\bibfnamefont {A.}~\bibnamefont
  {Bazavov}}, \bibinfo {author} {\bibfnamefont {Y.}~\bibnamefont {Meurice}},
  \bibinfo {author} {\bibfnamefont {S.-W.}\ \bibnamefont {Tsai}}, \bibinfo
  {author} {\bibfnamefont {J.}~\bibnamefont {Unmuth-Yockey}}, \ and\ \bibinfo
  {author} {\bibfnamefont {J.}~\bibnamefont {Zhang}},\ }\href {\doibase
  10.1103/PhysRevD.92.076003} {\bibfield  {journal} {\bibinfo  {journal} {Phys.
  Rev.}\ }\textbf {\bibinfo {volume} {D92}},\ \bibinfo {pages} {076003}
  (\bibinfo {year} {2015})},\ \Eprint {http://arxiv.org/abs/1503.08354}
  {arXiv:1503.08354 [hep-lat]} \BibitemShut {NoStop}%
\bibitem [{\citenamefont {Zohar}\ \emph {et~al.}(2016)\citenamefont {Zohar},
  \citenamefont {Cirac},\ and\ \citenamefont {Reznik}}]{Zohar:2015hwa}%
  \BibitemOpen
  \bibfield  {author} {\bibinfo {author} {\bibfnamefont {E.}~\bibnamefont
  {Zohar}}, \bibinfo {author} {\bibfnamefont {J.~I.}\ \bibnamefont {Cirac}}, \
  and\ \bibinfo {author} {\bibfnamefont {B.}~\bibnamefont {Reznik}},\ }\href
  {\doibase 10.1088/0034-4885/79/1/014401} {\bibfield  {journal} {\bibinfo
  {journal} {Rept. Prog. Phys.}\ }\textbf {\bibinfo {volume} {79}},\ \bibinfo
  {pages} {014401} (\bibinfo {year} {2016})},\ \Eprint
  {http://arxiv.org/abs/1503.02312} {arXiv:1503.02312 [quant-ph]} \BibitemShut
  {NoStop}%
\bibitem [{\citenamefont {Mezzacapo}\ \emph {et~al.}(2015)\citenamefont
  {Mezzacapo}, \citenamefont {Rico}, \citenamefont {Sabín}, \citenamefont
  {Egusquiza}, \citenamefont {Lamata},\ and\ \citenamefont
  {Solano}}]{Mezzacapo:2015bra}%
  \BibitemOpen
  \bibfield  {author} {\bibinfo {author} {\bibfnamefont {A.}~\bibnamefont
  {Mezzacapo}}, \bibinfo {author} {\bibfnamefont {E.}~\bibnamefont {Rico}},
  \bibinfo {author} {\bibfnamefont {C.}~\bibnamefont {Sabín}}, \bibinfo
  {author} {\bibfnamefont {I.~L.}\ \bibnamefont {Egusquiza}}, \bibinfo {author}
  {\bibfnamefont {L.}~\bibnamefont {Lamata}}, \ and\ \bibinfo {author}
  {\bibfnamefont {E.}~\bibnamefont {Solano}},\ }\href {\doibase
  10.1103/PhysRevLett.115.240502} {\bibfield  {journal} {\bibinfo  {journal}
  {Phys. Rev. Lett.}\ }\textbf {\bibinfo {volume} {115}},\ \bibinfo {pages}
  {240502} (\bibinfo {year} {2015})},\ \Eprint
  {http://arxiv.org/abs/1505.04720} {arXiv:1505.04720 [quant-ph]} \BibitemShut
  {NoStop}%
\bibitem [{\citenamefont {Dalmonte}\ and\ \citenamefont
  {Montangero}(2016)}]{Dalmonte:2016alw}%
  \BibitemOpen
  \bibfield  {author} {\bibinfo {author} {\bibfnamefont {M.}~\bibnamefont
  {Dalmonte}}\ and\ \bibinfo {author} {\bibfnamefont {S.}~\bibnamefont
  {Montangero}},\ }\href {\doibase 10.1080/00107514.2016.1151199} {\bibfield
  {journal} {\bibinfo  {journal} {Contemp. Phys.}\ }\textbf {\bibinfo {volume}
  {57}},\ \bibinfo {pages} {388} (\bibinfo {year} {2016})},\ \Eprint
  {http://arxiv.org/abs/1602.03776} {arXiv:1602.03776 [cond-mat.quant-gas]}
  \BibitemShut {NoStop}%
\bibitem [{\citenamefont {Zohar}\ \emph {et~al.}(2017)\citenamefont {Zohar},
  \citenamefont {Farace}, \citenamefont {Reznik},\ and\ \citenamefont
  {Cirac}}]{Zohar:2016iic}%
  \BibitemOpen
  \bibfield  {author} {\bibinfo {author} {\bibfnamefont {E.}~\bibnamefont
  {Zohar}}, \bibinfo {author} {\bibfnamefont {A.}~\bibnamefont {Farace}},
  \bibinfo {author} {\bibfnamefont {B.}~\bibnamefont {Reznik}}, \ and\ \bibinfo
  {author} {\bibfnamefont {J.~I.}\ \bibnamefont {Cirac}},\ }\href {\doibase
  10.1103/PhysRevA.95.023604} {\bibfield  {journal} {\bibinfo  {journal} {Phys.
  Rev.}\ }\textbf {\bibinfo {volume} {A95}},\ \bibinfo {pages} {023604}
  (\bibinfo {year} {2017})},\ \Eprint {http://arxiv.org/abs/1607.08121}
  {arXiv:1607.08121 [quant-ph]} \BibitemShut {NoStop}%
\bibitem [{\citenamefont {Martinez}\ \emph {et~al.}(2016)\citenamefont
  {Martinez} \emph {et~al.}}]{Martinez:2016yna}%
  \BibitemOpen
  \bibfield  {author} {\bibinfo {author} {\bibfnamefont {E.~A.}\ \bibnamefont
  {Martinez}} \emph {et~al.},\ }\href {\doibase 10.1038/nature18318} {\bibfield
   {journal} {\bibinfo  {journal} {Nature}\ }\textbf {\bibinfo {volume}
  {534}},\ \bibinfo {pages} {516} (\bibinfo {year} {2016})},\ \Eprint
  {http://arxiv.org/abs/1605.04570} {arXiv:1605.04570 [quant-ph]} \BibitemShut
  {NoStop}%
\bibitem [{\citenamefont {Bermudez}\ \emph {et~al.}(2017)\citenamefont
  {Bermudez}, \citenamefont {Aarts},\ and\ \citenamefont
  {Müller}}]{Bermudez:2017yrq}%
  \BibitemOpen
  \bibfield  {author} {\bibinfo {author} {\bibfnamefont {A.}~\bibnamefont
  {Bermudez}}, \bibinfo {author} {\bibfnamefont {G.}~\bibnamefont {Aarts}}, \
  and\ \bibinfo {author} {\bibfnamefont {M.}~\bibnamefont {Müller}},\ }\href
  {\doibase 10.1103/PhysRevX.7.041012} {\bibfield  {journal} {\bibinfo
  {journal} {Phys. Rev.}\ }\textbf {\bibinfo {volume} {X7}},\ \bibinfo {pages}
  {041012} (\bibinfo {year} {2017})},\ \Eprint
  {http://arxiv.org/abs/1704.02877} {arXiv:1704.02877 [quant-ph]} \BibitemShut
  {NoStop}%
\bibitem [{\citenamefont {Gambetta}\ \emph {et~al.}(2017)\citenamefont
  {Gambetta}, \citenamefont {Chow},\ and\ \citenamefont
  {Steffen}}]{gambetta2017building}%
  \BibitemOpen
  \bibfield  {author} {\bibinfo {author} {\bibfnamefont {J.~M.}\ \bibnamefont
  {Gambetta}}, \bibinfo {author} {\bibfnamefont {J.~M.}\ \bibnamefont {Chow}},
  \ and\ \bibinfo {author} {\bibfnamefont {M.}~\bibnamefont {Steffen}},\ }\href
  {\doibase 10.1038/s41534-016-0004-0} {\bibfield  {journal} {\bibinfo
  {journal} {npj Quantum Information}\ }\textbf {\bibinfo {volume} {3}},\
  \bibinfo {pages} {2} (\bibinfo {year} {2017})}\BibitemShut {NoStop}%
\bibitem [{\citenamefont {Krinner}\ \emph {et~al.}(2018)\citenamefont
  {Krinner}, \citenamefont {Stewart}, \citenamefont {Pazmi{\~n}o},
  \citenamefont {Kwon},\ and\ \citenamefont
  {Schneble}}]{krinner2018spontaneous}%
  \BibitemOpen
  \bibfield  {author} {\bibinfo {author} {\bibfnamefont {L.}~\bibnamefont
  {Krinner}}, \bibinfo {author} {\bibfnamefont {M.}~\bibnamefont {Stewart}},
  \bibinfo {author} {\bibfnamefont {A.}~\bibnamefont {Pazmi{\~n}o}}, \bibinfo
  {author} {\bibfnamefont {J.}~\bibnamefont {Kwon}}, \ and\ \bibinfo {author}
  {\bibfnamefont {D.}~\bibnamefont {Schneble}},\ }\href {\doibase
  10.1038/s41586-018-0348-z} {\bibfield  {journal} {\bibinfo  {journal}
  {Nature}\ }\textbf {\bibinfo {volume} {559}},\ \bibinfo {pages} {589}
  (\bibinfo {year} {2018})}\BibitemShut {NoStop}%
\bibitem [{\citenamefont {Macridin}\ \emph {et~al.}(2018)\citenamefont
  {Macridin}, \citenamefont {Spentzouris}, \citenamefont {Amundson},\ and\
  \citenamefont {Harnik}}]{Macridin:2018gdw}%
  \BibitemOpen
  \bibfield  {author} {\bibinfo {author} {\bibfnamefont {A.}~\bibnamefont
  {Macridin}}, \bibinfo {author} {\bibfnamefont {P.}~\bibnamefont
  {Spentzouris}}, \bibinfo {author} {\bibfnamefont {J.}~\bibnamefont
  {Amundson}}, \ and\ \bibinfo {author} {\bibfnamefont {R.}~\bibnamefont
  {Harnik}},\ }\href {\doibase 10.1103/PhysRevLett.121.110504} {\bibfield
  {journal} {\bibinfo  {journal} {Phys. Rev. Lett.}\ }\textbf {\bibinfo
  {volume} {121}},\ \bibinfo {pages} {110504} (\bibinfo {year} {2018})},\
  \Eprint {http://arxiv.org/abs/1802.07347} {arXiv:1802.07347 [quant-ph]}
  \BibitemShut {NoStop}%
\bibitem [{\citenamefont {Zache}\ \emph {et~al.}(2018)\citenamefont {Zache},
  \citenamefont {Hebenstreit}, \citenamefont {Jendrzejewski}, \citenamefont
  {Oberthaler}, \citenamefont {Berges},\ and\ \citenamefont
  {Hauke}}]{Zache:2018jbt}%
  \BibitemOpen
  \bibfield  {author} {\bibinfo {author} {\bibfnamefont {T.~V.}\ \bibnamefont
  {Zache}}, \bibinfo {author} {\bibfnamefont {F.}~\bibnamefont {Hebenstreit}},
  \bibinfo {author} {\bibfnamefont {F.}~\bibnamefont {Jendrzejewski}}, \bibinfo
  {author} {\bibfnamefont {M.~K.}\ \bibnamefont {Oberthaler}}, \bibinfo
  {author} {\bibfnamefont {J.}~\bibnamefont {Berges}}, \ and\ \bibinfo {author}
  {\bibfnamefont {P.}~\bibnamefont {Hauke}},\ }\href {\doibase
  10.1088/2058-9565/aac33b} {\bibfield  {journal} {\bibinfo  {journal} {Sci.
  Technol.}\ }\textbf {\bibinfo {volume} {3}},\ \bibinfo {pages} {034010}
  (\bibinfo {year} {2018})},\ \Eprint {http://arxiv.org/abs/1802.06704}
  {arXiv:1802.06704 [cond-mat.quant-gas]} \BibitemShut {NoStop}%
\bibitem [{\citenamefont {Zhang}\ \emph {et~al.}(2018)\citenamefont {Zhang},
  \citenamefont {Unmuth-Yockey}, \citenamefont {Zeiher}, \citenamefont
  {Bazavov}, \citenamefont {Tsai},\ and\ \citenamefont
  {Meurice}}]{Zhang:2018ufj}%
  \BibitemOpen
  \bibfield  {author} {\bibinfo {author} {\bibfnamefont {J.}~\bibnamefont
  {Zhang}}, \bibinfo {author} {\bibfnamefont {J.}~\bibnamefont
  {Unmuth-Yockey}}, \bibinfo {author} {\bibfnamefont {J.}~\bibnamefont
  {Zeiher}}, \bibinfo {author} {\bibfnamefont {A.}~\bibnamefont {Bazavov}},
  \bibinfo {author} {\bibfnamefont {S.~W.}\ \bibnamefont {Tsai}}, \ and\
  \bibinfo {author} {\bibfnamefont {Y.}~\bibnamefont {Meurice}},\ }\href
  {\doibase 10.1103/PhysRevLett.121.223201} {\bibfield  {journal} {\bibinfo
  {journal} {Phys. Rev. Lett.}\ }\textbf {\bibinfo {volume} {121}},\ \bibinfo
  {pages} {223201} (\bibinfo {year} {2018})},\ \Eprint
  {http://arxiv.org/abs/1803.11166} {arXiv:1803.11166 [hep-lat]} \BibitemShut
  {NoStop}%
\bibitem [{\citenamefont {Klco}\ \emph {et~al.}(2018)\citenamefont {Klco},
  \citenamefont {Dumitrescu}, \citenamefont {McCaskey}, \citenamefont {Morris},
  \citenamefont {Pooser}, \citenamefont {Sanz}, \citenamefont {Solano},
  \citenamefont {Lougovski},\ and\ \citenamefont {Savage}}]{Klco:2018kyo}%
  \BibitemOpen
  \bibfield  {author} {\bibinfo {author} {\bibfnamefont {N.}~\bibnamefont
  {Klco}}, \bibinfo {author} {\bibfnamefont {E.~F.}\ \bibnamefont
  {Dumitrescu}}, \bibinfo {author} {\bibfnamefont {A.~J.}\ \bibnamefont
  {McCaskey}}, \bibinfo {author} {\bibfnamefont {T.~D.}\ \bibnamefont
  {Morris}}, \bibinfo {author} {\bibfnamefont {R.~C.}\ \bibnamefont {Pooser}},
  \bibinfo {author} {\bibfnamefont {M.}~\bibnamefont {Sanz}}, \bibinfo {author}
  {\bibfnamefont {E.}~\bibnamefont {Solano}}, \bibinfo {author} {\bibfnamefont
  {P.}~\bibnamefont {Lougovski}}, \ and\ \bibinfo {author} {\bibfnamefont
  {M.~J.}\ \bibnamefont {Savage}},\ }\href {\doibase
  10.1103/PhysRevA.98.032331} {\bibfield  {journal} {\bibinfo  {journal} {Phys.
  Rev.}\ }\textbf {\bibinfo {volume} {A98}},\ \bibinfo {pages} {032331}
  (\bibinfo {year} {2018})},\ \Eprint {http://arxiv.org/abs/1803.03326}
  {arXiv:1803.03326 [quant-ph]} \BibitemShut {NoStop}%
\bibitem [{\citenamefont {Klco}\ and\ \citenamefont
  {Savage}(2019{\natexlab{a}})}]{Klco:2018zqz}%
  \BibitemOpen
  \bibfield  {author} {\bibinfo {author} {\bibfnamefont {N.}~\bibnamefont
  {Klco}}\ and\ \bibinfo {author} {\bibfnamefont {M.~J.}\ \bibnamefont
  {Savage}},\ }\href {\doibase 10.1103/PhysRevA.99.052335} {\bibfield
  {journal} {\bibinfo  {journal} {Phys. Rev.}\ }\textbf {\bibinfo {volume}
  {A99}},\ \bibinfo {pages} {052335} (\bibinfo {year} {2019}{\natexlab{a}})},\
  \Eprint {http://arxiv.org/abs/1808.10378} {arXiv:1808.10378 [quant-ph]}
  \BibitemShut {NoStop}%
\bibitem [{\citenamefont {Lu}\ \emph {et~al.}(2019)\citenamefont {Lu} \emph
  {et~al.}}]{Lu:2018pjk}%
  \BibitemOpen
  \bibfield  {author} {\bibinfo {author} {\bibfnamefont {H.-H.}\ \bibnamefont
  {Lu}} \emph {et~al.},\ }\href {\doibase 10.1103/PhysRevA.100.012320}
  {\bibfield  {journal} {\bibinfo  {journal} {Phys. Rev.}\ }\textbf {\bibinfo
  {volume} {A100}},\ \bibinfo {pages} {012320} (\bibinfo {year} {2019})},\
  \Eprint {http://arxiv.org/abs/1810.03959} {arXiv:1810.03959 [quant-ph]}
  \BibitemShut {NoStop}%
\bibitem [{\citenamefont {Klco}\ and\ \citenamefont
  {Savage}(2019{\natexlab{b}})}]{Klco:2019xro}%
  \BibitemOpen
  \bibfield  {author} {\bibinfo {author} {\bibfnamefont {N.}~\bibnamefont
  {Klco}}\ and\ \bibinfo {author} {\bibfnamefont {M.~J.}\ \bibnamefont
  {Savage}},\ }\href@noop {} {\  (\bibinfo {year} {2019}{\natexlab{b}})},\
  \Eprint {http://arxiv.org/abs/1904.10440} {arXiv:1904.10440 [quant-ph]}
  \BibitemShut {NoStop}%
\bibitem [{\citenamefont {Lamm}\ and\ \citenamefont
  {Lawrence}(2018)}]{Lamm:2018siq}%
  \BibitemOpen
  \bibfield  {author} {\bibinfo {author} {\bibfnamefont {H.}~\bibnamefont
  {Lamm}}\ and\ \bibinfo {author} {\bibfnamefont {S.}~\bibnamefont
  {Lawrence}},\ }\href {\doibase 10.1103/PhysRevLett.121.170501} {\bibfield
  {journal} {\bibinfo  {journal} {Phys. Rev. Lett.}\ }\textbf {\bibinfo
  {volume} {121}},\ \bibinfo {pages} {170501} (\bibinfo {year} {2018})},\
  \Eprint {http://arxiv.org/abs/1806.06649} {arXiv:1806.06649 [quant-ph]}
  \BibitemShut {NoStop}%
\bibitem [{\citenamefont {Gustafson}\ \emph {et~al.}(2019)\citenamefont
  {Gustafson}, \citenamefont {Meurice},\ and\ \citenamefont
  {Unmuth-Yockey}}]{Gustafson:2019mpk}%
  \BibitemOpen
  \bibfield  {author} {\bibinfo {author} {\bibfnamefont {E.}~\bibnamefont
  {Gustafson}}, \bibinfo {author} {\bibfnamefont {Y.}~\bibnamefont {Meurice}},
  \ and\ \bibinfo {author} {\bibfnamefont {J.}~\bibnamefont {Unmuth-Yockey}},\
  }\href {\doibase 10.1103/PhysRevD.99.094503} {\bibfield  {journal} {\bibinfo
  {journal} {Phys. Rev.}\ }\textbf {\bibinfo {volume} {D99}},\ \bibinfo {pages}
  {094503} (\bibinfo {year} {2019})},\ \Eprint
  {http://arxiv.org/abs/1901.05944} {arXiv:1901.05944 [hep-lat]} \BibitemShut
  {NoStop}%
\bibitem [{\citenamefont {Klco}\ \emph {et~al.}(2019)\citenamefont {Klco},
  \citenamefont {Stryker},\ and\ \citenamefont {Savage}}]{Klco:2019evd}%
  \BibitemOpen
  \bibfield  {author} {\bibinfo {author} {\bibfnamefont {N.}~\bibnamefont
  {Klco}}, \bibinfo {author} {\bibfnamefont {J.~R.}\ \bibnamefont {Stryker}}, \
  and\ \bibinfo {author} {\bibfnamefont {M.~J.}\ \bibnamefont {Savage}},\
  }\href@noop {} {\  (\bibinfo {year} {2019})},\ \Eprint
  {http://arxiv.org/abs/1908.06935} {arXiv:1908.06935 [quant-ph]} \BibitemShut
  {NoStop}%
\bibitem [{\citenamefont {Alexandru}\ \emph
  {et~al.}(2019{\natexlab{a}})\citenamefont {Alexandru}, \citenamefont
  {Bedaque}, \citenamefont {Lamm},\ and\ \citenamefont
  {Lawrence}}]{Alexandru:2019ozf}%
  \BibitemOpen
  \bibfield  {author} {\bibinfo {author} {\bibfnamefont {A.}~\bibnamefont
  {Alexandru}}, \bibinfo {author} {\bibfnamefont {P.~F.}\ \bibnamefont
  {Bedaque}}, \bibinfo {author} {\bibfnamefont {H.}~\bibnamefont {Lamm}}, \
  and\ \bibinfo {author} {\bibfnamefont {S.}~\bibnamefont {Lawrence}} (\bibinfo
  {collaboration} {NuQS}),\ }\href {\doibase 10.1103/PhysRevLett.123.090501}
  {\bibfield  {journal} {\bibinfo  {journal} {Phys. Rev. Lett.}\ }\textbf
  {\bibinfo {volume} {123}},\ \bibinfo {pages} {090501} (\bibinfo {year}
  {2019}{\natexlab{a}})},\ \Eprint {http://arxiv.org/abs/1903.06577}
  {arXiv:1903.06577 [hep-lat]} \BibitemShut {NoStop}%
\bibitem [{\citenamefont {Alexandru}\ \emph
  {et~al.}(2019{\natexlab{b}})\citenamefont {Alexandru}, \citenamefont
  {Bedaque}, \citenamefont {Harmalkar}, \citenamefont {Lamm}, \citenamefont
  {Lawrence},\ and\ \citenamefont {Warrington}}]{Alexandru:2019nsa}%
  \BibitemOpen
  \bibfield  {author} {\bibinfo {author} {\bibfnamefont {A.}~\bibnamefont
  {Alexandru}}, \bibinfo {author} {\bibfnamefont {P.~F.}\ \bibnamefont
  {Bedaque}}, \bibinfo {author} {\bibfnamefont {S.}~\bibnamefont {Harmalkar}},
  \bibinfo {author} {\bibfnamefont {H.}~\bibnamefont {Lamm}}, \bibinfo {author}
  {\bibfnamefont {S.}~\bibnamefont {Lawrence}}, \ and\ \bibinfo {author}
  {\bibfnamefont {N.~C.}\ \bibnamefont {Warrington}} (\bibinfo {collaboration}
  {NuQS}),\ }\href {\doibase 10.1103/PhysRevD.100.114501} {\bibfield  {journal}
  {\bibinfo  {journal} {Phys. Rev.}\ }\textbf {\bibinfo {volume} {D100}},\
  \bibinfo {pages} {114501} (\bibinfo {year} {2019}{\natexlab{b}})},\ \Eprint
  {http://arxiv.org/abs/1906.11213} {arXiv:1906.11213 [hep-lat]} \BibitemShut
  {NoStop}%
\bibitem [{\citenamefont {Mueller}\ \emph {et~al.}(2019)\citenamefont
  {Mueller}, \citenamefont {Tarasov},\ and\ \citenamefont
  {Venugopalan}}]{Mueller:2019qqj}%
  \BibitemOpen
  \bibfield  {author} {\bibinfo {author} {\bibfnamefont {N.}~\bibnamefont
  {Mueller}}, \bibinfo {author} {\bibfnamefont {A.}~\bibnamefont {Tarasov}}, \
  and\ \bibinfo {author} {\bibfnamefont {R.}~\bibnamefont {Venugopalan}},\
  }\href@noop {} {\  (\bibinfo {year} {2019})},\ \Eprint
  {http://arxiv.org/abs/1908.07051} {arXiv:1908.07051 [hep-th]} \BibitemShut
  {NoStop}%
\bibitem [{\citenamefont {Lamm}\ \emph {et~al.}(2019)\citenamefont {Lamm},
  \citenamefont {Lawrence},\ and\ \citenamefont {Yamauchi}}]{Lamm:2019uyc}%
  \BibitemOpen
  \bibfield  {author} {\bibinfo {author} {\bibfnamefont {H.}~\bibnamefont
  {Lamm}}, \bibinfo {author} {\bibfnamefont {S.}~\bibnamefont {Lawrence}}, \
  and\ \bibinfo {author} {\bibfnamefont {Y.}~\bibnamefont {Yamauchi}} (\bibinfo
  {collaboration} {NuQS}),\ }\href@noop {} {\  (\bibinfo {year} {2019})},\
  \Eprint {http://arxiv.org/abs/1908.10439} {arXiv:1908.10439 [hep-lat]}
  \BibitemShut {NoStop}%
\bibitem [{\citenamefont {Magnifico}\ \emph
  {et~al.}(2019{\natexlab{a}})\citenamefont {Magnifico}, \citenamefont
  {Dalmonte}, \citenamefont {Facchi}, \citenamefont {Pascazio}, \citenamefont
  {Pepe},\ and\ \citenamefont {Ercolessi}}]{Magnifico:2019kyj}%
  \BibitemOpen
  \bibfield  {author} {\bibinfo {author} {\bibfnamefont {G.}~\bibnamefont
  {Magnifico}}, \bibinfo {author} {\bibfnamefont {M.}~\bibnamefont {Dalmonte}},
  \bibinfo {author} {\bibfnamefont {P.}~\bibnamefont {Facchi}}, \bibinfo
  {author} {\bibfnamefont {S.}~\bibnamefont {Pascazio}}, \bibinfo {author}
  {\bibfnamefont {F.~V.}\ \bibnamefont {Pepe}}, \ and\ \bibinfo {author}
  {\bibfnamefont {E.}~\bibnamefont {Ercolessi}},\ }\href@noop {} {\  (\bibinfo
  {year} {2019}{\natexlab{a}})},\ \Eprint {http://arxiv.org/abs/1909.04821}
  {arXiv:1909.04821 [quant-ph]} \BibitemShut {NoStop}%
\bibitem [{\citenamefont {Chakraborty}\ \emph {et~al.}(2020)\citenamefont
  {Chakraborty}, \citenamefont {Honda}, \citenamefont {Izubuchi}, \citenamefont
  {Kikuchi},\ and\ \citenamefont {Tomiya}}]{Chakraborty:2019}%
  \BibitemOpen
  \bibfield  {author} {\bibinfo {author} {\bibfnamefont {B.}~\bibnamefont
  {Chakraborty}}, \bibinfo {author} {\bibfnamefont {M.}~\bibnamefont {Honda}},
  \bibinfo {author} {\bibfnamefont {T.}~\bibnamefont {Izubuchi}}, \bibinfo
  {author} {\bibfnamefont {Y.}~\bibnamefont {Kikuchi}}, \ and\ \bibinfo
  {author} {\bibfnamefont {A.}~\bibnamefont {Tomiya}},\ }\href@noop {} {\
  (\bibinfo {year} {2020})},\ \Eprint {http://arxiv.org/abs/2001.00485}
  {arXiv:2001.00485 [hep-lat]} \BibitemShut {NoStop}%
\bibitem [{\citenamefont {Kharzeev}\ and\ \citenamefont
  {Kikuchi}(2020)}]{Kharzeev:2020kgc}%
  \BibitemOpen
  \bibfield  {author} {\bibinfo {author} {\bibfnamefont {D.~E.}\ \bibnamefont
  {Kharzeev}}\ and\ \bibinfo {author} {\bibfnamefont {Y.}~\bibnamefont
  {Kikuchi}},\ }\href {\doibase 10.1103/PhysRevResearch.2.023342} {\bibfield
  {journal} {\bibinfo  {journal} {Phys. Rev. Res.}\ }\textbf {\bibinfo {volume}
  {2}},\ \bibinfo {pages} {023342} (\bibinfo {year} {2020})},\ \Eprint
  {http://arxiv.org/abs/2001.00698} {arXiv:2001.00698 [hep-ph]} \BibitemShut
  {NoStop}%
\bibitem [{\citenamefont {Florio}\ \emph {et~al.}(2023)\citenamefont {Florio},
  \citenamefont {Frenklakh}, \citenamefont {Ikeda}, \citenamefont {Kharzeev},
  \citenamefont {Korepin}, \citenamefont {Shi},\ and\ \citenamefont
  {Yu}}]{PhysRevLett.131.021902}%
  \BibitemOpen
  \bibfield  {author} {\bibinfo {author} {\bibfnamefont {A.}~\bibnamefont
  {Florio}}, \bibinfo {author} {\bibfnamefont {D.}~\bibnamefont {Frenklakh}},
  \bibinfo {author} {\bibfnamefont {K.}~\bibnamefont {Ikeda}}, \bibinfo
  {author} {\bibfnamefont {D.}~\bibnamefont {Kharzeev}}, \bibinfo {author}
  {\bibfnamefont {V.}~\bibnamefont {Korepin}}, \bibinfo {author} {\bibfnamefont
  {S.}~\bibnamefont {Shi}}, \ and\ \bibinfo {author} {\bibfnamefont
  {K.}~\bibnamefont {Yu}},\ }\href {\doibase 10.1103/PhysRevLett.131.021902}
  {\bibfield  {journal} {\bibinfo  {journal} {Phys. Rev. Lett.}\ }\textbf
  {\bibinfo {volume} {131}},\ \bibinfo {pages} {021902} (\bibinfo {year}
  {2023})}\BibitemShut {NoStop}%
\bibitem [{\citenamefont {Shaw}\ \emph {et~al.}(2020)\citenamefont {Shaw},
  \citenamefont {Lougovski}, \citenamefont {Stryker},\ and\ \citenamefont
  {Wiebe}}]{Shaw:2020udc}%
  \BibitemOpen
  \bibfield  {author} {\bibinfo {author} {\bibfnamefont {A.~F.}\ \bibnamefont
  {Shaw}}, \bibinfo {author} {\bibfnamefont {P.}~\bibnamefont {Lougovski}},
  \bibinfo {author} {\bibfnamefont {J.~R.}\ \bibnamefont {Stryker}}, \ and\
  \bibinfo {author} {\bibfnamefont {N.}~\bibnamefont {Wiebe}},\ }\href
  {\doibase 10.22331/q-2020-08-10-306} {\  (\bibinfo {year} {2020}),\
  10.22331/q-2020-08-10-306},\ \Eprint {http://arxiv.org/abs/2002.11146}
  {arXiv:2002.11146 [quant-ph]} \BibitemShut {NoStop}%
\bibitem [{\citenamefont {Şahinoğlu}\ and\ \citenamefont
  {Somma}(2020)}]{sahinoslu2020hamiltonian}%
  \BibitemOpen
  \bibfield  {author} {\bibinfo {author} {\bibfnamefont {B.}~\bibnamefont
  {Şahinoğlu}}\ and\ \bibinfo {author} {\bibfnamefont {R.~D.}\ \bibnamefont
  {Somma}},\ }\href@noop {} {\  (\bibinfo {year} {2020})},\ \Eprint
  {http://arxiv.org/abs/2006.02660} {arXiv:2006.02660 [quant-ph]} \BibitemShut
  {NoStop}%
\bibitem [{\citenamefont {Paulson}\ \emph {et~al.}(2020)\citenamefont {Paulson}
  \emph {et~al.}}]{Paulson:2020zjd}%
  \BibitemOpen
  \bibfield  {author} {\bibinfo {author} {\bibfnamefont {D.}~\bibnamefont
  {Paulson}} \emph {et~al.},\ }\href@noop {} {\  (\bibinfo {year} {2020})},\
  \Eprint {http://arxiv.org/abs/2008.09252} {arXiv:2008.09252 [quant-ph]}
  \BibitemShut {NoStop}%
\bibitem [{\citenamefont {Mathis}\ \emph {et~al.}(2020)\citenamefont {Mathis},
  \citenamefont {Mazzola},\ and\ \citenamefont {Tavernelli}}]{Mathis:2020fuo}%
  \BibitemOpen
  \bibfield  {author} {\bibinfo {author} {\bibfnamefont {S.~V.}\ \bibnamefont
  {Mathis}}, \bibinfo {author} {\bibfnamefont {G.}~\bibnamefont {Mazzola}}, \
  and\ \bibinfo {author} {\bibfnamefont {I.}~\bibnamefont {Tavernelli}},\
  }\href {\doibase 10.1103/PhysRevD.102.094501} {\bibfield  {journal} {\bibinfo
   {journal} {Phys. Rev. D}\ }\textbf {\bibinfo {volume} {102}},\ \bibinfo
  {pages} {094501} (\bibinfo {year} {2020})},\ \Eprint
  {http://arxiv.org/abs/2005.10271} {arXiv:2005.10271 [quant-ph]} \BibitemShut
  {NoStop}%
\bibitem [{\citenamefont {Ji}\ \emph {et~al.}(2020)\citenamefont {Ji},
  \citenamefont {Lamm},\ and\ \citenamefont {Zhu}}]{Ji:2020kjk}%
  \BibitemOpen
  \bibfield  {author} {\bibinfo {author} {\bibfnamefont {Y.}~\bibnamefont
  {Ji}}, \bibinfo {author} {\bibfnamefont {H.}~\bibnamefont {Lamm}}, \ and\
  \bibinfo {author} {\bibfnamefont {S.}~\bibnamefont {Zhu}} (\bibinfo
  {collaboration} {NuQS}),\ }\href@noop {} {\  (\bibinfo {year} {2020})},\
  \Eprint {http://arxiv.org/abs/2005.14221} {arXiv:2005.14221 [hep-lat]}
  \BibitemShut {NoStop}%
\bibitem [{\citenamefont {Raychowdhury}\ and\ \citenamefont
  {Stryker}(2020)}]{Raychowdhury:2019iki}%
  \BibitemOpen
  \bibfield  {author} {\bibinfo {author} {\bibfnamefont {I.}~\bibnamefont
  {Raychowdhury}}\ and\ \bibinfo {author} {\bibfnamefont {J.~R.}\ \bibnamefont
  {Stryker}},\ }\href {\doibase 10.1103/PhysRevD.101.114502} {\bibfield
  {journal} {\bibinfo  {journal} {Phys. Rev. D}\ }\textbf {\bibinfo {volume}
  {101}},\ \bibinfo {pages} {114502} (\bibinfo {year} {2020})},\ \Eprint
  {http://arxiv.org/abs/1912.06133} {arXiv:1912.06133 [hep-lat]} \BibitemShut
  {NoStop}%
\bibitem [{\citenamefont {Davoudi}\ \emph {et~al.}(2020)\citenamefont
  {Davoudi}, \citenamefont {Raychowdhury},\ and\ \citenamefont
  {Shaw}}]{Davoudi:2020yln}%
  \BibitemOpen
  \bibfield  {author} {\bibinfo {author} {\bibfnamefont {Z.}~\bibnamefont
  {Davoudi}}, \bibinfo {author} {\bibfnamefont {I.}~\bibnamefont
  {Raychowdhury}}, \ and\ \bibinfo {author} {\bibfnamefont {A.}~\bibnamefont
  {Shaw}},\ }\href@noop {} {\  (\bibinfo {year} {2020})},\ \Eprint
  {http://arxiv.org/abs/2009.11802} {arXiv:2009.11802 [hep-lat]} \BibitemShut
  {NoStop}%
\bibitem [{\citenamefont {Dasgupta}\ and\ \citenamefont
  {Raychowdhury}(2020)}]{Dasgupta:2020itb}%
  \BibitemOpen
  \bibfield  {author} {\bibinfo {author} {\bibfnamefont {R.}~\bibnamefont
  {Dasgupta}}\ and\ \bibinfo {author} {\bibfnamefont {I.}~\bibnamefont
  {Raychowdhury}},\ }\href@noop {} {\  (\bibinfo {year} {2020})},\ \Eprint
  {http://arxiv.org/abs/2009.13969} {arXiv:2009.13969 [hep-lat]} \BibitemShut
  {NoStop}%
\bibitem [{\citenamefont {Magnifico}\ \emph
  {et~al.}(2019{\natexlab{b}})\citenamefont {Magnifico}, \citenamefont
  {Vodola}, \citenamefont {Ercolessi}, \citenamefont {Kumar}, \citenamefont
  {M\"uller},\ and\ \citenamefont {Bermudez}}]{Magnifico:2018wek}%
  \BibitemOpen
  \bibfield  {author} {\bibinfo {author} {\bibfnamefont {G.}~\bibnamefont
  {Magnifico}}, \bibinfo {author} {\bibfnamefont {D.}~\bibnamefont {Vodola}},
  \bibinfo {author} {\bibfnamefont {E.}~\bibnamefont {Ercolessi}}, \bibinfo
  {author} {\bibfnamefont {S.~P.}\ \bibnamefont {Kumar}}, \bibinfo {author}
  {\bibfnamefont {M.}~\bibnamefont {M\"uller}}, \ and\ \bibinfo {author}
  {\bibfnamefont {A.}~\bibnamefont {Bermudez}},\ }\href {\doibase
  10.1103/PhysRevD.99.014503} {\bibfield  {journal} {\bibinfo  {journal} {Phys.
  Rev. D}\ }\textbf {\bibinfo {volume} {99}},\ \bibinfo {pages} {014503}
  (\bibinfo {year} {2019}{\natexlab{b}})},\ \Eprint
  {http://arxiv.org/abs/1804.10568} {arXiv:1804.10568 [cond-mat.quant-gas]}
  \BibitemShut {NoStop}%
\bibitem [{\citenamefont {Rico}\ \emph {et~al.}(2014)\citenamefont {Rico},
  \citenamefont {Pichler}, \citenamefont {Dalmonte}, \citenamefont {Zoller},\
  and\ \citenamefont {Montangero}}]{Rico:2013qya}%
  \BibitemOpen
  \bibfield  {author} {\bibinfo {author} {\bibfnamefont {E.}~\bibnamefont
  {Rico}}, \bibinfo {author} {\bibfnamefont {T.}~\bibnamefont {Pichler}},
  \bibinfo {author} {\bibfnamefont {M.}~\bibnamefont {Dalmonte}}, \bibinfo
  {author} {\bibfnamefont {P.}~\bibnamefont {Zoller}}, \ and\ \bibinfo {author}
  {\bibfnamefont {S.}~\bibnamefont {Montangero}},\ }\href {\doibase
  10.1103/PhysRevLett.112.201601} {\bibfield  {journal} {\bibinfo  {journal}
  {Phys. Rev. Lett.}\ }\textbf {\bibinfo {volume} {112}},\ \bibinfo {pages}
  {201601} (\bibinfo {year} {2014})},\ \Eprint {http://arxiv.org/abs/1312.3127}
  {arXiv:1312.3127 [cond-mat.quant-gas]} \BibitemShut {NoStop}%
\bibitem [{\citenamefont {Buyens}\ \emph {et~al.}(2014)\citenamefont {Buyens},
  \citenamefont {Haegeman}, \citenamefont {Van~Acoleyen}, \citenamefont
  {Verschelde},\ and\ \citenamefont {Verstraete}}]{Buyens:2013yza}%
  \BibitemOpen
  \bibfield  {author} {\bibinfo {author} {\bibfnamefont {B.}~\bibnamefont
  {Buyens}}, \bibinfo {author} {\bibfnamefont {J.}~\bibnamefont {Haegeman}},
  \bibinfo {author} {\bibfnamefont {K.}~\bibnamefont {Van~Acoleyen}}, \bibinfo
  {author} {\bibfnamefont {H.}~\bibnamefont {Verschelde}}, \ and\ \bibinfo
  {author} {\bibfnamefont {F.}~\bibnamefont {Verstraete}},\ }\href {\doibase
  10.1103/PhysRevLett.113.091601} {\bibfield  {journal} {\bibinfo  {journal}
  {Phys. Rev. Lett.}\ }\textbf {\bibinfo {volume} {113}},\ \bibinfo {pages}
  {091601} (\bibinfo {year} {2014})},\ \Eprint {http://arxiv.org/abs/1312.6654}
  {arXiv:1312.6654 [hep-lat]} \BibitemShut {NoStop}%
\bibitem [{\citenamefont {Buyens}\ \emph {et~al.}(2016)\citenamefont {Buyens},
  \citenamefont {Verstraete},\ and\ \citenamefont
  {Van~Acoleyen}}]{Buyens:2016ecr}%
  \BibitemOpen
  \bibfield  {author} {\bibinfo {author} {\bibfnamefont {B.}~\bibnamefont
  {Buyens}}, \bibinfo {author} {\bibfnamefont {F.}~\bibnamefont {Verstraete}},
  \ and\ \bibinfo {author} {\bibfnamefont {K.}~\bibnamefont {Van~Acoleyen}},\
  }\href {\doibase 10.1103/PhysRevD.94.085018} {\bibfield  {journal} {\bibinfo
  {journal} {Phys. Rev. D}\ }\textbf {\bibinfo {volume} {94}},\ \bibinfo
  {pages} {085018} (\bibinfo {year} {2016})},\ \Eprint
  {http://arxiv.org/abs/1606.03385} {arXiv:1606.03385 [hep-lat]} \BibitemShut
  {NoStop}%
\bibitem [{\citenamefont {Ba\~nuls}\ \emph {et~al.}(2013)\citenamefont
  {Ba\~nuls}, \citenamefont {Cichy}, \citenamefont {Jansen},\ and\
  \citenamefont {Cirac}}]{Banuls:2013jaa}%
  \BibitemOpen
  \bibfield  {author} {\bibinfo {author} {\bibfnamefont {M.~C.}\ \bibnamefont
  {Ba\~nuls}}, \bibinfo {author} {\bibfnamefont {K.}~\bibnamefont {Cichy}},
  \bibinfo {author} {\bibfnamefont {K.}~\bibnamefont {Jansen}}, \ and\ \bibinfo
  {author} {\bibfnamefont {J.~I.}\ \bibnamefont {Cirac}},\ }\href {\doibase
  10.1007/JHEP11(2013)158} {\bibfield  {journal} {\bibinfo  {journal} {JHEP}\
  }\textbf {\bibinfo {volume} {11}},\ \bibinfo {pages} {158} (\bibinfo {year}
  {2013})},\ \Eprint {http://arxiv.org/abs/1305.3765} {arXiv:1305.3765
  [hep-lat]} \BibitemShut {NoStop}%
\bibitem [{\citenamefont {Ba\~nuls}\ \emph {et~al.}(2015)\citenamefont
  {Ba\~nuls}, \citenamefont {Cichy}, \citenamefont {Cirac}, \citenamefont
  {Jansen},\ and\ \citenamefont {Saito}}]{Banuls:2015sta}%
  \BibitemOpen
  \bibfield  {author} {\bibinfo {author} {\bibfnamefont {M.~C.}\ \bibnamefont
  {Ba\~nuls}}, \bibinfo {author} {\bibfnamefont {K.}~\bibnamefont {Cichy}},
  \bibinfo {author} {\bibfnamefont {J.~I.}\ \bibnamefont {Cirac}}, \bibinfo
  {author} {\bibfnamefont {K.}~\bibnamefont {Jansen}}, \ and\ \bibinfo {author}
  {\bibfnamefont {H.}~\bibnamefont {Saito}},\ }\href {\doibase
  10.1103/PhysRevD.92.034519} {\bibfield  {journal} {\bibinfo  {journal} {Phys.
  Rev. D}\ }\textbf {\bibinfo {volume} {92}},\ \bibinfo {pages} {034519}
  (\bibinfo {year} {2015})},\ \Eprint {http://arxiv.org/abs/1505.00279}
  {arXiv:1505.00279 [hep-lat]} \BibitemShut {NoStop}%
\bibitem [{\citenamefont {Ba\~nuls}\ \emph {et~al.}(2020)\citenamefont
  {Ba\~nuls} \emph {et~al.}}]{Banuls:2019bmf}%
  \BibitemOpen
  \bibfield  {author} {\bibinfo {author} {\bibfnamefont {M.~C.}\ \bibnamefont
  {Ba\~nuls}} \emph {et~al.},\ }\href {\doibase 10.1140/epjd/e2020-100571-8}
  {\bibfield  {journal} {\bibinfo  {journal} {Eur. Phys. J. D}\ }\textbf
  {\bibinfo {volume} {74}},\ \bibinfo {pages} {165} (\bibinfo {year} {2020})},\
  \Eprint {http://arxiv.org/abs/1911.00003} {arXiv:1911.00003 [quant-ph]}
  \BibitemShut {NoStop}%
\bibitem [{\citenamefont {Kokail}\ \emph {et~al.}(2019)\citenamefont {Kokail}
  \emph {et~al.}}]{Kokail:2018eiw}%
  \BibitemOpen
  \bibfield  {author} {\bibinfo {author} {\bibfnamefont {C.}~\bibnamefont
  {Kokail}} \emph {et~al.},\ }\href {\doibase 10.1038/s41586-019-1177-4}
  {\bibfield  {journal} {\bibinfo  {journal} {Nature}\ }\textbf {\bibinfo
  {volume} {569}},\ \bibinfo {pages} {355} (\bibinfo {year} {2019})},\ \Eprint
  {http://arxiv.org/abs/1810.03421} {arXiv:1810.03421 [quant-ph]} \BibitemShut
  {NoStop}%
\bibitem [{\citenamefont {Ikeda}\ \emph
  {et~al.}(2023{\natexlab{c}})\citenamefont {Ikeda}, \citenamefont {Kharzeev},\
  and\ \citenamefont {Shi}}]{Ikeda:2023vfk}%
  \BibitemOpen
  \bibfield  {author} {\bibinfo {author} {\bibfnamefont {K.}~\bibnamefont
  {Ikeda}}, \bibinfo {author} {\bibfnamefont {D.~E.}\ \bibnamefont {Kharzeev}},
  \ and\ \bibinfo {author} {\bibfnamefont {S.}~\bibnamefont {Shi}},\ }\href
  {\doibase 10.1103/PhysRevD.108.074001} {\bibfield  {journal} {\bibinfo
  {journal} {Phys. Rev. D}\ }\textbf {\bibinfo {volume} {108}},\ \bibinfo
  {pages} {074001} (\bibinfo {year} {2023}{\natexlab{c}})},\ \Eprint
  {http://arxiv.org/abs/2305.05685} {arXiv:2305.05685 [hep-ph]} \BibitemShut
  {NoStop}%
\bibitem [{\citenamefont {Ikeda}(2023{\natexlab{d}})}]{Ikeda:2023uqy}%
  \BibitemOpen
  \bibfield  {author} {\bibinfo {author} {\bibfnamefont {K.}~\bibnamefont
  {Ikeda}},\ }\href@noop {} {\  (\bibinfo {year} {2023}{\natexlab{d}})},\
  \Eprint {http://arxiv.org/abs/2311.16297} {arXiv:2311.16297 [hep-th]}
  \BibitemShut {NoStop}%
\bibitem [{\citenamefont {Grieninger}\ \emph {et~al.}(2024)\citenamefont
  {Grieninger}, \citenamefont {Ikeda}, \citenamefont {Kharzeev},\ and\
  \citenamefont {Zahed}}]{Grieninger:2023ufa}%
  \BibitemOpen
  \bibfield  {author} {\bibinfo {author} {\bibfnamefont {S.}~\bibnamefont
  {Grieninger}}, \bibinfo {author} {\bibfnamefont {K.}~\bibnamefont {Ikeda}},
  \bibinfo {author} {\bibfnamefont {D.~E.}\ \bibnamefont {Kharzeev}}, \ and\
  \bibinfo {author} {\bibfnamefont {I.}~\bibnamefont {Zahed}},\ }\href
  {\doibase 10.1103/PhysRevD.109.016023} {\bibfield  {journal} {\bibinfo
  {journal} {Phys. Rev. D}\ }\textbf {\bibinfo {volume} {109}},\ \bibinfo
  {pages} {016023} (\bibinfo {year} {2024})}\BibitemShut {NoStop}%
\bibitem [{\citenamefont {Mishra}\ \emph {et~al.}(2020)\citenamefont {Mishra},
  \citenamefont {Thompson}, \citenamefont {Pooser},\ and\ \citenamefont
  {Siopsis}}]{Mishra:2019xbh}%
  \BibitemOpen
  \bibfield  {author} {\bibinfo {author} {\bibfnamefont {C.}~\bibnamefont
  {Mishra}}, \bibinfo {author} {\bibfnamefont {S.}~\bibnamefont {Thompson}},
  \bibinfo {author} {\bibfnamefont {R.}~\bibnamefont {Pooser}}, \ and\ \bibinfo
  {author} {\bibfnamefont {G.}~\bibnamefont {Siopsis}},\ }\href {\doibase
  10.1088/2058-9565/ab8f63} {\bibfield  {journal} {\bibinfo  {journal} {Quantum
  Sci. Technol.}\ }\textbf {\bibinfo {volume} {5}},\ \bibinfo {pages} {035010}
  (\bibinfo {year} {2020})},\ \Eprint {http://arxiv.org/abs/1912.07767}
  {arXiv:1912.07767 [quant-ph]} \BibitemShut {NoStop}%
\bibitem [{\citenamefont {Czajka}\ \emph {et~al.}(2022)\citenamefont {Czajka},
  \citenamefont {Kang}, \citenamefont {Tee},\ and\ \citenamefont
  {Zhao}}]{Czajka:2022plx}%
  \BibitemOpen
  \bibfield  {author} {\bibinfo {author} {\bibfnamefont {A.~M.}\ \bibnamefont
  {Czajka}}, \bibinfo {author} {\bibfnamefont {Z.-B.}\ \bibnamefont {Kang}},
  \bibinfo {author} {\bibfnamefont {Y.}~\bibnamefont {Tee}}, \ and\ \bibinfo
  {author} {\bibfnamefont {F.}~\bibnamefont {Zhao}},\ }\href@noop {} {\
  (\bibinfo {year} {2022})},\ \Eprint {http://arxiv.org/abs/2210.03062}
  {arXiv:2210.03062 [hep-ph]} \BibitemShut {NoStop}%
\bibitem [{\citenamefont {Farrell}\ \emph {et~al.}(2023)\citenamefont
  {Farrell}, \citenamefont {Illa}, \citenamefont {Ciavarella},\ and\
  \citenamefont {Savage}}]{Farrell:2023fgd}%
  \BibitemOpen
  \bibfield  {author} {\bibinfo {author} {\bibfnamefont {R.~C.}\ \bibnamefont
  {Farrell}}, \bibinfo {author} {\bibfnamefont {M.}~\bibnamefont {Illa}},
  \bibinfo {author} {\bibfnamefont {A.~N.}\ \bibnamefont {Ciavarella}}, \ and\
  \bibinfo {author} {\bibfnamefont {M.~J.}\ \bibnamefont {Savage}},\
  }\href@noop {} {\  (\bibinfo {year} {2023})},\ \Eprint
  {http://arxiv.org/abs/2308.04481} {arXiv:2308.04481 [quant-ph]} \BibitemShut
  {NoStop}%
\bibitem [{\citenamefont {Farrell}\ \emph {et~al.}(2024)\citenamefont
  {Farrell}, \citenamefont {Illa}, \citenamefont {Ciavarella},\ and\
  \citenamefont {Savage}}]{Farrell:2024fit}%
  \BibitemOpen
  \bibfield  {author} {\bibinfo {author} {\bibfnamefont {R.~C.}\ \bibnamefont
  {Farrell}}, \bibinfo {author} {\bibfnamefont {M.}~\bibnamefont {Illa}},
  \bibinfo {author} {\bibfnamefont {A.~N.}\ \bibnamefont {Ciavarella}}, \ and\
  \bibinfo {author} {\bibfnamefont {M.~J.}\ \bibnamefont {Savage}},\
  }\href@noop {} {\  (\bibinfo {year} {2024})},\ \Eprint
  {http://arxiv.org/abs/2401.08044} {arXiv:2401.08044 [quant-ph]} \BibitemShut
  {NoStop}%
\bibitem [{\citenamefont {Johansson}\ \emph {et~al.}(2012)\citenamefont
  {Johansson}, \citenamefont {Nation},\ and\ \citenamefont
  {Nori}}]{JOHANSSON20121760}%
  \BibitemOpen
  \bibfield  {author} {\bibinfo {author} {\bibfnamefont {J.}~\bibnamefont
  {Johansson}}, \bibinfo {author} {\bibfnamefont {P.}~\bibnamefont {Nation}}, \
  and\ \bibinfo {author} {\bibfnamefont {F.}~\bibnamefont {Nori}},\ }\href
  {\doibase https://doi.org/10.1016/j.cpc.2012.02.021} {\bibfield  {journal}
  {\bibinfo  {journal} {Computer Physics Communications}\ }\textbf {\bibinfo
  {volume} {183}},\ \bibinfo {pages} {1760} (\bibinfo {year}
  {2012})}\BibitemShut {NoStop}%
\bibitem [{\citenamefont {Streltsov}\ \emph {et~al.}(2011)\citenamefont
  {Streltsov}, \citenamefont {Kampermann},\ and\ \citenamefont
  {Bru\ss{}}}]{PhysRevLett.107.170502}%
  \BibitemOpen
  \bibfield  {author} {\bibinfo {author} {\bibfnamefont {A.}~\bibnamefont
  {Streltsov}}, \bibinfo {author} {\bibfnamefont {H.}~\bibnamefont
  {Kampermann}}, \ and\ \bibinfo {author} {\bibfnamefont {D.}~\bibnamefont
  {Bru\ss{}}},\ }\href {\doibase 10.1103/PhysRevLett.107.170502} {\bibfield
  {journal} {\bibinfo  {journal} {Phys. Rev. Lett.}\ }\textbf {\bibinfo
  {volume} {107}},\ \bibinfo {pages} {170502} (\bibinfo {year}
  {2011})}\BibitemShut {NoStop}%
\bibitem [{\citenamefont {Modi}(2014)}]{doi:10.1142/S123016121440006X}%
  \BibitemOpen
  \bibfield  {author} {\bibinfo {author} {\bibfnamefont {K.}~\bibnamefont
  {Modi}},\ }\href {\doibase 10.1142/S123016121440006X} {\bibfield  {journal}
  {\bibinfo  {journal} {Open Systems \& Information Dynamics}\ }\textbf
  {\bibinfo {volume} {21}},\ \bibinfo {pages} {1440006} (\bibinfo {year}
  {2014})}\BibitemShut {NoStop}%
\bibitem [{\citenamefont {Brodutch}\ and\ \citenamefont
  {Terno}(2011)}]{PhysRevA.83.010301}%
  \BibitemOpen
  \bibfield  {author} {\bibinfo {author} {\bibfnamefont {A.}~\bibnamefont
  {Brodutch}}\ and\ \bibinfo {author} {\bibfnamefont {D.~R.}\ \bibnamefont
  {Terno}},\ }\href {\doibase 10.1103/PhysRevA.83.010301} {\bibfield  {journal}
  {\bibinfo  {journal} {Phys. Rev. A}\ }\textbf {\bibinfo {volume} {83}},\
  \bibinfo {pages} {010301} (\bibinfo {year} {2011})}\BibitemShut {NoStop}%
\bibitem [{\citenamefont {Fukushima}(2008)}]{PhysRevD.77.114028}%
  \BibitemOpen
  \bibfield  {author} {\bibinfo {author} {\bibfnamefont {K.}~\bibnamefont
  {Fukushima}},\ }\href {\doibase 10.1103/PhysRevD.77.114028} {\bibfield
  {journal} {\bibinfo  {journal} {Phys. Rev. D}\ }\textbf {\bibinfo {volume}
  {77}},\ \bibinfo {pages} {114028} (\bibinfo {year} {2008})}\BibitemShut
  {NoStop}%
\bibitem [{\citenamefont {Costa}\ \emph {et~al.}(2007)\citenamefont {Costa},
  \citenamefont {{de Sousa}}, \citenamefont {Ruivo},\ and\ \citenamefont
  {Kalinovsky}}]{COSTA2007431}%
  \BibitemOpen
  \bibfield  {author} {\bibinfo {author} {\bibfnamefont {P.}~\bibnamefont
  {Costa}}, \bibinfo {author} {\bibfnamefont {C.}~\bibnamefont {{de Sousa}}},
  \bibinfo {author} {\bibfnamefont {M.}~\bibnamefont {Ruivo}}, \ and\ \bibinfo
  {author} {\bibfnamefont {Y.}~\bibnamefont {Kalinovsky}},\ }\href {\doibase
  https://doi.org/10.1016/j.physletb.2007.02.045} {\bibfield  {journal}
  {\bibinfo  {journal} {Physics Letters B}\ }\textbf {\bibinfo {volume}
  {647}},\ \bibinfo {pages} {431} (\bibinfo {year} {2007})}\BibitemShut
  {NoStop}%
\bibitem [{\citenamefont {Korepin}(1979)}]{Korepin:1979qq}%
  \BibitemOpen
  \bibfield  {author} {\bibinfo {author} {\bibfnamefont {V.~E.}\ \bibnamefont
  {Korepin}},\ }\href {\doibase 10.1007/BF01028501} {\bibfield  {journal}
  {\bibinfo  {journal} {Theor. Math. Phys.}\ }\textbf {\bibinfo {volume}
  {41}},\ \bibinfo {pages} {953} (\bibinfo {year} {1979})}\BibitemShut
  {NoStop}%
\bibitem [{\citenamefont {Klevansky}(1992)}]{Klevansky:1992qe}%
  \BibitemOpen
  \bibfield  {author} {\bibinfo {author} {\bibfnamefont {S.~P.}\ \bibnamefont
  {Klevansky}},\ }\href {\doibase 10.1103/RevModPhys.64.649} {\bibfield
  {journal} {\bibinfo  {journal} {Rev. Mod. Phys.}\ }\textbf {\bibinfo {volume}
  {64}},\ \bibinfo {pages} {649} (\bibinfo {year} {1992})}\BibitemShut
  {NoStop}%
\bibitem [{\citenamefont {Bernard}\ \emph {et~al.}(1987)\citenamefont
  {Bernard}, \citenamefont {Meissner},\ and\ \citenamefont
  {Zahed}}]{PhysRevD.36.819}%
  \BibitemOpen
  \bibfield  {author} {\bibinfo {author} {\bibfnamefont {V.}~\bibnamefont
  {Bernard}}, \bibinfo {author} {\bibfnamefont {U.-G.}\ \bibnamefont
  {Meissner}}, \ and\ \bibinfo {author} {\bibfnamefont {I.}~\bibnamefont
  {Zahed}},\ }\href {\doibase 10.1103/PhysRevD.36.819} {\bibfield  {journal}
  {\bibinfo  {journal} {Phys. Rev. D}\ }\textbf {\bibinfo {volume} {36}},\
  \bibinfo {pages} {819} (\bibinfo {year} {1987})}\BibitemShut {NoStop}%
\bibitem [{\citenamefont {Lawrence}\ and\ \citenamefont
  {Yamauchi}(2023)}]{PhysRevD.107.114505}%
  \BibitemOpen
  \bibfield  {author} {\bibinfo {author} {\bibfnamefont {S.}~\bibnamefont
  {Lawrence}}\ and\ \bibinfo {author} {\bibfnamefont {Y.}~\bibnamefont
  {Yamauchi}},\ }\href {\doibase 10.1103/PhysRevD.107.114505} {\bibfield
  {journal} {\bibinfo  {journal} {Phys. Rev. D}\ }\textbf {\bibinfo {volume}
  {107}},\ \bibinfo {pages} {114505} (\bibinfo {year} {2023})}\BibitemShut
  {NoStop}%
\bibitem [{\citenamefont
  {Ikeda}(2023{\natexlab{e}})}]{Ikeda_Quantum_Energy_Teleportation_2023}%
  \BibitemOpen
  \bibfield  {author} {\bibinfo {author} {\bibfnamefont {K.}~\bibnamefont
  {Ikeda}},\ }\href
  {https://github.com/IKEDAKAZUKI/Quantum-Energy-Teleportation.git} {\enquote
  {\bibinfo {title} {{Quantum Energy Teleportation with Quantum Computers}},}\
  } (\bibinfo {year} {2023}{\natexlab{e}})\BibitemShut {NoStop}%
\end{thebibliography}%

\if{
\appendix
\section{Continuous NJL Transition}
The resulting figures show the NJL model for $m=10$. For this scenario it is clear that the phase transition is continuous and qualitatively displays similar behaviour to the $N=4$ Schwinger model. As a final emphasis of our conclusions, there is a direct link between the quantum correlation, and the teleported energy. Interestingly for this scenario, the asymmetry in quantum discord is more prevalent than for the Schwinger model. 
\begin{figure}[H]
    \centering
    \includegraphics[width=\linewidth]{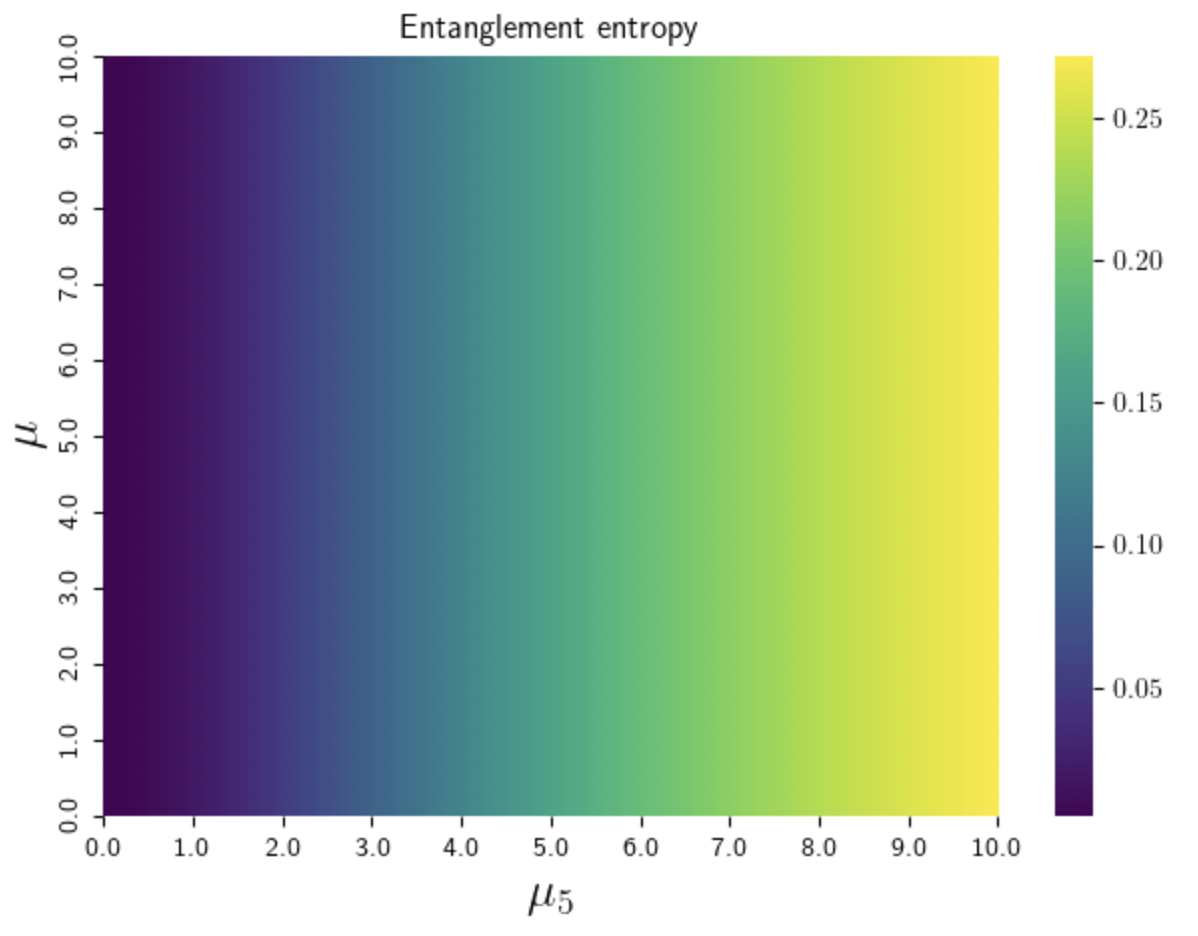}
    \caption{Phase diagram of the NJL, drawn with the entanglement entropy for $m=10$. There is a smooth transition as $\mu_5$ increases.}
    \label{fig:enter-label}
\end{figure}

\begin{figure}[H]
    \centering
    \includegraphics[width=\linewidth]{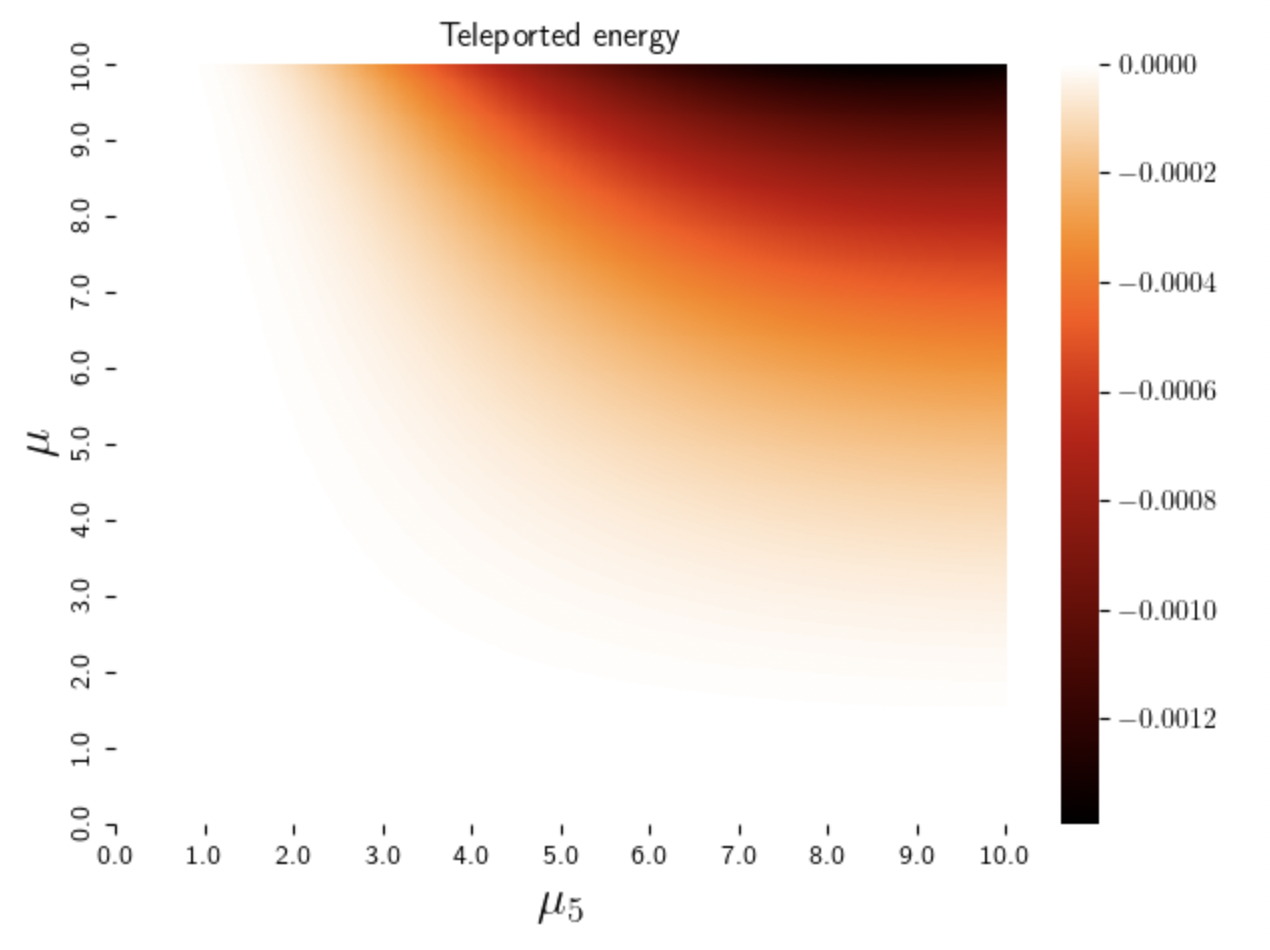}
    \caption{Theoretical teleported energy of the NJL for $m=10$. A similarly smooth transition is seen here.}
    \label{fig:enter-label}
\end{figure}
It is seen that for small $\mu$ the teleported energy does not reproduce the transition seen in the entanglement entropy. This is an interesting deviation from the qualitative behaviour observed.

\begin{figure}[H]
    \centering
    \includegraphics[width=\linewidth]{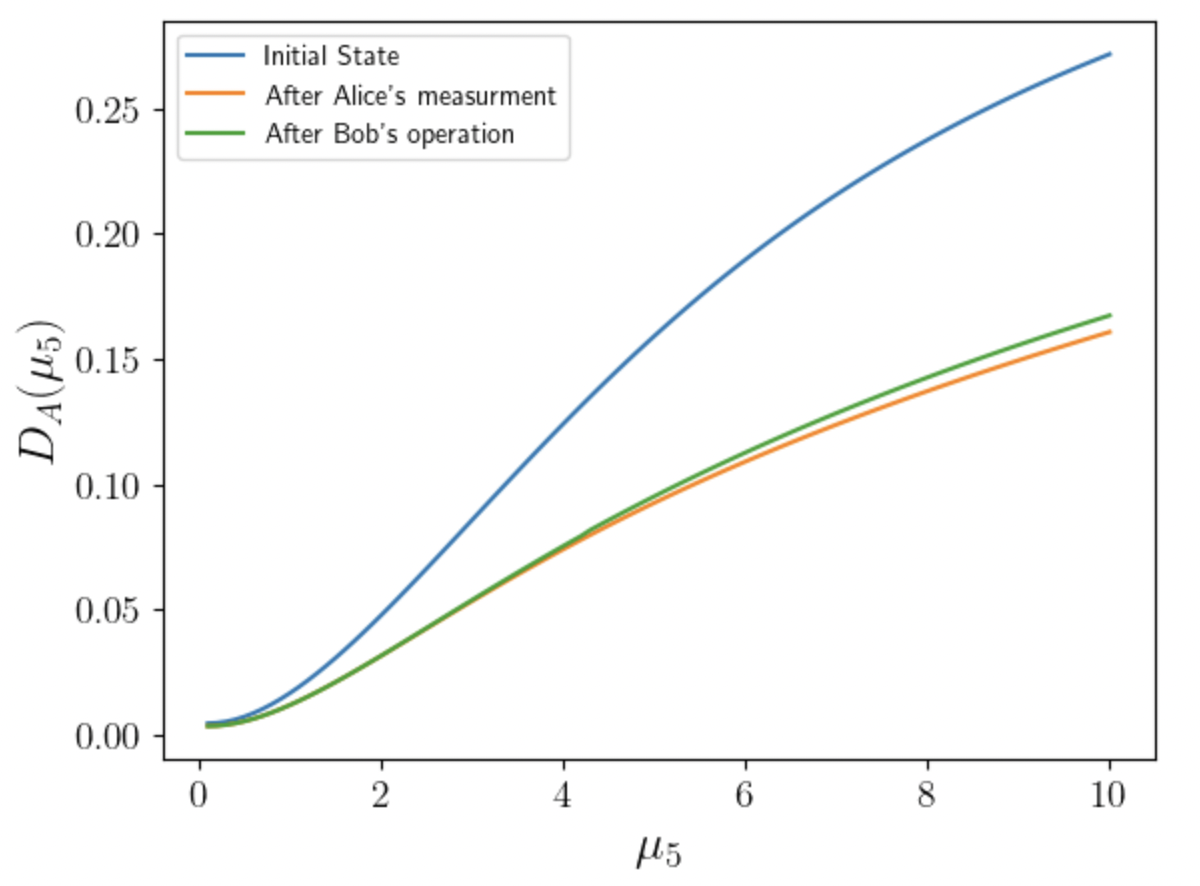}
    \caption{$D_A(\mu_5)$ for $m=10$, $a=g=1$ and $\mu=5$. This plot behaves displays the observed behaviour of quantum discord in the QET protcol throughout the paper.}
    \label{NJL_DA_m10}
\end{figure}

\begin{figure}[H]
    \centering
    \includegraphics[width=\linewidth]{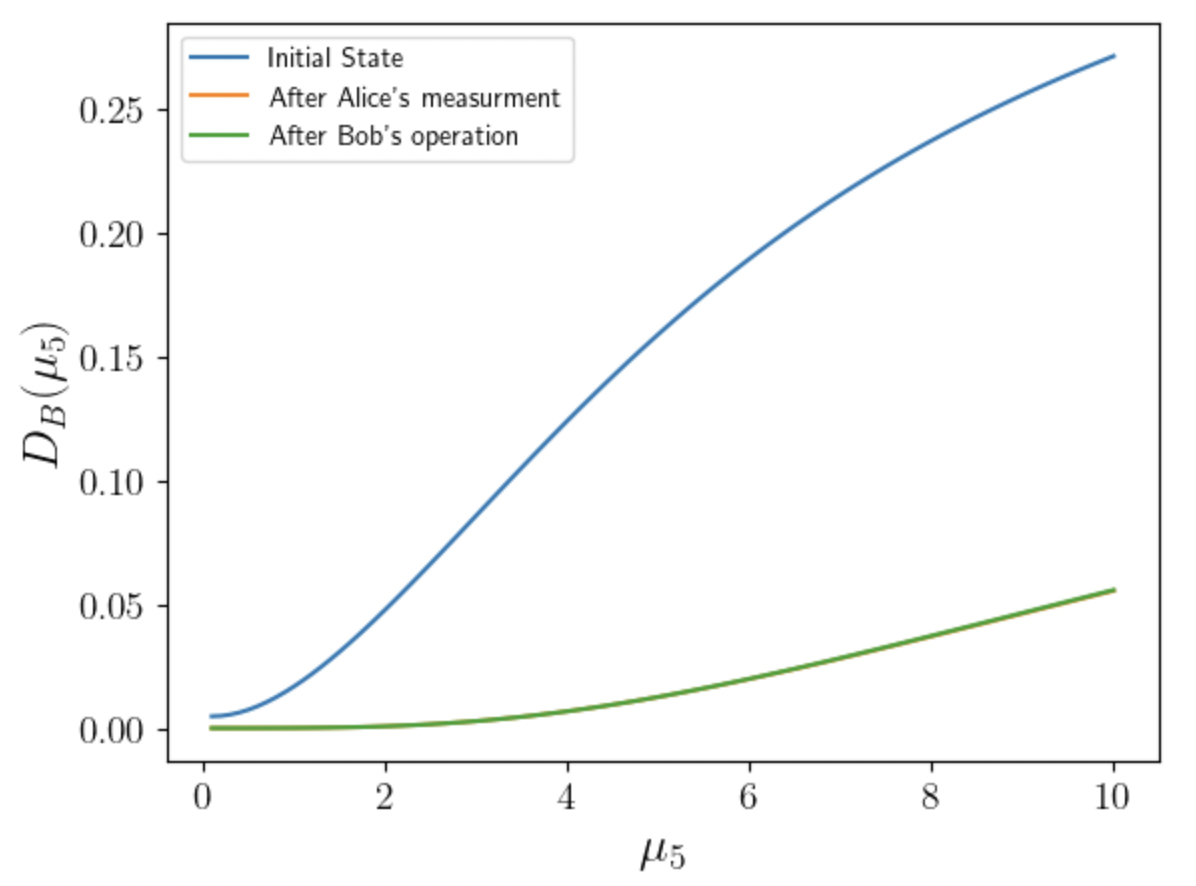}
    \caption{$D_B(\mu_5)$ for $m=10$, $a=g=1$ and $\mu=5$. Notice the asymmetry relative to $D_A(\mu_5)$ and the non-increase in quantum correlation after Bob's operation.}
    \label{NJL_DB_m10}
\end{figure}
The asymmetry in the quantum discord reveals the change in growth of quantum correlation, as changes from concave to convex depending on which system is being measured.
}\fi
\end{document}